Evidence for a New Family of 2-D Honeycomb Surface Reconstructions on Si(111)

J. E. Demuth

Naples, Florida 34114

ABOUT THIS DOCUMENT:

This manuscript and its supplement have been vetted but were ultimately rejected. As a result I decided not to waste another 9 months of re-editing and peer review by submitting it to another journal, but to simply place it in arXiv as a way for others to access these findings.

This paper is a longer version of an earlier short manuscript ( arXiv:1905.12416v1 ) that was submitted as a regular article ( not a letter or rapid communication.) This longer paper also considered additional systems as well as a broader perspective of these new findings. Last year's short manuscript was brief and highlighted the possibility of a new topologically identical structure to the long established Si111 7x7 structure. This new structure was never considered before but was very similar to the structure of a monolayer of silicon on a Ag111 surface which is widely believed to be a silicene.

Several referees provided useful comments to this short paper: in particular, that I needed to present more definitive results to prove this structure was correct. One referee also indicated the need for theoretical calculations to validate this structure.

As a result I expanded the original paper to address these concerns as well as include this family structure not just the 7x7 and insights into this old system that would be useful to a new generation of researchers. Included in this was an expanded discussion of many of the most plausible reasons why the earlier calculations did not predict this new structure and what sort of calculations beyond DFT would be needed. It also quoted earlier authors who noted the paradoxical aspects and limitations of their results. This resulted in a greatly expanded paper and more figures to make these points. A supplement was added to present much of this additional detail as well as further analyses outside the points made in the original shorter paper. This included a new Patterson analysis that confirms this new structure as well as my analysis of many of the earlier measurements and discussed their shortcomings. This supplement now appears on page 38 at the end of main manuscript as required by arXiv.org

While a majority of referees felt that I had met their concerns, there was some descent and the paper was not accepted on the basis that it was a review paper and not suitable for publication in this journal.

I appealed this decision and my appeal was rejected . Although disappointed by that decision, I am grateful for the referees' suggestions which improved this paper.

This paper should be a useful reference and background for future work. I eventually discovered why DFT calculations found the wrong structure as well as the fundamental properties of the Si111 surface that allow this new structure to occurs. These are being submitted shortly.

J. E. Demuth,

February 20, 2020

Evidence for a New Family of 2-D Honeycomb Surface Reconstructions on Si(111) *

J. E. Demuth**

Naples, Florida 34114

ABSTRACT: A new silicene-like family of reconstructed surfaces on Si111 are discussed which appears to be a polymorph to the well know family of surface reconstructions best epitomized by the 7x7 surface. Several experimental features are discussed which lead to this new conclusion, as are several recently established limitations of density functional theory that may currently limit its ability to predict such 2-D surface structures. The atomic locations of the surface state charge densities from several Scanning Tunneling Spectroscopy studies provide the basis for this new structure which is supported by several previous measurements. The new structure appears to favor a faulted supercell honeycomb motif having a delocalized 2-D state that enables greater sp$^2$ character. This top layer has an unusual periodic p-orbital structure that also interdigitates with the terminal bulk 'dangling bonds' to create a 2-D π-bonded structure. This new polymorphic structure resolves many long standing paradoxes of the 7x7 surface while its unusual bonding and structure may help better understand some silicenes as well as the behavior of other 2-D phases on 111 surfaces.

Renewed interest in (111) surfaces has intensified with the recent discovery of a variety of hexagonal 2-D phases that are of potential technological importance.[1] Graphene[2] and silicene[3] are leading examples of 2-D materials whose hexagonal symmetry and planarity can lead to the formation of Dirac cones and Dirac fermions, i.e., mass less carriers, as well as other unusual properties.[4] Such properties in silicene are attributed to a more planar sp2 hybridized 2-D structure than exists for crystalline Silicon where tetrahedral (sp3) bonding occurs. In contrast the reconstructed Si surfaces studied decades earlier were explained based on the same bulk-like tetrahedral bonding. The silicenes were thereby thought to be a distinctly different 2-D system than the reconstructed Si surfaces. Evidence is presented here that the top layer of Si on a 111 surface may not behave as a bulk-like layer as believed, but reflects 2-D properties similar to freestanding and mono-layer silicenes. A new structural polymorph is proposed for the 7x7 and its corresponding 2n+1 x 2n+1 family of reconstructions that has an underlying honeycomb structure and which is topologically identical to the currently accepted structure(s). The wide body of evidence presented here and in the Supplement provides evidence for a new structure, and as such, provides a starting point for further investigations.

This new perspective arose initially out of an effort to better understand the surface states of Si111 7x7, based on new theoretical insights and more recent experimental results, and in particular, one surface state that has never been understood and which has been largely forgotten. As such, it is referred as the 'forgotten surface state' or FSS. A new holistic examination of the measured surface geometric and electronic structure together with theoretical results suggest a different structure and leads to the conclusion that a different mode of bonding occurs in the 7x7 than is currently believed. Instead of having a bulk like covalently bonded surface reconstruction, experiments suggest an unusual 2-D supercell type of honeycomb layer that has enhanced sp2 character. This new honeycomb ad-layer has an unanticipated type of 2-D π−bonding that also produces the FSS.

* jedemuth7x7@gmail.com

In relating the 7x7 surface reconstructions to alternate structures, it is important to understand the evolution of its now widely accepted structure. The 7x7 had been a subject of interest for many years but rapidly escalated in the early 1980's. The breakthrough came with the observation of the topology of the Si 7x7 using Scanning Tunneling Microscopy, STM. This revealed 12 atomic scale features in the unit cell, called adatoms, and an inhomogeniously distorted top layer.[5] Later, diffraction indicated a faulted unit cell: and a bulk-like stacking fault on one side of the unit cell was proposed.[6] An analysis of the TEM diffraction intensities by Takayanagi led him to propose the Dimer-Adatom-Stacking fault model, DAS. [7,8] This model was evaluated against other leading models, and of all the models considered was found to provide the best overall agreement. It was also evaluated using many other approaches, for example, ion scattering,[9] model and semi-empirical electronic structure calculations[10,11] to name a few. Shortly thereafter, atom-resolved spectroscopy using STM indicated the locations of the surface states to confirm the nature of electron redistribution and bonding in the DAS structure.[12] Full local density, LD, calculations of the DAS structure finally confirmed these features and further indicated that the 7x7 DAS structure produced the lowest system energy. [13-22] Such LD calculations of surface structures are effective one-electron models that have successfully worked for bulk materials, but as discussed later may have starting configurations or contain approximations that may not be suitable for certain 2-D structures.

The widely accepted DAS structure arises from a terminal layer with covalent, tetrahedrally bonded silicon atoms that minimizes the number of dangling $sp^3$ bonds. Dimers enable a bulk-like stacking fault, i.e., a 180 ° rotation of the top layer that allows two side of the top layer of Si atoms to rebond to one another. Such dimer rebonding produces large strain which is offset by strong electron pair bonds, i.e., covalent bonding.[10,11,22] The local building blocks of the DAS model, the adatoms, dimers and corner holes, describe a family of 2n+1 structures called the DAS family (3x3, 5x5, 7x7, 9x9, etc.) [23] which show similar geometric features. The 7x7 is the easiest to form and has been the most widely studied. Fig. 1 shows an overview of the main features of a few members of the DAS family of structures.

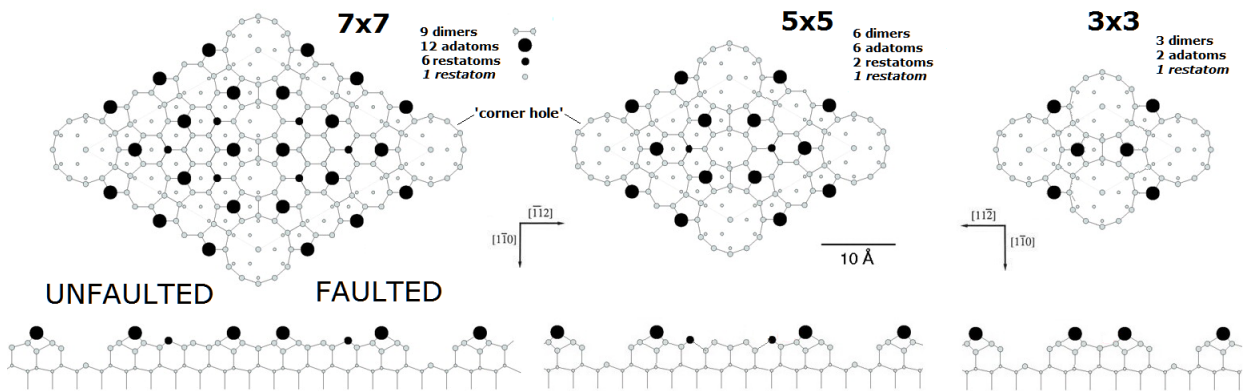

FIG. 1: Schematic of the DAS family of structures up to the 7x7 structure. Note that while the 3x3 is frequently proposed as the least energetically favored of all the structures, experimentally it is not observed as a complete or standalone structure: usually only 2 corner holes form.[24] Note that the DAS structure does not have inversion symmetry across the stacking faults where the dimers are.

CURRENT STATUS OF THE GEOMETRIC STRUCTURE:

The atomic structure is central to understanding the nature of the DAS structure, and for that matter, any other 2-D surface structure. In retrospect however, the structure of the 7x7 is not as clear as it appeared 34 years ago in view of numerous theoretical and experimental advances.[19, 21, 25-29] Various structural as well as electronic paradoxes are discussed here to give the reader a sense of the current level of issues and uncertainties in the DAS structure. As presented here such paradoxes provide clues for an alternative structures, in particularly, a new family of structure that has never been considered.

Fig. 2 shows a detailed portion of the 7x7 structure recently calculated within the LD approximation using plane waves and the PBE exchange functional.[22] This spin unrestricted calculation is perhaps the most advanced and accurate of all 7x7 calculations to date. This side view confirms the complexity of the 7x7, particularly in the vertical direction. In addition to the lateral distortions arising from the dimers, the top layer shows 6 distinct layers of atoms on each side of the unit cell. The underlying bi-layers (layers 4 and 5) are still displaced 0.13 A from their bulk positions. However, that still leaves 5 distinct planes of atoms in the top layer that differ slight on each side of the unit cell. Experimental determining these displacements is important but complicated by the large number of structural parameters involved. Ion scattering,[9] X-ray studies,[31,32] Low Energy Electron Diffraction, LEED[30], Reflection High Energy Electron Diffraction. RHEED[33,34] and Reflection High Energy Positron Diffraction, RHEPD[34,35] measurements all suggest similar expansions in the top layers. The experimental accuracy of some probes, such as RHEED, have been questioned due to their limited sensitivity to the top surface layer.[34] As a result it is believed that HREPD provides more accurate vertical heights for the outermost layers[34] than RHEED as discussed later.

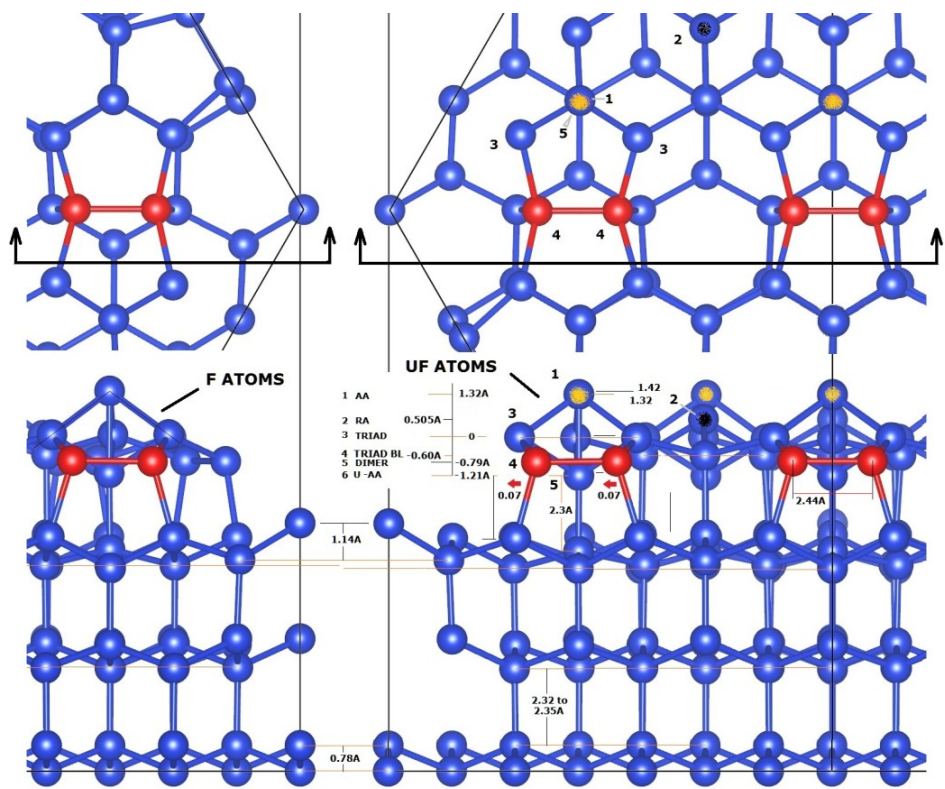

FIG. 2: ( Color online) Top and side view of a recent relaxed, fully converged DAS structure.[22] The overall accuracy of the dimensions show is to within 0.02A. The red atoms along the unit cell boundary that form dimers are 2.44A apart, versus 2.35A in the bulk, and have larger distortions in tetrahedral bond angles. This creates the large stain and distortions that propagate into the subsurface.

An important distinction in experimental structural analyses especially for such complex systems is whether one is seeking a approximate structure to understand the scheme of bonding, or a detailed structure suitable for extracting accurate bond lengths and bond angles, for example, to determine the state of hybridization of certain atoms.  Detailed structures have generally been solved via some sort of refinement process essentially 'testing' proposed structures involving some type of regression analysis: the final structure being the one providing the 'best' agreement . Finding an optimized structure requires a starting point based or a general scheme of bonding, intuition or other information that defines atom locations. Theoretical optimization that calculate the forces and move atoms to improve their energy can be more direct so long as all atoms are unrestricted in their motions or symmetry, the calculation is accurate and energy barriers to alternative structures can be overcome.

The first paradox of the 7x7 was that a Si bi-layer stacking fault was not observed in x-ray standing wave measurements in contradiction to the DAS model.[36]  A bulk like stacking fault with mixed cubic symmetry was only consistent with diffraction.[6]  Based on this, a stacking fault together with a faulted arrangement of adatoms were part of the structural model that also included dimers.  A summary of all such models at the time is provided by Takayanagi.[7]  His TEM diffraction measurements and a Patterson analysis of them were used to validate the 7x7 DAS structures over other proposed structures.[8]  He then optimized the DAS structure using his measured TEM diffraction intensities.  The new model proposed here is a faulted honeycomb model and does not have a bulk-like  stacking fault. Reconsidering this Patterson analysis and applying it to such a hexagonal honeycomb structure is thereby warranted .

An underlying issue in this TEM based diffraction analysis is that at normal incidence  an 'almost' 6 fold symmetry was observed with evidence of strong (non-kinematic) scattering by the electrons. To mitigate this the sample was rotated 8 ° and the data of 'equivalent' 6 fold symmetries averaged. This symmetry assumption has been used in most subsequent analyses as it greatly simplifies the large number of free parameters in any structural model. However, as discussed here,  this is an important limitation.

The  Patterson function is a Fourier transform of the diffraction intensities that produces N(N-1) features from each unique atom, N, of a structure  that is used to produce a Patterson Map, PM.[37,38]  The phase information that arises in diffraction from the positions of different atoms of the original structure is lost and cannot be recovered.  However the PM creates a complex set of  correlation pairs of atoms that are mapped  relative to the unit cell. This arises  from the nature of the pair correlation function generated by the Fourier transform.  The periodicity and relative locations of the PM peaks can be  examined to define possible atom locations.   For the top 5 layers shown in Fig. 2, these atoms would produce 5550 Patterson peaks  or when reduced by 6-fold symmetry corresponds to ~600 peaks. The Fourier synthesis also typically produces relatively broad peaks since only a fraction of all the diffraction peaks are measured and used in such analyses. Also an insufficient number of peaks used in the PM can also lead to 'ghost' features due to incomplete Fourier synthesis.[37]

It is for these reasons that strategies are practiced that reduce the uncertainties of a Patterson analysis. In particular, pairs of features that fall along particular symmetry lines called Harker planes in 3D or lines in 2D are used to isolate particular atom features.[39] Also heavy atoms can be readily identified since their high electron density leads to more intense Patterson peaks. Since neither of these arise for the Si DAS structure, interpreting the PM features becomes problematical. In such cases other methods such as a Keating analysis are used as an intermediate tool to model and identify possible structures. The final structure is then confirmed by a detailed x-ray structural refinement now that the general features have been determined. However, the new structure proposed here does has a unique symmetry properties that can aid in a Patterson analysis. However, as discussed later the

Takayanagi used his measured 7x7 TEM diffraction intensities to create a 'partial' Patterson function but excluded the 1/7th and 2/7 fractional order beams (and strong integral order beams) due to the strong background intensity of the bulk diffraction features. The analysis included the other 460 fractional order beams many of which were averaged for 6-fold symmetry.[8] The peaks in the Patterson function were then compared to various pairs of atoms of the 7x7 models to validate the structure. The intensities of the Patterson peaks can be positive or negative depending on the various diffraction intensities and their relative locations.

Fig. 3 (a) shows the calculated 7x7 DAS structure [22] and (b) shows Takayanagi's derived Patterson Map. [8] The small peak with a spacing of 2/3a was associated with the existence of dimers and corresponds to the peak with an intensity of '74' in the enlarged 'reduced' section of the PM shown in (c). However, the many peaks in a PM, makes selecting unique periodic features or atom pair vectors problematical. The presence of a dimer in the 7x7 was chemically attractive and this small feature in the PM could be identified with it. The dimer was then included in a structural model, from which a new set of diffraction intensities were calculated, then optimized by displacing atoms slightly. This optimized diffraction pattern was then Patterson mapped again and is shown enlarged in (d). The PMs were also created from the calculated Intensities of the other proposed structures: all those with dimers produced nearly identical PMs to the optimized DAS in Fig. 3 (d). All assumed models with ordered adatoms as seen in the STM image produced an intense feature/peak at 2a in their PMs.

In comparing the experimental PMs of the DAS structure in (c) to the calculated PM for the optimized DAS structure shown in (d), two distinguishing features are not found in the optimized structure. The second most intense experimental PM peak circled in (c) is not found as well as several weak pairs of peaks in the [11] directions that are repeated frequently as indicated by a shorter lines. These pairs of PM peaks are 1/3a apart and can reflect the lateral displacement of the abc stacking in the substrate.

Overall the optimized PM in (d) does maintain the intense adatom feature observed experimentally but does not improve the intensity and presence of the feature associated with the dimer. A few other PM peaks roughly correspond to features in the experimental map but were never discussed or related to features in the DAS structure. Overall the correspondence to the optimized PM is reasonable but not compelling.

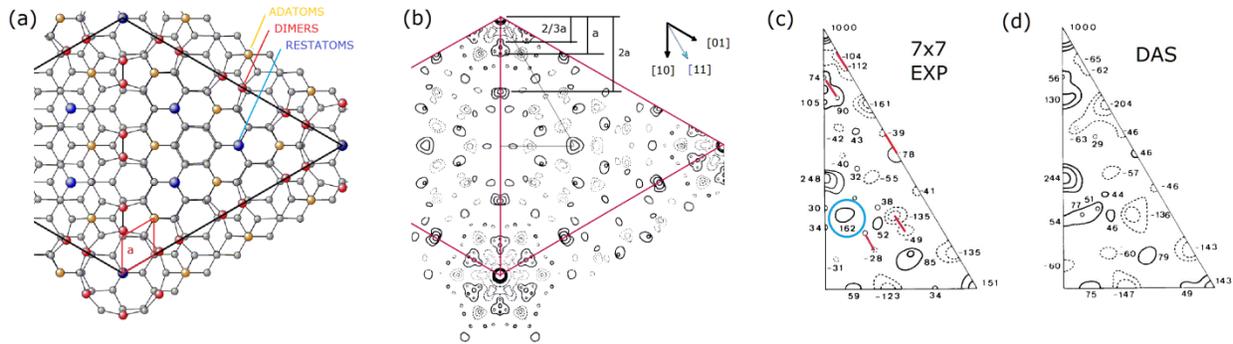

FIG. 3: (Color online) Section of the calculated DAS structure (a) compared to the experimentally derive 7x7 Patterson map (b) and enlarged section in(c). (d) is the PM from the 'optimized' DAS structure which best matches the TEM diffraction intensities.[8] The second most intense peak in the PM was never identified and is circled in blue in 3(c).

The origin of the circled PM feature in Fig. 3(c) and several other related features can be understood by considering the symmetry and structure of the new honeycomb based 7x7 polymorph which is discussed in more detail later. Fig. 4(a) shows the new 7x7 polymorph and Takayanagi's experimentally determined PM in (b). This 2-D honeycomb lattice with adatoms consists of two supercells with complete mirror symmetry across the unit cell boundary. The experimental PM in (b) again reflects the same repeat patterns of the substrate and adatoms as found for the DAS shown at the top in red. The 9 horizontal blue lines in (b) show pair correlations of the lower (blue) atoms along one diagonal of the unit cell boundaries. They correspond to the down atoms of the alternating up and down atoms of the 2-D honeycomb shown in (c) that are mirrored on the other side of the unit cell. This reflection symmetry creates a supercell honeycomb structure whose unit cell boundaries demark Harker planes.[39]

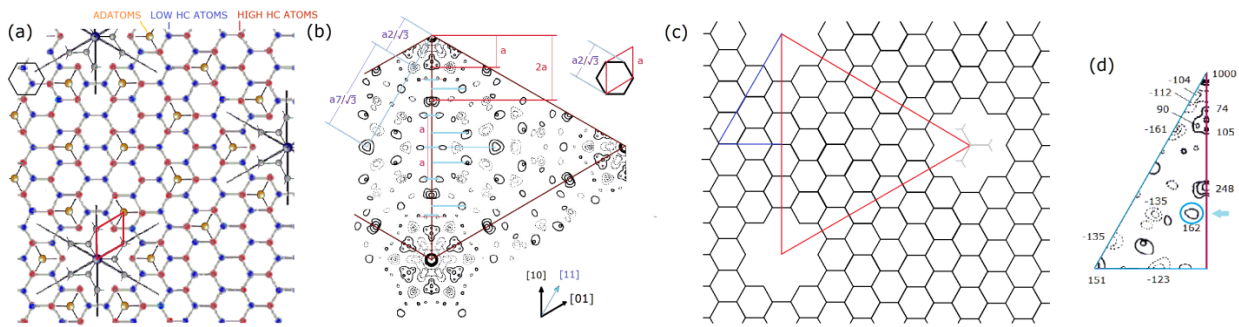

FIG. 4: (Color online) Section of a new polymorph of the 7x7 (a) compared to the experimental 7x7 Patterson map[8] (b). The blue horizontal lines about the <1-10> direction correspond to (blue) atom positions in (a) and half the atoms in the 2-D honeycomb lattice shown in (c). (d) shows the reduced PM [8] where the blue circle again shows a prominent Patterson peak not found for the DAS model.

The presence of Harker planes now allow a simplified interpretation of the PM features. The 9 (blue) pairs of features across each triangular supercell boundary are with atom pairs of the three Harker planes. One of these correspond to the prominent circled peak in (d) found experimentally but missing for the DAS model. The honeycomb motif of the polymorph strikingly fits the experimental PM with ~82% of the peaks showing a correspondence to atom positions of the polymorph. Such 'Harker' symmetry planes were experimentally identified at the same time from the non-linear surface optical response of the 7x7 [64] but went unnoticed in all subsequent work.

The variations of the pair separations and intensity along the [10] direction ( i.e., the unit cell boundary) can reflect  structural variations that modulate the diffraction intensities along this direction.  For the atoms in and around the corner hole,  this can cause the  PM peak to broaden and shifted due to distortions that create phase changes and modulation to the  experimental diffraction intensities.  As discussed later  the atoms around the corner hole are likely shifted down and closer to the corner hole to facilitate bonding as well as surface stress.  Overall, the Harker planes of the polymorph provide a more  detailed atomic solution to the PM which is discussed and illustrated further in the Supplement.

 As found in solving more complex  structures, e.g., the DAS, such PMs may not be unique and other methods  are usually relied upon to distinguish which structure class is correct.  One such refinement approach uses the Patterson analysis to define a trial structure and then a Keating analysis to refine the structure.[38]  In attempting to refine the detailed structure  as discussed next, there are two additional structural paradoxes that are significant.  One is in the height of the adatom atop the base of 3 atoms upon which it sits. The other is in the failure of the Keating optimized structure of the DAS model to improve the agreement found by LEED. [30]  In that work, the Keating optimized structure actually degrades the agreement !  There is also the overall conflict between various structural methods whose atomic positions vary by as much as 0.2-0.4 A.  And finally, the DAS structure derived from detailed DFT calculations disagree with experimentally determined lateral positions in excess of known variations as established for bulk structures. Thus, the DAS model provides a structure with many ambiguities!

While early analyses and calculations were limited by the overall complexity of the 7x7, more recent calculations are not, but yet these structural discrepancies for the 7x7  continue to exist.  For example. DFT benchmarks for semiconductors bond lengths are expected to be accurate to within ~0.05A and off by at most 0.08 A for the PBE exchange functional. [40]  The differences found for the 7x7 go well beyond these ranges !

DETAILED STRUCTURAL INFORMATION  FOR THE 7x7:

 The height of the adatom  determined from the first semi empirical and ab-initio calculations initially ranged from 1.14A - 1.24A. [8, 31]  Later, more realistic calculations within density functional theory, DFT using the local density approximation  involving bi-layer slabs of atoms, determined  heights of 1.18 - 1.34 A [15-21] and most recently 1.32 and 1.42 A for the different adatoms.[22]  The smallest values arose in the calculations employing  pseudo-potentials.[11,17, 19]  In comparison, several  experimental structural methods such as MEIS [9], XRAY [31] and RHEPD [34]  that physically measure something, all reveal a height of 1.48- 1.58 A.  Such differences between experiment  and LD calculations are generally considered unreasonable for bulk materials. [40]   One must also recognize that in deriving an experimental structure most approaches assume a model structure, in most cases the DAS model, to determine the 'best' structural parameters.

Dynamical LEED analysis, perhaps the most successful and widely used quantitative surface structural method to date, has shown an extreme sensitivity to surface atomic positions and has resolved the detailed structures of may surface systems.  However, its limitation is that one has to find a specific arrangements of atoms to within 0.1 A  to achieve good agreement with experiment. The complexity of the 7x7  makes it essential to start with a generally correct structure from the onset!  To overcome part of this complexity, the LEED analysis of the 7x7 surface[30] identified a few repeating units of the 7x7 DAS structure that were taken as a group. Then each atom or group was varied independently for both sides of the unit cell. Of the many structure examined, a reasonable level of agreement was found for a LEED refined DAS structure. This 'LEED' optimized DAS structure had a reliability, $R_{T-vH}$ - factor, of ~ 0.34. In contrast the original TEM based DAS structure provided an R-factor of 0.54 - not an acceptable structure by dynamical LEED standards. The optimized LEED structure also found large vertical and lateral displacements from the DAS model by as much as 0.67 A and 0.2A, respectively. Similarly, the LEED analysis of the Keating optimized structure also had large vertical and lateral spacings of 0.38A and 0.18A. Thus, more conflicts are found for the Keating optimized structure as well as the X-ray structure discussed shortly.

In contrast to the 7x7, a dynamical LEED analysis of the 4x4 Si/Ag monolayer 2-D silicene is an exemplary system in which the determined LEED structure provides an R factor of 0.17 [41] and provides a detailed match to both integral and many fractional order beams. A later LEED analysis of the 4√3 x 4√√3 R 30º multilayer structure produces a low R factor of 0.14. [42] These analysis also included many more fractional order beams and phase shifts than used in the cited 7x7 LEED analysis which may also limit agreement. The poorer agreement of the Keating optimized LEED structure from the DAS model is significant given the role that Keating methods play in refining Patterson derived structures.

The first x-ray structural analysis [31] performed in the same time frame as the LEED work confirmed the overall general TEM DAS structure but also indicated several atoms with 0.1- 0.25A displacements from it. Since this first X-ray work looked at the reciprocal lattice rods normal to the surface, only the lateral positions of the atoms could be analyzed. This analysis also considered 6-fold symmetry as assumed earlier and, due to the large number of structural parameters, also constrained the adatom and the atom directly below them to move together. Thus, it is not surprising that the X-ray diffraction patterns and derived structures are slightly different than the original TEM structures.

Figure 5 shows a comparisons of the original X-ray determination of the lateral structure of the 7x7 [31] to the most recent spin polarized LD calculations. [22] This shows the differences from the original optimized DAS structure and sheds light on the structural differences and reliability of these methods.

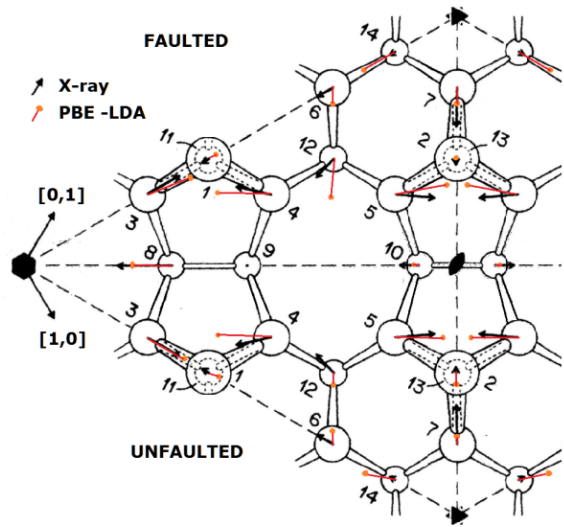

FIG. 5: (Color online) The TEM derived DAS structure and distortions 'refined' from the first x-ray measurements and analysis,[31] and the corresponding distortions from this DAS structure found in the most recent calculations of the 7x7 [22] as shown in Figure 2. The arrows and dumbbells show the distortions found from the x-ray and calculated structures respectively, both multiplied by 10.

The x-ray structure starts with the TEM diffraction derived DAS structure (as done in LEED) and then based on a structure factor analysis of the x-ray data, determines the lateral distortions from the original DAS structure. These x-ray distortions are shown in Fig. 5 (multiplied by 10) by arrows and compared to the calculated distortions as indicated by (orange) dumbbells. Both structures differ from the original DAS model on average by the same amounts (~ 0.1A), *and some atoms have displacements in totally different directions!* More specifically the mean of the scalar variation of the DFT variations from the DAS structure are 0.098A while than the x-ray variations are 0.090A, with a small variation of 0.02 between atoms on the faulted side versus the unfaulted sides of the unit cell. Such variations show that the 'refined' DAS structure is equally in error for either the x-ray of theoretical refinement and cast doubts as to the meaning of such 'refinements'.

As discussed in the Supplement, significant questions exist for this first x-ray analysis that may account for these structural differences. These arise from sample perfection, surface contamination and exceptionally high, unphysical vibrational amplitudes used in extracting this x-ray structure not to mention the 6-fold symmetry assumptions utilized. Similarly, HR-TEM studies to resolve specific atom locations of the 7x7 structure [78] are similarly fraught with many of these same limitations, again due to low signal to noise, 6 fold and translational averaging of the data as well as other simplifications in this early analysis noted by these workers. These are both discussed further in the Supplement.

Vertical displacements of atoms in the 7x7 could be determined in a second x-ray study by an analysis of the truncation rod intensities up to the 333 reflection.[32] This later study also used improved sample preparations involving high temperature oxide flash-off versus sputter cleaning and lower temperature annealing [31] but still required days for x-ray data acquisition. In this later work the authors state that the overall accuracy in determining vertical displaces is to within 0.03 A. This new study found significant expansions of vertical separations of the first two bi-layers similar to those found in LEED, but also an unusual adatom height. The author's state:

*"the most glaring inconsistency between the best-fit parameters and those determined previously is in the height of the "..Ad-atom... " Our Ad-atom layer spacing is bigger than any of the previous values by 0.3 to 0.4 A to date."*

This x-ray determined adatom height of 1.58 A was well above heights of ~1.10 to 1.20A found at that time, or the 1.32-1.42A found in the latest calculations.[22] The small adatom height of 1.20A determined by LEED probably represents a DAS structure that allows constructive multiple scattering interferences that reduce the R-factor, which still is too high to confirm the DAS structure.

However, this larger x-ray height does agrees with the adatom height of 1.52A using MEIS, [9] and later confirmed as 1.54A in RHEPD measurements.[34] While the MEIS values are based on fitting models having a DAS configuration, the RHEPD results are less model specific. Comparably large vertical adatom displacements of 1.8 ± 0.2 A were also found in a third x-ray study[93, 94] that is discussed further in the Supplement. This improved x-ray measurement also found larger dimer separations, of 2.89A versus 2.49A in the earlier x-ray study which compare to ~2.34-2.44A in theoretical studies. The structure of the DAS model does not appear to be converging !

The subsurface vertical layer distortions/displacements and stacking fault shown in the DF calculations of Fig. 2 are also significant since they can create a new state and bond charge as occurs for an intrinsic bulk stacking fault, or ISF. [43] The stacking fault in the DAS model is a surface analogue to the ISF and can produce an analogous surface state, a stacking fault surface state, SFSS. The ISF and a possible SFSS are thereby important in any analysis of the DAS structure or STM images, but were never needed nor considered into account for any of the experimental results for the 7x7 DAS model. As a result, a possible SFSS is important to reconsider given the existence of the forgotten surface state, FSS.

The decay length of the calculated ISF is shown in Fig. 6 and reflects the vertical layer displacements or 'healing' cause by the changes arising from the fault. This fault creates the additional charge density between certain atoms that are split off from the bulk valance band and decay gradually into the bulk reaching half its value after 2 bi- layers below the top layer. This suggests that at a minimum, 3 bi-layers must be considered below the faulted top layer in any calculation. These charge densities also suggest that the unoccupied SFSS is more intense above the surface than the filled SFSS.

The charge densities shown in Fig. 6 can also be projected onto the surface of the 7x7 to predict where they may be observed in STM. This is discussed and shown later in Fig. 16. While all DAS calculations show no evidence of the observed FSS, but they do show some evidence for a SFSS localized around the adatom. In contrast, the experimentally observed FSS appears to have a different symmetry, arises at a different location and has significantly lower energy than the calculated SFSS.

In principle, the unusual stresses from the dimers could also lead to a localized type of surface state split off from the bulk states analogous to the ISF. However, no evidence for such a state in the vicinity of the dimers, has yet been established.

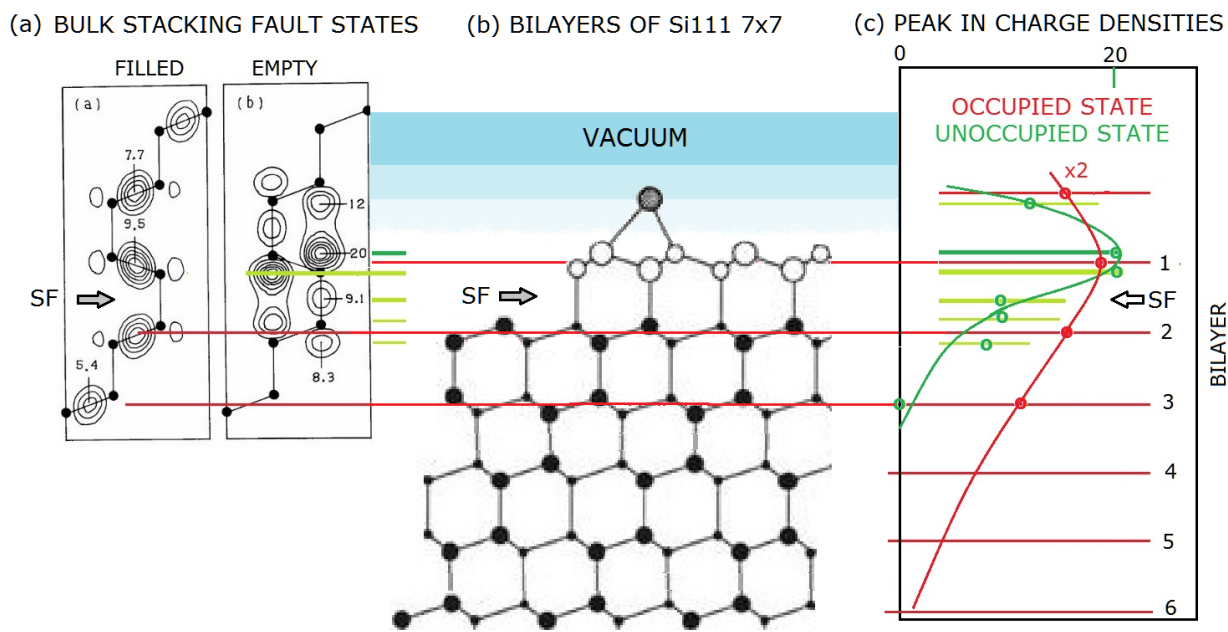

FIG 6: (Color online) (a) Charge densities of the filled and empty states of the intrinsic stacking fault, ISF, in Si 111,[43] (b) the stacking sequence of the top and underlying layers and (c) the decay of their charge densities into the bulk. The filled state is calculated as 0.1 eV above the VBE , and the empty ISF 0.3 eV below the CB max, i.e. both around the $\Gamma$-point of the Brillouin zone. The more localized empty state CD of the ISF  with its 2x higher density should make it easier to detect via STS of the empty states.

In summary, a variety of structural methods indicate an adatom height for the DAS structure  that represents a glaring inconsistency with all ab initio calculations. The lateral displacements as well as the dimer bond lengths while only examined by a few experimental methods  also differ significantly from calculations.  In addition several of the features of the original Patterson MAP of the 7x7  can be shown to be consistent with an entirely different class of 7x7 surface based on a honeycomb type of 7x7 polymorph proposed here.

In the debate as to the conflict between the calculated structure and the experimentally determined features,  yet another issue looms. Even today's best DFT calculations contain LD approximations that may limit finding the lowest energy  configuration for certain 2-D structures on 111 surfaces. These limitations are discussed in the concluding discussion of this paper.

THE ROLE OF ELECTRONIC STRUCTURE IN UNDERSTRANDING SURFACE STRUCTURES:

The electronic structure of surfaces has long been a key element in confirming or ruling out structural models, as epitomized by the structure of the π-bonded 2x1 structure of the cleaved Si111 surface. [44] Insight into the 2x1 structure involved comparing the measured electron energy band dispersions of the surface state using angle resolved photoemission spectroscopy, AR-PES, to various theoretical models. Ab initio calculations can now be used to predict  the  CDs of the surface electronic states,  that can be directly compared with those observed with the STM.[45]

Such surface CDs are a fundamental aspect of quantum mechanics that reflect stationary states of the system. Namely, the atomic crystalline potential together with the 2-D potential at the surface defines the stationary and evanescent states that arise on the surface. As the Schrödinger equation prescribes, the electrons respond to these potentials to find a configuration that minimizes the total system energy. The underlying symmetry of the resulting wave-function and its spatial and energy structure on the surface provides a resulting probability distribution that can be measured through its energy and spatially dependent charge densities, CDs. In particular, symmetry plays a very important role in defining the stationary states of all systems.

For the 7x7, the surface CDs of the observed surface states have not been fully considered. As noted earlier while three surface states of the 7x7 have been understood, [10-22] the forth experimentally observed surface state has not. These three surface states include the adatom state (near $E_F$), the restatom state ( ~0.8eV) and the backbond state ( ~1.8 eV). The forth surface state peaking near ~0.4eV, is identified as the 'forgotten surface state' or FSS. It has been repeatedly observed: first, in temperature-dependent Photoelectron Spectroscopy (PES),[46] then reported in STS images,[12, 47] band mapped by angle resolved AR- PES,[48] seen in polarization depended AR- PES,[25] again found in STM measurements with an InAs tip [26] and later in STS measurements at T=78°K. [27.28] All calculations of the DAS structure clearly show three DAS surface states, but the features of the observed fourth state, the FSS has not been found in any calculation. The possibility that the FSS could be an analogue to the bulk stacking fault state, ISF, shown in Fig. 6, can also be excluded for several reasons discussed later. .

Since all calculations appear to confirm the DAS model, a new structure may exist in nature with similar electronic structure and CD's as found for the 7x7, but which for some reason cannot currently be calculated accurately. Finding such a structure has involved reverse engineering of experimental results together with the believe that nature, and specifically electrons, know more than we do !

After an long evaluation of possible alternatives and modifications to the DAS model, a related system - the silicenes - was found that provides a framework to explain these results and many other findings. Several features of a well characterized 2-D monolayer phase of silicene [41,42] provide a basis for this new structure, and a blueprint for the 7x7. This also generates a new polymorphic family of 2n+1 reconstructions that are topologically indistinguishable from the DAS family.

The remainder of this paper is organized as follows: the scanning tunneling spectroscopy, STS, ambiguities are presented first, followed by a description of the new model. Next, the features of the FSS are presented which distinguish its symmetry, its structure and its similarities to monolayer silicenes. Then the relation to and the formation of this polymorph from the 2x1 structure is discussed. Finally, an independent confirmation is shown based on the experimentally deduced valence charge densities derived from x-ray and electron scattering measurements. The theoretical uncertainties and physical properties that may be limiting LD calculations on Si111 are also briefly discussed from a pedagogical perspective.

WHAT SURFACE SPECTROSCOPY TELLS US:

Fig. 7 shows a comparison of experimental spectral measurements of the surface DOS states, SDOS, for the 7x7 surface as determined by STS measurements of dI/dV normalized by I/V. The SDOS in (a) was obtained by averaging all the STS pixels over the unit cell when a tip of the highest resolution arose.[49, 50]

The spectra in (a) was measured at 298 °K and reveals the adatom (AA), restatom (RA) and FSS as marked. The lower panel (b) shows atom-resolved experimental STS spectra obtained at 78° K [27] utilizing the type of same tip reformation process used for the averaged SDOS above. This tip process involves gently picking up Si atoms on the tip which produce high STM resolution and large corrugations.[9,47,50,] The calculated SDOS [27] for the DAS model by these same authors is also indicated in (b) by the dashed lines. In order to directly compare with experiment, the calculated energy scale is shifted to adjust for widely known electron self energy corrections, [51] as done in all following comparisons between calculated and experimental spectra. It should be noted that non- equilibrium surface charging due to limited electron transport to the surface is expected at low temperatures.[28] Comparisons to other temperature dependent studies, [28,29] suggest small shifts in these filled state energies of at most ~0.2 eV.

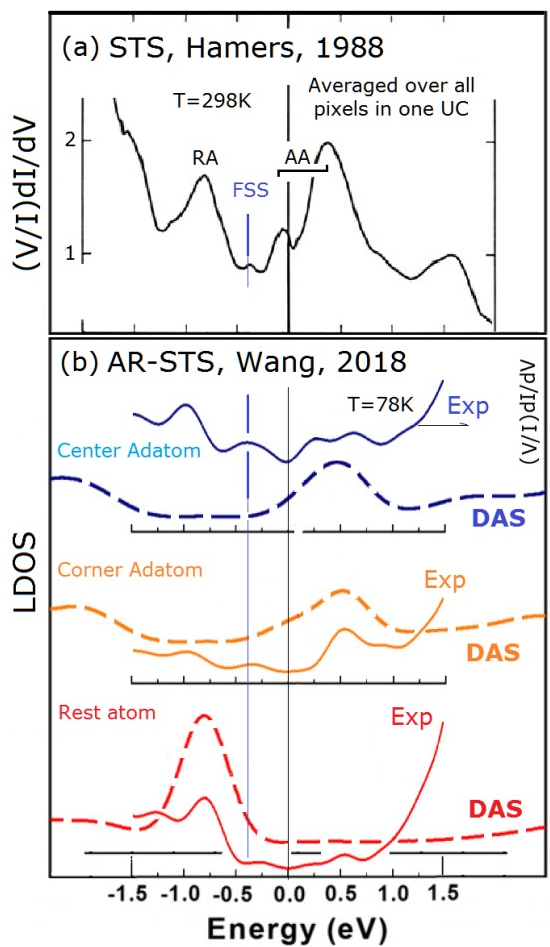

FIG. 7. (Color online) Comparison of the surface states of the 7x7 from (a) area averaged STS [49] and (b) DFT calculations and atom resolved STS. [27] In (a) the restatom (RA), the SFF and the Adatom (AA)

manifold of states are indicated. A simple self energy correction [51] of x1.6 is made to the calculated energy scale using the well defined restatom energy as a reference. The theoretical spectra in (b) have been further broadened by its authors by ~0.2eV[27] from the spectra found in other calculations.[19-22]

The atom resolved STS results for the restatom and adatoms[27] in the lower panels of (b) show marked differences between theory and experiment. First, the calculations displayed do not show the FSS state, but the FSS shows up in the experiments. The calculated rest atom state has the same locations as found in other calculations [14-22] and appears to be only weakly coupled to the adatom states. This weak admixing of the adatom and restatom states is discussed later.

The experiments in Fig. 7b, show that the FSS is not just associated with a particular atomic location but delocalized over the unit cell. Its energy varies slightly with surface location. The restatom state at ~0.8 eV and adatom states nearer $E_f$ are also slightly admixed. These admixed states suggest that the basis states of the wave-function comprising these CDs, e.g., atomic orbitals or Slater determinants, occur on each of these atoms. Such admixed atomic charge densities will contribute to the delocalization of these states over its structure. The honeycomb lattice of free standing silicenes contain energy bands spread out over their honeycomb lattice, [52-54] essentially delocalized, that may provide such admixing. Also, the suppression of dispersion for the FSS [25, 48, 67,68] can be associated with the faulted honeycomb lattice and their supercells, that create a non-traditional Surface Brillouin Zone, SBZ [25] that is discussed later.

A comparison of different earlier calculations of the DAS electronic structure [19-22] are shown in Fig. 8 and confirm the absence of the experimentally observed FSS. These calculations all use hydrogen termination of the back layer with a 249 atom 4 layer slab in (b) to a 347 atom, 7 layer slab in (d) where 2 such layers represent bi-layers. These are all DFT calculations within the LD approximation that produce an energy optimized DAS structure, but each reflect a slightly different methodology. Fig. (b) uses the local density approximation of Perdew and Zinger, with norm conserving pseudo potentials and an atomic basis to represent the wave functions (using the SIESTA program[19]), (b) is a spin polarized LD calculation using plane waves and the PBE exchange functional (using the QUANTUM EXPRESSO program [22]), and (d) is an ab-initio tight binding calculation with atomic orbital basis functions (using the FIREBALL program [20]). Further details of each calculation can be found within the cited references. It should be noted that all calculations use a limited number of atoms, thereby limiting the number of eigenvalues [14] and are broadened typically by a Fermi broadening function to produce the density of states, DOS or the projected DOS, PDOS.

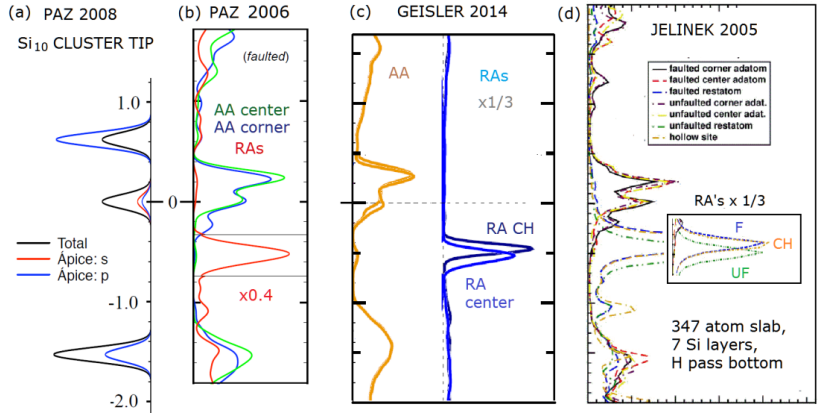

FIG 8: (color online). Calculated Projected DOS for (a) a Si10 cluster tip and for the Si111 7x7 DAS surface from various types of LD calculations (b)-(d). Calculation (d) is different in that it use an ab-initio tight binding method that used atomic orbital basis in constructing the wave-function. The calculated energies scale are those calculated in each method without any self energy corrections.

One question that clouds all STM studies, and in particular STS measurements, is the electronic structure of the tip which is shown in panel (a) for a $Si_{10}$ cluster tip.[19] This cluster tip was also used in the calculations of the STS images (discussed later) which will provide a clean window for imaging the filled 7x7 states down to ~-0.8 eV and will rule out any tip electronic structure effects in the STS images in this range. Such a Si cluster tip is also consistent with the tip reformation process [27, 50] and, in addition, when used to calculate the STS images, reproduces experiment images for all but the FSS feature.[19] In contrast, W, Pt and PtIr STM tips widely used can have a more complex set of tip states as found for cluster calculations of a $W_{20}$ cluster tip. [19] This W tip produces smaller corrugations and lower contrast STS images than the $Si_{10}$ tip.[19]

The projected DOS or PDOS from the calculations in Fig. 8 are all very similar and show a manifold of adatom states straddling $E_F$ and the lower lying restatom states which vary slightly in energy. In the adatom DOS of (c) and (d) there appears to be a feature corresponding to the energy of the restatom state that can be associated with the admixture of these states into each other but not the formation of a new state between them. These are all similar and find no feature that can be associated with the FSS.

Fig. 9 shows results from the most complete LD calculation to date which is spin unrestricted and includes larger k-space sampling, large APW cutoffs as well as the PBE exchange-correlation functional. [22, 55] All the restatoms have nearly identical spin states and again show a small admixture (less than 4%) with the lowest lying corner adatom state at 0.14eV. A similar feature at -0.13 eV is also found in earlier DFT calculations..[56] Correcting for self energy effects,[51] the 0.14 eV feature should peak at -0.23 eV and tail out to near zero by -0.35 eV. This is too low to account for the FSS unless there were significant dispersion which has not been found in AR-PES measurements.[48] This feature likely corresponds to the SFSS of the faulted DAS layer as noted earlier.

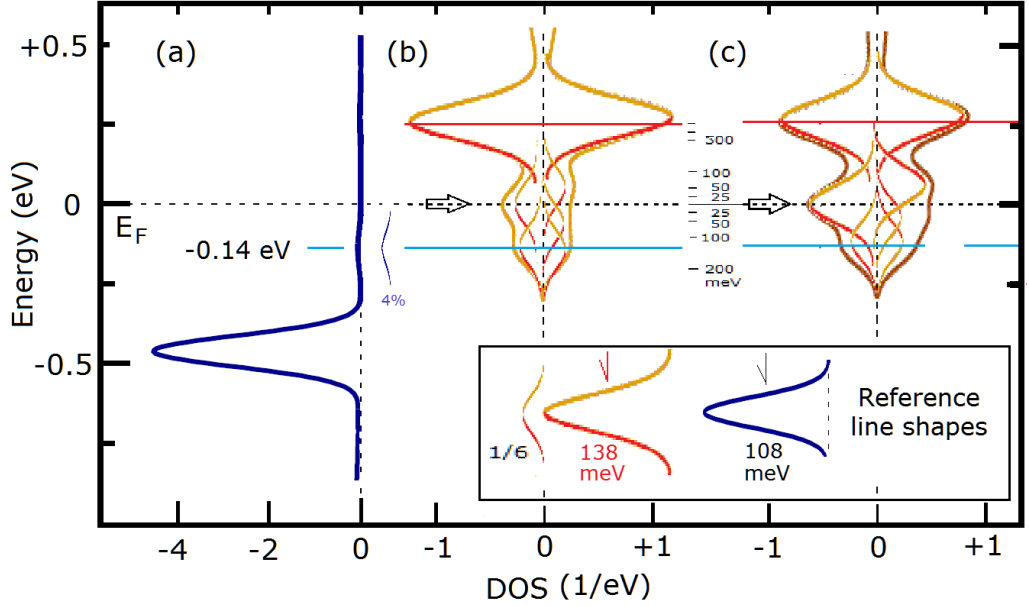

FIG. 9: (Color online) Spin polarized PDOS calculated in LDA with the PBE functional for the optimized DAS structure.[55] (a) is for the restatom ( identical for both spins) while (b) and (c) represent the corner adatoms on the unfaulted and faulted side of the unit cell. The restatom peak shown is the higher lying of the two restatom states from Fig. 8(c) that are identical for each spin. Also observed is a state at -0.14 eV on the restatoms that corresponded to 4% of the intensity of this restatom peak. The deconvolved features within the adatom manifold are shown for the corner adatom on the unfaulted (c) and faulted sides (c) of the unit cell using the referenced line shapes.

To further determine if any such features of the FSS might be visible in the calculated STM image for these most recent LDA DAS calculations,[55] one can compare the differences in the calculated images at different bias voltages, in particular at energies just after tunneling turns on for the FSS. For the experimentally observed FSS at 0.4 eV which is later shown to be observed as a broad feature between -0.25 to -0.65 eV, the FSS should start to turn on for calculated STM energies of ~ -0.25 eV. Fig. 10 show the result of subtracting STM images calculated at -0.3 and -0.5 eV. Here image (a) represents the additional tunneling (DOS) below -0.3 eV arising from the restatom and any other state, while (d) represents removing the -0.5 eV features from the -0.3 eV image to search for any additional states between 0.3 to 0.5 eV below $E_F$. In (a) there is additional tunneling from the restatom states which appears to spread out towards its three nearest neighbors and is partially suppressed by the adatom. Image (d) looks like a noisy background with the restatoms missing. However, as shown next, the experimentally observed CD location of the FSS has a different orientation, i.e. rotated by 60° from this so as to arise between the restatom and the adatom. The energy dependent barrier transmission is not accounted for in such difference images but are included in the calculated STS images[19] discussed next.

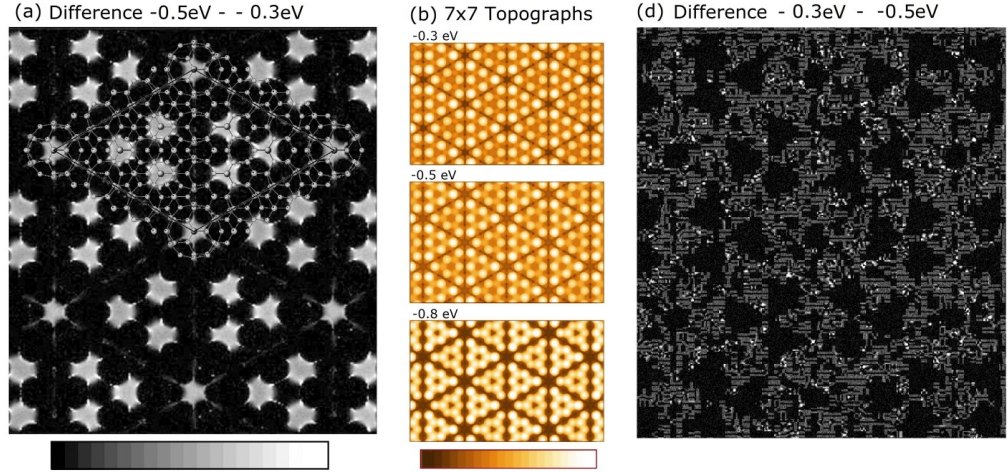

FIG. 10: (Color online) Differential calculated STM images above -0.3eV (a) and below (c) -0.5 eV with the calculated DAS structure superimposed in (a). (b) shows the calculated filled state topographs for different bias voltages. [55]

A comparison of the spatial and energy characteristic of STS images are shown in Fig.11 and are more revealing. Here (a) shows the calculated images [19] and (b) those experimentally observed [19, 47] at an energy for each where the restatom and the adatom both show equally high intensity. Shown besides each image is a color coded schematic of how the adatom and restatom CDs appear as a function of energy. As observed experimentally the additional CD structure of the FSS appears visible at -0.6 eV to -0.95 with a maxima at ~ -0.75 eV. This feature is missing from the DAS calculations. For the DAS structure the restatom and adatom CDs appear to be well separated with a 'node' between them and observed together over a narrow energy range of 0.15 eV centered at -0.44 eV [19]. Image subtraction of the calculated STS images at -0.44 and -0.58 eV or lower where the adatom state is significantly reduced in intensity, confirm negligible CD between the restatom and adatom for the DAS structure.

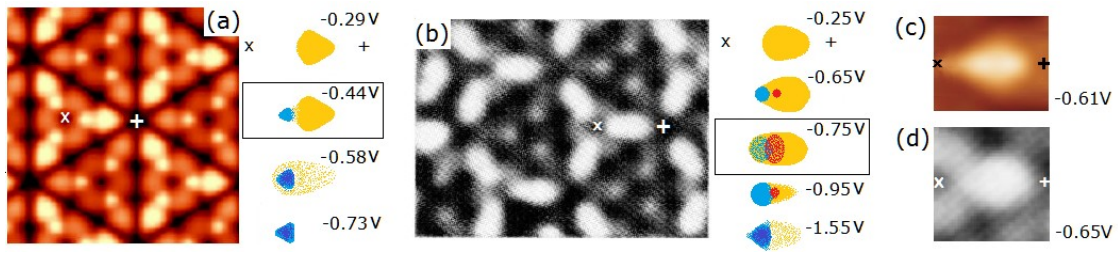

FIG. 11. (Color online) STS images of the surface state CDs as (a) calculated for -0.44 V [19] and (b) measured for -0.75 V. [47] These calculated energies have been corrected for self-energy effects as discussed. X is the center of the unit cell and + is the corner hole, both for the 'faulted' side of the unit cell. Alongside each image is an enlarged schematic of the CDs between these two points. Here the restatom CD is on the left, the adatom on the right and the FSS in between. Panels (c) and (d) compare two independent experimental STS CDs [19, 47] averaged over equivalent directions to improve signal to noise.

The elongated features in (b) at -0.75eV, together with the calculated DAS CD's, suggest an additional CD between the restatom and adatom and their symmetry equivalent locations that make the FSS look like a 'propeller'.  Constant height, current imaging based STS measurements using lock-in techniques, also shows a  propeller like feature starting around at  -0.5eV but is noisy due to the low current measured at these locations associated with the constant height scanning mode.[57]   Another experiment using an InAs tip that allows energy filtered imaging [26] produces  a feature  identical to that shown in Fig. 11b. These results demonstrates that the FSS CD is not an artifact of either the tip or the STS procedures. Four different measurements [19, 26, 47, 58] confirm this new CD  which is not found in DAS calculations and also show where the FSS arises within the surface unit cell. The difference in the FSS energy from the low temperature results in Fig. 7 may reflect non-equilibrium carrier band shift at low temperatures. [28,29]

The experimentally determined location of the FSS  is between the adatom and restatom, with a possible shift between them with energy. This has many implications.  First, for a different structure and secondly, possible tunneling via higher momentum tip states for the DAS model structure as initially proposed. [26] If such tunneling by higher momentum states were to occur they are important and will limit the value of the widely used Tersoff-Hammond[58] approximation to simulate STM images for comparisons to STM measurements.

AN ALTERNATIVE STRUCTURAL MODEL:

An evaluation of many possible modifications to the DAS model was initially considered to produce this new experimental CD and the symmetry of the FSS. This included changes to the dimer and atoms along the cell boundary (as well as earlier proposed models and various modifications.)  The new structure was  derived from the well characterized  4x4 Si monolayer on Ag111  whose detailed structure has been theoretically predicted and confirmed by LEED.[41,42, 52,53]  The new structure is designated as the DFA structure for reasons that will become obvious later. The DFA essentially starts from the same type of 3x3 honeycomb superstructure  but lattice matched  to Si111, adding adatoms similar to the dumbbell structures of free standing silicene[59] and finally extending this faulted honeycomb and adatom arrangement to larger meshes. This produces a DFA family of structures analogous to the DAS family. This is illustrated in Fig. 12 for the smaller 5x5 structure which is considered to have similar electronic structure and bonding as the 7x7.[14, 56]   However, while all the main structural features of the 5x5 correspond well to the 7x7, the 5x5 appears to be electronically different with a energy gap and zero DOS at $E_F$ at 300 °K. A small  surface state gap feature is also observed [60] which  may correspond to its 'FSS' . These unusual features of the 5x5 are discussed in further detail in the Supplement.

The  top layer of the DFA structure is very different from the bulk Si bi-layer and does not possess a stacking fault.  Instead, the unit cell of this top layer has supercells with mirror symmetry across the unit cell boundaries which creates a new type of honeycomb/adatom structure.  The DAS and DFA structures both have  corner holes and identical 2-D arrangements of adatoms. This make these models difficult to distinguish via STM alone since strong tunneling from the adatoms states near $E_F$ dominate tunneling. [13-22,27]  This domination of the STM image by the adatoms can also create the charge density depressions along the domain wall that have been taken as indirect evidence for the presence of dimers.[61] However, as discussed further in the Supplement, these depressions do not require dimers but instead appear to reflect the charge density contours of the adatoms and their arrangement on the surface.

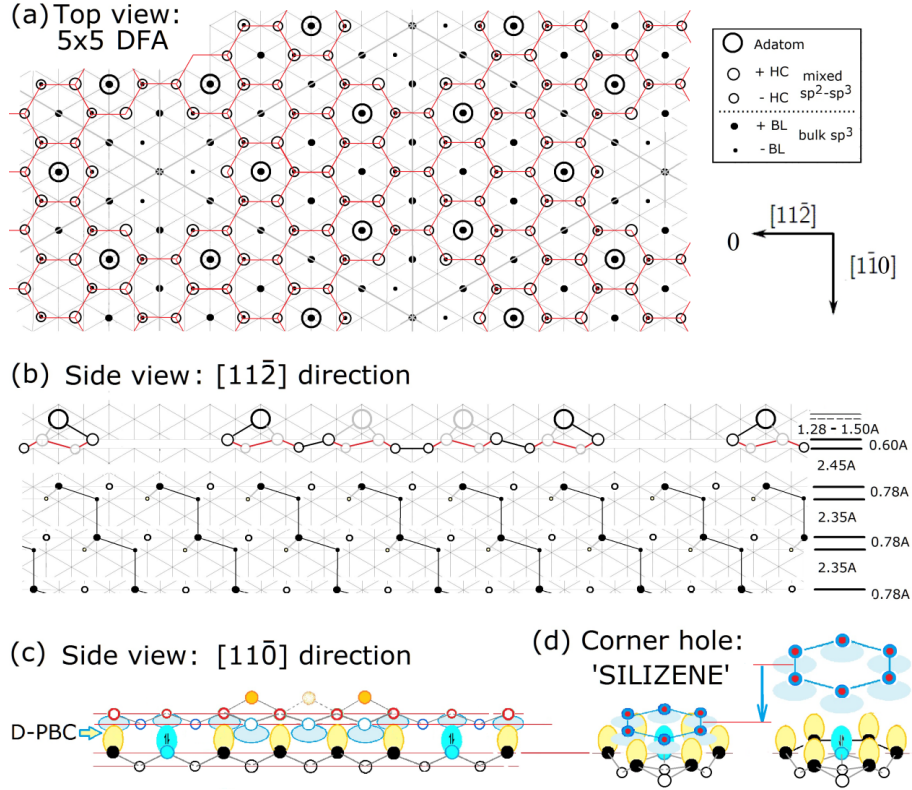

FIG. 12. (Color online) Structural model of the 5x5 DFA: (a) top view and side views (b) along the long diagonal and (c) the domain boundary. The hexagons shown in (a) reflect the honeycomb layer. The spacings indicated in (b) are nominal bulk and silicene layer spacings with a 1.28A adatom height drawn and 1.5A indicated. The top layer is slightly expanded from the bulk interlayer spacing. (d) shows a perspective view of the atoms comprising the corner hole with the edge atoms around the honeycomb layer nestling in to form the cyclic 'silizene' structure. The relative sizes for the p-orbitals in (c) use the peak in the radial charge density of 2p hydrogenic wave-functions whose spatial extent is scaled to produce the correct bulk Si-Si bond distance. (The trigonal construction lines are shown to aid readers in constructing this model.)

Thus, the new '7x7' family of structure consists of two honeycomb cells forming the two halves of the unit cell with adatoms atop each of them. In Fig. 12, each 2-D cell and the resulting 5x5 supercell has mirror symmetry on each side of the unit cell. This type of symmetry and faulting of the honeycomb provides a structure which reduces strain that would otherwise arise from the DAS dimers. Instead, this allows each cell to maintain the strong intra-layer interactions associated with sp2-sp3 hybridization. Such an arrangement enables the top layer to be very flexible, readily form larger unit cell as well as accommodate surface stress. Now the 2n in this (2n+1) x (2n+1) family of structures refers to the honeycomb cell configuration. The 2n represents the length of the unit cells defining the each cell, i.e., (2n+1) -1. i.e. or one lattice vector smaller than total unit cell size. 2n also corresponds to the number of substrate atoms between the central atoms of the corner hole.

Another distinguishing feature of the DFA structure is that the adatoms have a different local bonding geometry than in the $T_4$ site of the DAS model. The DFA adatoms are three fold coordinated $H_3$ sites with another Si atom in the substrate layer far below. This substrate atom may be pushed up to form a 'pseudo' dumbbell atom with the adatoms as occurs for the 4x4 Si monolayer on Ag111. [41,42] Such a site difference may account for the larger adatom heights determined by x-ray measurements [32] and RHEPD from DAS calculations as noted earlier. These particular methods are more sensitive to vertical displacements and not the local bonding site of the adatoms as some other structural methods are.

(Note that a related honeycomb structure was also considered which had the same atomic structure around the corner holes as for the DAS model. This was excluded for several reasons: failure to replicate the adatom topography on the 7x7, features in the Patterson map, as well as limited CD overlap with the subsurface p-orbitals associated with it being an incommensurate structure.

In general the honeycomb supercells of the DFA structure have Si atoms that are positioned in the folds of the bulk bi-layer due to a registry shift and change in stacking from the that on the substrate. This creates an ad-layer whose atoms are nestled into the hollows of the underlying bi-layer. As illustrated in Fig.12 (c) the new structure has an unusual array of p-orbitals in both the ad-layer and underlying substrate atoms. As viewed from above in (a) the ad-layer p-orbitals have a configuration similar to those of the $\pi$-bonded chains of the 2x1 si111 surface. However, these chains exist in pairs that straddle the two supercells of the DFA and have a substrate atom directly in between and below them on the unit cell boundary. These pairs of chains and underlying substrate atoms essentially define the unit cell boundary where they link up to the corner hole.

This supercell arrangement of the DFA creates an interesting stereochemistry arising from the up and down pattern of the atoms that nestle into the underlying 1x1 bulk Si bi-layer. As shown in Fig. 12 (c), this allows the interdigitation of the sub-surface $sp^3$ 'dangling bonds' facing up with the periodic p-orbitals of the 2-D honeycomb that face down. This gives rise to designating this new structure as the Digitaled Faulted adatom or DFA structure. These p-orbital interactions are herein defined as interdigitated $\pi$–bonding or D-PB. In extended $\pi$-bonding systems, $\pi$-bonding is estimated to be ~0.39 eV for hexasilabenzene [62] or ~0.5 eV as found for the dispersion bonding of benzene's $\pi$-bonds to Ag111.[63]

A simple rational for the formation of such D-PBs on Si arises from strong intra- layer interaction within the top layer and a change in symmetry from the bulk layers that favors $\pi$-bonds over the bulk-like sp3 bonding. This strong intra-layer bonding and weaker bonding to the substrate would allow the top layer to relax more into the vacuum as is indicated by the larger interlayer spacing shown in Fig. 12(b).

A pedagogical illustration of how such interactions may arise is shown in Fig. 13 which compares the distribution of the charge densities for the outer p shells for C and Si based on hydrogenic wave-functions. For C, shown in (a), the charge density is more compact and facilitates direct overlap of the s- and p- atomic charge densities, as shown in blue for the $2p_x$ orbital. This allows a contraction for both the C-C bond and $\pi$-bond formation, as shown in blue and red, respectively. Silicon's nodes in both the 3s and 3p wave-functions push out the electrons further lengthening the Si-Si bond. Si's node also changes the phase of the inner and outer regions of these wave-functions. This increased separation for Si and nodal character reduces the overlap of the $p_z$ orbitals relative to the $p_x$ orbitals, thereby reducing the degree of $\pi$-bond formation in Si that can arise for C. Enhanced Si $\pi$-overlap can arise from the interdigitation of the concaved wedges of 3p charge density from the p-orbital of substrate atoms below as shown in (c). This becomes an extended π–bonded system when periodically repeated and spread out these electrons to reduce electron correlation and exchange interactions. In the DFA the top two silicon atoms in (c) are replaced with a chain of Si atoms with substrate atoms below them.

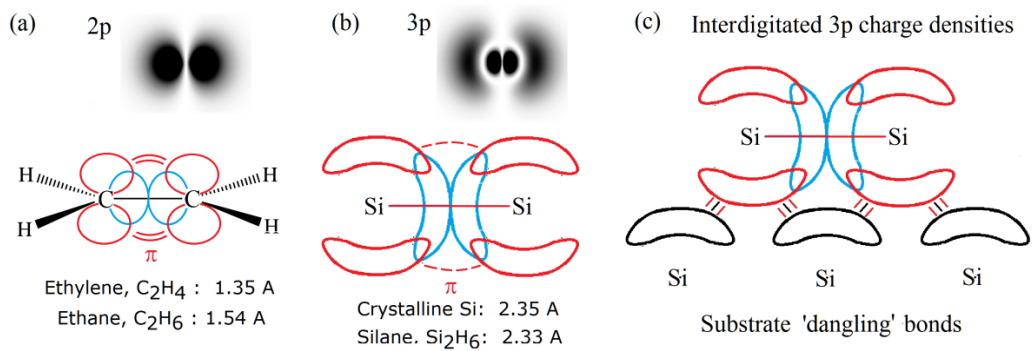

FIG. 13. (Color online) Hydrogenic 2p and 3p charge densities of the Carbon and Si atoms. (See text.)

Similar interactions between the atoms around the corner hole with the substrate atoms there can allow these π-bonds to wrap around the boundary atoms of the corner holes. This creates a 6 member 'cyclic' ring of rehybridized Si atoms that can better inter-digitate with the 6 upward facing Si111 sp3 'dangling bonds' directly below. This forms a more strongly bound cyclic D-PB 'sandwich' structure, defined here as 'silizene'. This conjugated structure essentially diverts the π- bonding of a 1-D π - bonded chain[44-45] or the D-PBs to form an energetically favored cyclic structure. This interaction may cause the DFA atoms around the corner hole to drop down and compress slightly around it. A less attractive alternative to achieve P-BD is to have the up and down atoms of the honeycomb ( Fig. 12a) reversed so as to "invert" the honeycomb. In either case the top atoms of silizene atop the corner hole are likely pulled down and compressed to interact with the dangling substrate charge densities below.

This stereo-chemical silizene compression provides another important function by relieving some of the tensile stresses arising from the smaller buckling possible for each sp2-sp3 honeycomb cell. As a result the D-PB interactions along the unit cell boundary together with those in silizene, act to minimize the overall tensile stresses to stabilize the 2-D layer. The creation of a pseudo dumbbell-like structure by the adatom with the Si atom below it, can also contribute to stabilize and bond this 2-D silicene layer to the substrate. It would not be surprising if this 2-D silicene layer could be peeled off with a 'sticky tape' that somehow substituted or altered the adatom interaction with the substrate!

In the DFA model the upward facing substrate $sp^3$ 'dangling bonds' shown in Fig. 12c and 13c can be likened to a 'spine' that can interact with pairs of down facing $p_z$ components of the honeycomb CDs, to form this new form of π- bond, the D-PB. This arrangement results in a regular array of substrate 'spines' along equivalent [1-10] directions that interact with the pairs of atoms along the unit cell boundary as if these latter atoms were 'ribs'. Stereochemically, the 'ribs' have folded or zippered onto the periodic row of substrate dangling bonds with minimal stress to the lattice. The periodic substrate dangling bonds not only allow bonding to the substrate but also allow flexibility to expand this structure to larger unit cells. Such 'folding' is a very simplified analogue to the more complex folding in biological systems.

This new D-PB mode of interaction has direct implications to the details of how the 2x1 structure breaks its 2-fold local symmetry starting at ~245°C [64] and converts to either the 5x5 above 280 °C [60] or the 7x7 at higher temperatures. [64] First, the symmetry directions of the 2x1 $\pi$- bonded chains (PBCs) and the D-PBCs are identical. In annealing the 2x1, an impurity or defect 'center' is proposed to create an instability that initiates the ejection of a pair of atoms to form two adatoms. This starts the first domain boundary and a precursor to the corner hole. As this precursor expands its cyclic $\pi$-bonds, another pair of Si atoms are ejected and a second domain boundary forms via the zippering process. The production of adatoms and zippering proceeds radially around the silizene precursor to eventually form the corner hole and the domain boundaries. When complete, it has created a radial PBCs around silizene that eventually interlock with the P-BC along the unit cell boundary to form a network of 2-D, D-PBCs that cover the surface. As the structure grows it requires additional adatoms that are captured from diffusing surface adatoms which find new, more stable bonding sites in the DFA configuration. This is discussed in more detail in the Supplement.

Based on these structural features, the nature of the FSS can be considered further. The transition of adatom and restatom CDs to form the intermediate FSS shown in Fig. 11, is now shown in Fig. 14 with an overlay of the DAS or DFA structures. As noted earlier, the broadening and overlap of the CDs seen experimentally in Fig. (b) could arise from sampling higher momentum states due to the structure of the tip and its coupling to bulk derived block waves near the tunneling apex.[26] However, the PDOS calculations of the DAS shown earlier reveal no such state. If such a state were to occur, then it will disperse predominantly along the symmetry directions of nearby atoms or rows of atoms in the surface and near surface region. Such energy dispersion with k, arises from the scattering and interference of waves by the atoms in their periodic potential.[65] The DFA has periodic atoms along this direction that will support a new CD here as well as any possible dispersion in a direction between the adatom and restatom, whereas the DAS structure does not. As shown in the overlays of Fig. 14(b), the honeycomb DFA lattice will provide atom pairs aligned in the right direction to shift the probability density arising from any wave-function interferences with changing electron momentum. This region of additional tunneling from the FSS is indicated in (c). Line scans of highly resolved 7x7 STM images ( 0.4nA, -0.5) taken with exceptional W-tips.[66] also show a weak but discernible FSS CD between the adatom and restatom.

(a) Calculated DAS -0.44 $V_a$

(b) Experiment -0.63 V

(c) Extra Tunneling

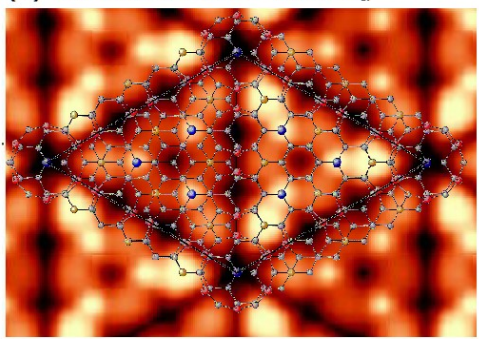
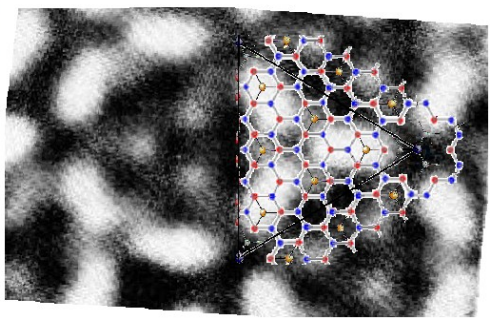
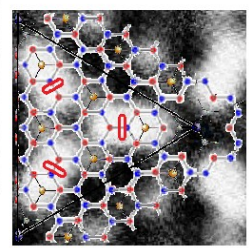

FIG.14. (Color online.) Comparison of the calculated STS images for the DAS model (a) and experimental measurements in (b) and (c) with an overlay of the DFA model. The energy of the calculated images is again adjusted for self energy corrections. [51] In (c) the additional tunneling observed experimentally is drawn in the region between the rest atom and adatom which corresponds to a 'valley' in (a). For calculated images as in (a) with biases of -0.58 eV and larger, the adatom CD has almost disappeared .

To further examine the nature of the FSS, Fig. 15 shows a comparison of (a) p- polarized AR-PES [25] for different angles (momentum) in the surface Brillouin zone, SBZ, of the 1x1 lattice to (b) the energy bands of a silicene. [67] (d) shows a schematic of the different phases of possible 2-D block waves along one of 3 equivalent M and K directions for the two supercells of the DFA structure.

The AR-PES signals from the adatom state is strongest as shown at the K point and disappears at normal emission where a restatom feature dominates the signal (not shown). These strong variations of the adatom and restatom features arise from p-polarized light and were not seen as clearly in earlier unpolarized AR-PES studies.[48] The FSS can be readily deconvolved from the polarized AR-PES studies shown in (a) from $0^0$ to $12.5°$ emission along the $\Gamma \rightarrow M$ direction and also at $38.5°$ again relative to the 1x1 SBZ . Earlier unpolarized measurements show small fluxuations in the dispersion of the FSS by ~ 0.1 eV that suggested weak critical point features in the dispersion of the FSS.[48] This implies it is part of the 7x7 SBZ, but as found in later, is more complicated and suggestive of a non-traditional SBZ.[48]

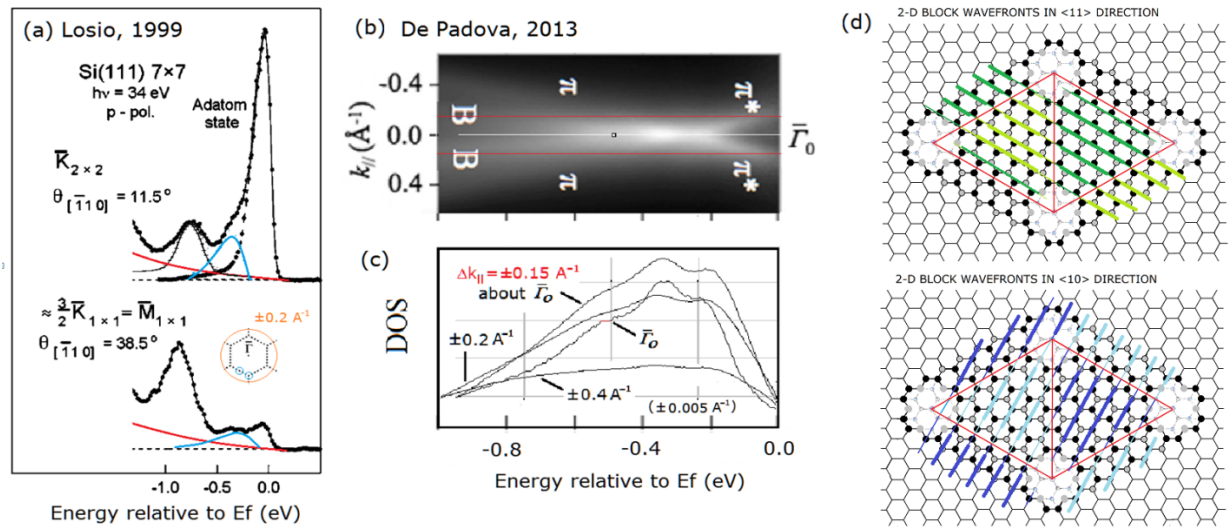

FIG. 15: ( Color online) Angle resolved photoemission [25] obtained at 16 $°K$ under flat-band conditions (plotted on different scales) that show the weak deconvolved FSS near ~0.4 eV (a) together with (b) band mapping for a silicene layer [65, 66] where the Ag substrate emission has been suppressed. (c) shows the integrated emission of (b) over different ranges of wave vectors. (d) shows how the faulting on each supercells has to de-phase the 2-D Block states in the K and M directions at their boundaries.

Fig. 15b shows the dispersion of the p- bands for a monolayer silicene phase that has been experimentally separated from the Ag energy bands, [67] and was later confirmed for a silicene layer formed on an intervening monolayer templete on Si111. [68] The observed silicene bands are proposed to arise from the Si p-band crossing at -0.25 eV. (It should be noted that recent linear dichroism angle-resolved photoemission spectroscopy and calculations [96] confirm that the E versus k bands shown in Fig.15b correspond to the modified Dirac cones of freestanding silicene that are derived from Si $p_z$ orbitals.)

Fig.15 (c) shows an integration of this silicene E versus k AR-PES spectra about $\Gamma=0$ to as far as +/- 0.4 $A^{-1}$ from $\Gamma=0$, so as to produce the DOS for this band. It reveals a broad peak at ~-0.4 eV that is almost identical to the FSS as deconvolved from the adatom and restatom AR-PES peaks as found in (a) for the 7x7.[25] This suggests that the FSS state represents a largely dispersion-less p-band of the honeycomb about the $\Gamma$-point and largely delocalized in k-space. A de-phasing of the Block waves propagating in all three of the directions shown in (d) can account this dispersionless state. Namely, there is a common boundary condition for the waves in the M and K directions due to reflection symmetry at the unit cell boundary that requires a matching of their magnitude and phase at these boundaries. This requires a de-phasing of the waves that would otherwise arise on either side of the unit cell. This restricts many possible wave and selects waves in certain directions, as occurs in other aperiodic 2-D systems.[95] Some coupling of the honeycomb $p_{x,y}$ states to the restatom (or adatom) likely introduces $p_z$ character that allows the FSS to be observed in STS and with p-polarized light in AR- PES. Clearly, such details need further investigation.

Fig. 16 (a) and (b) shows a schematic of the CDs of the three highest lying surface states experimentally observed near the corner hole for the DFA and DAS models. For the DFA structure this consists of the restatom, FSS and adatom states while for the DAS this consists of the restatom, an expected surface stacking fault state, SSFS and adatom states. The structure of the FSS in the 5x5 unit cell is unknown, but transferring its features from the 7x7, one might expect the arms of propeller like feature to be shorter placing all three around one central restatom versus three restatoms as in the 7x7. This may produce weaker features akin to the SFSS side lobes of the DAS model and be difficult to resolve. The STS gap feature observed for the 5x5 [60] may also correspond to the FSS for the 5x5.

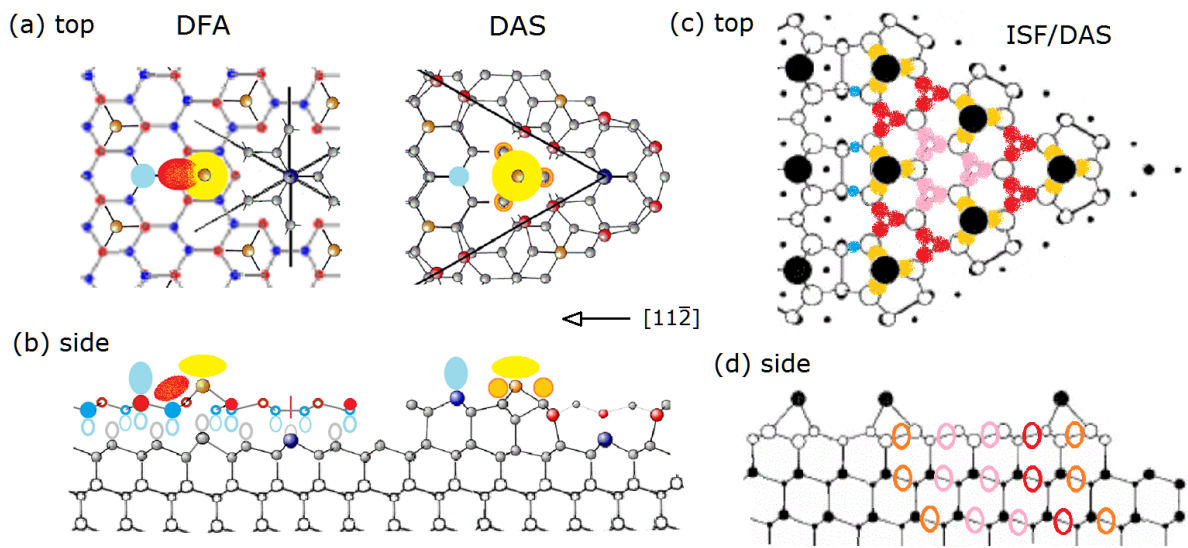

FIG. 16. (Color online) Top and side views of the proposed CDs on the 'faulted side' of a unit cell for the 7x7 DFA and DAS models. The elliptical lobes above the atoms in the DFA model represent p-like orbitals of the restatom, FSS and adatom surface states. The red (light) and blue (dark) atoms in the DFA honeycomb represent the higher and lower atoms respectively with an elevated substrate atom below the adatom in (b) representing the pseudo-dumbbell. The projected ISF states atop the 7x7 are shown in (c) where theoretical evidence for the SFSS occurs around the corner DAS adatoms. In the DAS model in (d) the CDs shown below the adatom correspond to the SFSS CDs anticipated based on an intrinsic stacking fault, ISF. [43] (In the side views, the smaller atoms represent atoms behind the cross section.)

The SFSSs expected from the DAS stacking fault are shown near the adatoms on the right of Fig. 16 (a) and (b). These three (orange) CDs can contribute to the triangular shape of the corner adatom's CD found in the calculated STS images [19] in Fig. 11a. The projections of other equivalent bulk-like ISF states are also shown in (c) and may form trimer like CD features in other regions of the unit cell. Again, these SSFSs are unique to the faulted DAS structure and do not arise in between the adatom and restatom states as observed experimentally. The FSS appears next to the adatom as a tear drop which is rotated $60^{\circ}$ from the triangular SFSS states of the DAS.

In the DFA structure the adatom, restatom and FSS state can all be involved to different degrees with in-plane bonding of the $sp^2$-$sp^3$ hybridized structure and share their electrons. Both the restatom and adatom states appear in Fig.11 to be mostly localized at certain atomic sites but the FSS being associated with the honeycomb lattice is more delocalized. Within a simple one electron model one can envision charge redistribution from the adatoms to the restatom and then to the FSS similar to that in the DAS model. This higher electron energy of the FSS relative to the restatom arises from its closer proximity to the adatom and the resulting larger Coulomb repulsion.

EXPERIMENTAL VERIFICATION OF THE CHARGE DENSITIES OF THE DFA:

An independent experimental result that supports the DFA model arises from an analysis of x-ray and electron scattering data to refine the valence CDs of the 7x7 surface.[21] Namely, a bond-centered pseudo-atom (BCPA) formalism was used to refine the valence CDs from scattering measurements. Fig. 17 (a) shows the CDs for the DFT calculation which agrees with all prior DFT results, and (b) the BCPA refined CDs. Due to the large number of parameters in BCPA, various forms of averaging are required to reduce parameter space which may average out some of the 'refined' CD features as well as spread them between other periodically related atoms.

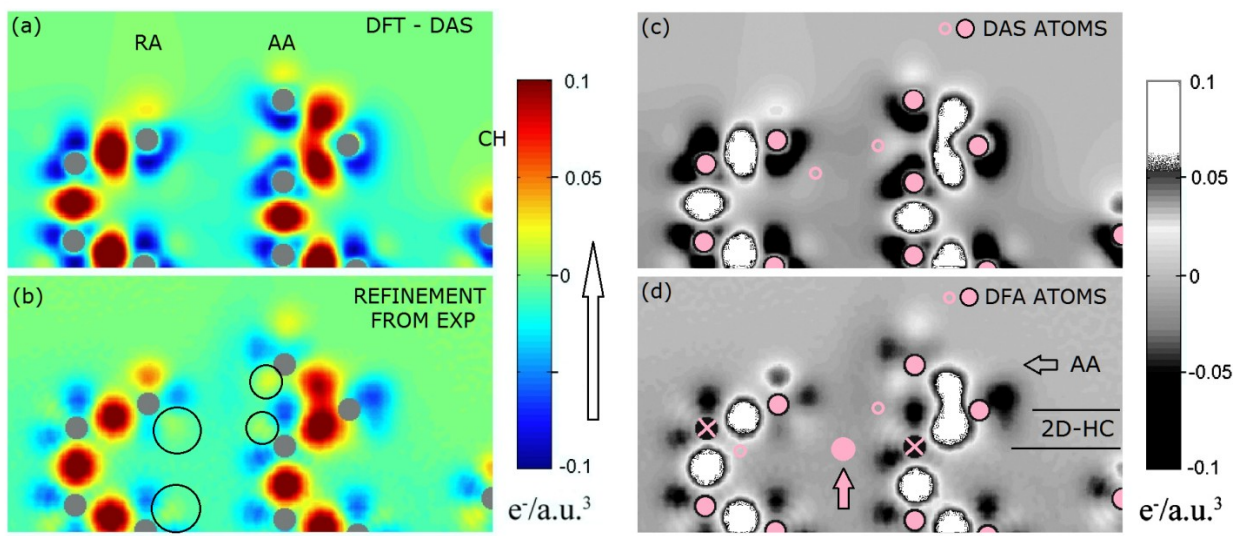

FIG. 17. (Color online.) Calculated CD differences for the restatom (RA), Adatom (AA) and corner hole atom (CH) on the faulted side of the 7x7 unit cell in (a)-(d) with (b) and (d) showing the experimentally refined CD differences from X-ray and electron scattering .[21] (c) and (d) more clearly indicate the atoms for the DAS and DFA structures. All differences reflect the change in charge density found from those of atomic Si. The arrow on the color scale shows the degree of change in the CDs in the refinement where those areas changed are circled in (b) The atom positions in (c) and (d) are a lighter tone (pink) while the open circles are for atoms behind this section. In (d) the non-existent DAS atoms are crossed out and DFA atoms are shown. Here the up arrow indicates an atom in the honeycomb in between the restatom and the adatom.

The differences in CDs are shown in Fig. 17 (a) and (b) after subtracting off the atomic charge density of each atom in the DAS structure. This is used to more clearly show changes in the CDs associated with bonding from the remaining atomic charge. The refined CDs in (b) include all dangling bonds as well as the charge transfer from the adatoms. The CD refinement shown in (b) indicates significant changes in the charge distribution from those found in the DAS calculation. For example, right below the restatom charge in (a), a dramatic increase in charge from about -0.8 e-/a.u. to +0.2 e-/a.u. is found and circled in (b). Fig. 17 (b) also shows several other low areas of CD for the DAS structure that similarly turn into areas of higher CD, again circled. The spatial extent of these circled CDs as well as their change in magnitude of ~0.1 e-/a.u. are comparable to those for the restatom and adatom in the original DAS structure shown in (a).

These are very significant changes from the DAS results and suggests either of three things: (1) that the LD calculation is grossly inaccurate, (2) that this methodology is over fitting noisy data or that (3) different CDs and bonds exist at the surface. In keeping with the DAS model, this change in the adatom CD was interpreted by the authors as due to greater anti-bonding character between the adatom and the atom below it. [21] This would explain why the measured adatom heights are larger than calculated. However changes in the CDs near the restatom and other atoms could not be explained. A different origin for these CD changes becomes evident in considering the DFA model.

The black and white images in Fig. 17(c) and (d) show the 'refined' CDs, but now with either the DAS atoms or the DFA atoms indicated, respectively. In (d) the DAS atoms are crossed out and the DFA atoms are indicated. As prescribed by the experimental diffraction data, the BCPA in (d) shows a redistribution of bond charge away from the DAS bonds shown in (c) to regions of the intra-layer bonds of the 2-D HC layer of the DFA. The presence of the honeycomb atom between the restatom and adatom in (d) indicated by the up arrow now has bond charge associated with it. The reduction in bond charge below the adatom for the DFA is simply due to no longer having an atom below it ! The additional lateral CDs to these 'new' DFA atom, represents bond charge to atoms in the honeycomb. Overall, these changes in CD appear to represent 'bonds' in the DFA structure and thereby support the DFA model.

There are many other experimental results that are problematical for the DAS model which the DFA model accounts for in addition to those already mentioned. A few include the reversible deconstruction of the Si111 7x7 with Si adatom adsorption at low temperatures,[69] the structural features of several other meta-stable Si structures,[70,71] and the unusual interaction potentials of He atoms[72] and positrons[35] in scattering from the 7x7 which are consistent with an $sp^2$ hybridized layer. The occurrence of [1-10] oriented P-BCs for 1-D metals[73] or and alkalis on Si111,[74] further suggest that the PBCs are part of a flexible honeycomb motif that occur in other linear or ribbon like 1-D structures.

POSSIBLE LIMITATIONS OF CURRENT THEORETICAL CALCULATIONS:

A preliminary calculation of the 5x5 DFA structure shown in Fig. 12 using GGA VASP calculations with the PBE functional and parameters used earlier[75] show bonding interactions, but fail to converge from the ideal starting structure shown in Fig. 12.[76] Further calculations were discontinued due to a change in research venue. One rational for this failure amongst others discussed here is that systematic defects of ~1-2 atom per surface unit cell, such as vacancies, dopants, or foreign atoms as reported for the 5x5 [59] (and 7x7 [77]) may change the nature of such ideal 'model' structures. Namely, changing the number of electron per unit cell from odd to even will alter the states near $E_F$ and assist in initiating the DFA structure. The remainder of this discussion focuses on current DFT calculated properties within the LD density approximation of 'ideal' structures without defects or doping. Evidence for such non-ideal structures are discussed in the supplement including their role in prior x-ray[31] and TEM analyses.[78]

As presented, there is significant disagreement between DF calculations of the DAS structure and several measured electronic and structural properties of the 7x7. Recent theoretical work has also shown several electronic interactions that may not be accounted for properly in LD calculations on Si111 surfaces.[63, 79-91] This includes the additional bonding from van der Waals, vdW, interactions in π-bonding, changes in electron screening arising from the discontinuity in translational symmetry at the surface (or any boundary layer), correlation effects associated with metallic screening or with electron phonon coupling as well as the role of stationary bulk-derived Bloch states/waves that extend and couple into the ad-layer. All may stem from LD approximations in DF theory that worked well for 3-D materials but which may fail for certain 2-D systems. These unusual effects may also arise in other 111 surfaces with 2-D layers on them. π-bonding would appear to be one of the simpler explanations arising from the unique stereochemistry of the terminal atoms of the DFA structure. On the other hand electronic interactions between bulk block states and the surface atoms or the use of bulk-like functionals could also favor certain registries and bonding on the surface. Differences in the symmetry of the terminal layer from the bulk may also favor different structures.

For example, a theoretical study of the bonding of Benzene to Ag111 using a new method to correct for dispersion interactions [63] provides about a 0.5 eV larger bonding energy relative to conventional 'PBE dispersion corrected' DFT calculations. Similar calculations for larger cyclic molecules on Ag111 and Au111 surfaces also show significant additional bonding ( ~1 eV) especially for extended cyclics. [79,80] Given the magnitude of vdW interactions, more accurate dispersion corrected calculations [79-81] could be a 'game changer' in predicting π-bonded silicon structures and their properties.

Further, correlation effects can be important on Si111 as the 111 symmetry leads to large inter-atomic interactions whose contributions to the electronic energy can dominate the ground state. [82, 83] The inclusion of electron-electron correlations is a very difficult problem and has not been done at an ab-initio level. The role of correlation effects have been modeled to qualitatively account for the metal-insulator( M-I) transition of the 7x7.[84] Here correlation effects allow a metallic surface to arise for the 7x7 as observed [46] but not for the 5x5. [60] This difference for the 5x5 is proposed to arise from the different arrangement of adatoms which results in a different filling of the adatom bands near EF. For the 5x5 the correlation effects are localized primarily on the corner hole [84] leading to differences in interactions that push down the half filled band at $E_f$ to create a surface state gap.

In this approximate theory, correlation effects in the 7x7, are found to arise from adatom pairs, excluding the adatoms around the corner hole, and found to play a role in forming a partially filled metallic band at Ef. In principle, the FSS could correspond to this new highly correlation filled band which simply does not arise in DFT calculations. For the DFA structure these same correlation effects can also arise between the different adatoms on the surface and be augmented by the π–interactions of D-PB. For either model, the larger RT gap of the 5x5 observed versus the small 40-70 meV 7x7 gap found at T ~ 20 °K, would argue that thermally activated carriers in the 5x5 (one proposed mechanism for the 7x7 M-I transition [46]) may not occur for the 5x5 before the thermally activated atomic rearrangements occur, converting it to a 7x7. Having a single mechanism for interactions and transitions of these structures is appealing whereas the aforementioned model for electron correlation is found to be structure dependent.

The 7x7 is also interesting in that this conversion exclusively occurs well above the temperatures where the surface will be metallic, [47] and stronger screening has developed. While such screening may enable vdW interactions in the 7x7, the 5x5 is not metallic at room temperature. [60] This lessens the appeal of vdW bonding and the likelihood of a universal mechanism that produces this family of structures.

In addition, any screening at the surface, is also expected to be very asymmetric due to the rapidly varying potential at the surface.[85] This alone can modify the electronic interactions in the top layer. Current functionals may suppress such asymmetric screening, and to counter this, one could envision use of a different functional for the top surface layer in a ' hybrid' type of local density calculation.

Recent experimental results suggest that the bulk wave-functions directly influence the interactions and structure in the top surface layer. This essentially reflects the wave nature of substrate electrons and their role in forming stable bonds of periodic structures above the substrate - the so called "dangling bonds". One example arises in experiments that found a low temperature phase of silicene forming atop a monolayer template of Ag in a $\sqrt{3} \times \sqrt{3}$ array on Si111.[68] This template has a half monolayer of Ag in the 'inequivalent Ag Trimer' (IET) structure that has an unusual symmetry . As a result it is referred to as a 'symmetry broken phase'.[86] Numerous DFT calculations of various Si adlayers atop this template shows convergence to a variety of stable non-bulk-like Si structures.[68] Earlier multilayer calculations of Si in a $\sqrt{3} \times \sqrt{3}$ array on Ag111 also found new non-bulk like silicene layers with low buckling.[87] Both of these examples contrast DF calculations of the reconstructed Si 111 surfaces which produces covalent structures with bulk-like tetrahedral bonding extending into the 2-D surface atomic layer.

One explanation for such behavior is that the 'broken' symmetry of the monolayer template dampens the bulk Block states/waves and allows the intra-layer interactions of a 2-D structure to prevail. This phenomenological explanation suggests that in DFT calculations of bulk terminated surfaces, the potentials or functionals used for the surface atoms do not allow sufficient changes to the surface wave-functions between iterations to allow other surface structures to develop. This may also reflect some sort of a energy or momentum barrier needed to be overcome converging to the DFA structure.

One way to evaluate the influence of such bulk states is by considering a model reconstructed system that will not have these bulk states. Here a 4 layer system could be constructed with the first two atomic layers consisting of the DFA structure and the below a single bulk-like bi-layer terminated with hydrogen atoms. This should allow the intra-layer bonding to develop within the top layer together without the bulk like inter-layer interactions from the bi-layer atom below, as found for the monolayer Si/Ag system.[41] Mapping of charge densities differences between this 4-layer calculation and one with more bulk bi-layers should provide insight into the role of bulk states in modifying the surface bonding.

Self-energy effects give rise to many electron self-energy interactions, SEIs, that are not included in conventional exchange-correlation functionals.[88-90] SEIs are known to alter electronic properties in 3-D systems, as recently found for the TTF-TCNQ system where a semi-local functional has been applied. This reveals a different, non-intuitive behavior than predicted by conventional functionals.[88] The occurrence of SEI effects has also been shown to be more likely in systems when localized bonds coexist with delocalized bonds.[87] That is certainly the case in the DFA model with localized adatoms bonded to the DFA with its delocalized 2-D π-bonding. The DFA layer with the delocalized bonds in the honeycomb also resides atop a covalently bonded substrate layer, i.e. with localized bonds.

It is also generally accepted [10-22] that the 3-fold coordinated surface adatoms produce a partially occupied localized $p_z$ 'dangling bond' from $sp^3$ hybridized adatoms by essentially satisfying the dangling bonds of the substrate. The adatoms of the DFA structure are expected to play the same role in balancing the charge within the unit cell. As a result of the faulted adatom motif, these localized adatom states on the two sides of the unit cell also have different electron occupations as experimentally observed (see Fig. 9c.) Such population differences between the two sides of the these relatively large unit cells, also creates a 'charge transfer' situation. Thus the DFA not only has localized and delocalized bonds but significant charge transfer within this large surface unit cell.

As a result, SEI effects should also be considered in calculations of either the DAS or DFA structures. It is established that SEI effects can alter both the energy levels and total system energies calculated in conventional DF calculations,[88-91] and could potentially even account for the FSS even within the DAS model. Nevertheless, many experimental results still favor the DFA model.

Very recently, studies of the very complex growth of graphene on SiC has shown that an unexpected intervening Si monolayer structure arises under high temperature growth conditions with Silane.[92] Thermodynamic ab-initio calculations that include the vibrational free energy show that these higher temperature Si structures become more stable than the T=0 °K structures by ~ 0.15 eV per 1x1 unit cell.[92] Such temperature-dependent energy changes may also allow the DFA family to form.

Related to such finite temperature effects, there may also be an energy barrier between the DAS structure with it half filled state at $E_F$ and the DFA structure that may affect the energetics of their formation. An energy or momentum barrier between these structures could trap atoms in certain regions and may prohibit T=0 °K DFT calculation from converging to a lower energy structure. The failure of the original 5x5 DFA structure to immediately converge[76] is disconcerting since it should avoid such a barrier. However, it must be noted that while the 5x5 structure is observed upon annealing of cleaved doped samples,[60] a 5x5 is not observed by LEED upon heating of cleaved ultra pure silicon samples. As discussed in the Supplement, many earlier LEED studies of cleaved surfaces never observed the 5x5 during the 2x1 conversion to the 7x7, and have conflicting results that depend on cleavage quality. This leads to the speculation that the formation of the larger domains of 5x5 may be aided by cleavage quality/ perfection and/or bulk dopants. The possible role of cleavage methods, defects and dopants in the reconstruction process is further discussed in the Supplement.

A more general symmetry effect may also arise that stems from the nature of the 2-D wave functions needed to adequately represent the behavior of electrons on both side of these DFA supercells. This is in addition to all the different electronic effects so far discussed. Namely, the wave functions used in conventional band calculations may not have sufficient flexibility to adequately describe the stationary 2-D states of such systems. Use of a Slater determinant type of wave-function comprised of one wave-function, $\psi(r)$ and its reflection, $\psi'(r) = \psi(-r)$, would allow more flexibility to define a solution to the stationary waves/states of the two cells that create the DFA supercell, i.e., both sides of the unit cell. Such multi-centered wave-functions are more typical of Hartree-Fock formulations that treat electron spin in solving the wave equation. Such a Slater determinant form of a Fermionic wave function can lower the system energy by producing an additional energy gap between two set of otherwise degenerate states. If one considers the FSS to be this split off state, then within a one electron model it can provide a significant contribution of ~0.4 eV to lower the system energy.

In summary, there are numerous physical explanations for why current DFT calculations may not be accurate for Si111 surfaces and, in particular, why the preliminary calculations for the 5x5 DFA structures have failed to converge. Yet a wide body of experimental evidence supports the DFA structure, a new 2-D honeycomb structure on Si111 and polymorph of the orginal DAS structure. If however, the DAS structure is correct and all the proposed theoretical limitations are inconsequential, then the most likely reason for the disagreement between calculated and experimental STM and STS images of the FSS, is the failure of the Tersoff-Hamann approximation.[46] This approximation is very important due to its wide use in comparing calculated STM images to experimental STM images to derive structure, particularly for semiconductor surfaces and silicene structures. However, the paradox of the unusual adatom height of the 7x7 as well as the 0.4A longer dimer bond lengths found experimentally relative to LD calculations of the DAS structure would still remain unresolved.

The Supplemental material for this paper can be found at: http:

SUMMARY:

A new 2-D structure for the 7x7 is proposed whose nature and interactions are very different from that in the DAS model. In the DAS model the terminal layer is driven by bulk-like covalent $sp^3$ bonds, with the dimers and corner holes formed to facilitate these covalent bonds. This results in a top layer with many stressed covalent bonds. In contrast, the DFA shows a change in stacking from that of the bulk bi-layers and the formation of two honeycomb cells having mirror symmetry to one another that make up the unit cell. Both cells appear to be bonded to the substrate by a new type of $\pi$–interaction that can occur around the corner hole and along the unit cell boundaries. The FSS appears to be a delocalized state of this type of supercell honeycomb motif that resides largely in the plane of the honeycomb and has accepted charge and interacts with the subsurface atoms and the adatoms. The corner hole, whose $\pi$-bonding creates the proposed silizene structure, together with the interdigitated $\pi$-interactions along the unit cell boundary dominate and drive the formation of the 2-D, $\pi$-bonded DFA structure. This combination of the delocalized FSS state and $\pi$-interactions will allow the DFA 2-D layer to have greater $sp^2$ character. The adatoms in the DFA likely play a similar role to those in the DAS model - to interact and balance the charge density of the highest lying states. In the DFA, the restatoms and the FSS both accept charge and lower the total electron energy as done in the DAS model.

The DFA structure and, in particular, extended 2-D $\pi$-bonding interactions may have important implications to the design of silicenes with low buckling and enhanced sp2 character, or for their use as a template for the growth of other novel layered structures.

Then there is also the broader question as to whether the experimental paradoxes discussed here simply reflect a limitation of current DFT calculations to treat Si surfaces due to the many approximations used therein. First, there is the strong asymmetry of the potential at the surface and the resulting asymmetry in screening within the terminal 2-D layer. Second, the existence of other long range interactions may not only change semi-local, many electron self- energy interactions of such systems, but produce additional bonding via longer range interactions from extended $\pi$-bonding or vander Waals bonding. Finally the initial geometry or momentum sampling selected for current local density calculations, or the form of the wavefunction used, may prevent such calculations from achieving its lowest energy state/structure.

Acknowledgements: The IBM Watson Laboratory and ONR are gratefully recognized for their support during the author's Research career, as well as the University of Pittsburgh for library access. Thanks to Ruud Tromp and Bob Jones for critical discussions as well as to Benjamin Geisler for providing unpublished results. Discussions with other colleagues about their work are acknowledged: O. Paz, J. M. Soler, L. D. Marks, J. Ciston, J. Ortega, R. M. Feenstra, N. Okabayashi, H. Ibach, M. Svec, E. W. Plummer, I. K. Robinson, H. Tochihara, N. Takagi, G. Benedek, F. J. Himpsel, P. A. Bennett, Ph. Avouris,

S. Y. Tong, W. A. Goddard, B. Voigtlander, P. Kocan, W. Jiang, P. Hansmann, E. G. C. P. van Loon, S. Meng, P. DePadova, M. Rohlfing, S. Kuemmel and E. Vlieg.

-------====-------

(69) J.E. Demuth, Surface Science Reports, (to appear).

(70) P. Kocan, O. Krejci and H. Tochihara, J. Vac. Sci. Tech. **33**(2) 021408 (2015).

(71) J. Ikeda, W. Shimada, S. Mizuno and H. Tochihara, Japanese J. of Surf. Sci. **22**(6), 382 (2001).

(72) G. Lange, J. P. Toennies, P. Ruggerone and G. Benedek , Europhys. Lett., **41** (6), 647 (1998).

(73) S. C. Erwin, I. Barke, and F. J. Himpsel, Phys. Rev. B **80**, 155409 (2009).

(74) L. Lottermoser, E. Landemark, D.-M. Smilgies, M. Nielsen, R. Feidenhansl, G. Falkenberg, R. L. Johnson, M. Gierer, A. P. Seitsonen, H. Kleine, H. Bludau, H. Over, S. K. Kim, and F. Jona, Phys. Rev. Lett. **80**(18), 3980 (1998)

(75) W. Jiang, Z. Liu, M. Zhou, X. Ni, and F. Liu, Phys. Rev. B **95**, 241405(R) (2017).

(76) W. Jiang, Univ. of Minnesota, private communication.

(77) Fourier transforms of large area STM images show diffuse backgrounds from the defects (Refr: 49) while analysis of unaveraged TEM data indicate consistent defect locations well above statistical noise. (Refr. 76).

(78) E. Bengu, R. Plass, L.D. Marks, T. Ichihashi, P.M. Ajayan, and S. Iijima, Phys. Rev. Lett. **77,** 4226 (1996).

(79) G. Li, I. Tamblyn, V. R. Cooper, H-J. Gao, and J. B. Neaton, Phys. Rev. B **85**, 121409(R) (2012).

(80) K.E.Riley, M. Pitonák, P. Jurecka, P. Hobza, Chem. Rev. **110**(9), 5023 (2010).

(81) J. Hermann, R.A. Di Stasio, A. Tkatchenko, Chem. Rev. **117**, 4714 (2017).

(82) E. G. C. P. van Loon and M I Katsnelson, J. Phys.: Conf. Ser. **1136,** 012006 (2018).

(83) P. Hansmann, L. Vaugier, H. Jiang, and S. Biermann, J. of Phys: Condensed Matter **25**(9), 094005 (2013).

(84) F. Flores, A. Levy Yeyati and J. Ortega, Surface Review and Letters, **4**(2), 281 (1997); J. Ortega, F. Flores,  R. Perez,  and A. Levy Yeyati, Prog. Surf. Sci 52, 229 (1998).

(85) M. Rohlfing, Phys. Rev. B **82**, 205127 (2010).

(86) H. Aizawa, M. Tsukada, N. Sato, and S. Hasegawa, Surface Science **429**, L509 (1999).

(87) H. Fu, L. Chen, J. Chen, J. Qiu, Z. Ding, J. Zhang, K. Wu, H. Li and S. Meng, Nanoscale **7**, 15880 (2015)

(88) T. Schmidt and S. Kümmel, Phys. Rev B **93**, 165129 (2016).

(89) S. Jana, B. Patra, H. Myneni, P. Samal, Chemical Physics Letters **713**, 1 (2018).

(90) C. Park, V. Atalla, S. Smith, M Yoon, ACS Applied Materials and Interfaces, **9**(32), 27266 (2017).

(91) T. Körzdörfer, S. Kümmel, N. Marom and L. Kronik, Phys. Rev. B **79**, 201205R (2009).

(92) J. Li, Q. Wang, G. He, M.Widom, L. Nemec, V. Blum, M. Kim, P. Rinke; R.M. Feenstra, Phys. Rev. Matls. **3**, 084006 (2019).§

Supplemental Material for:

Evidence for a New Family of 2-D Honeycomb Surface Reconstructions on Si(111)
J. E. Demuth   (jedemuth7x7@gmail.com)
Naples, Florida 34114

This section discusses additional features of the DFA and earlier work that will be useful to the reader in understanding (1) the new Digitated Faulted Adatom, DFA, structure and its relation to the Dimer Adatom Stacking fault, DAS, structure[1], (2) the unusual features of the DFA that may provide insights as to its formation, (3) additional ways to distinguish between the two structures, (4) the possible role of intrinsic defects or dopants in its formation as well as (5) limitations of early xray and TEM work that supported the DAS structure.  (Note that all images adopted from earlier work here and in the main paper are done so with the author's written consent and referenced to their original work.)

THE 2-D LAYERS AND THEIR STACKING IN THE DFA and DAS STRUCTURES:

Fig. 1S shows top views of the unreconstructed 1x1 surface in (a) and (b) the arrangement of atoms in the DFA unit cell. The registry of the DFA to the bulk substrate can be seen in the corner holes. The honeycomb atoms of the DFA are shown as light (higher) and dark (lower) circles on a honeycomb mesh. The adatoms of the DFA are the yellow circles. All rest atoms are blue and their position in (a) are marked with blue cross hairs. The 1x1 substrate layer shown in (a) is to allow the reader to overlay the two layers for further inspection. The side view in (c) also shows two rows of atoms whose stationary states, i.e., Bloch waves, project upward along bulk bonded atomic chains onto the surface area.

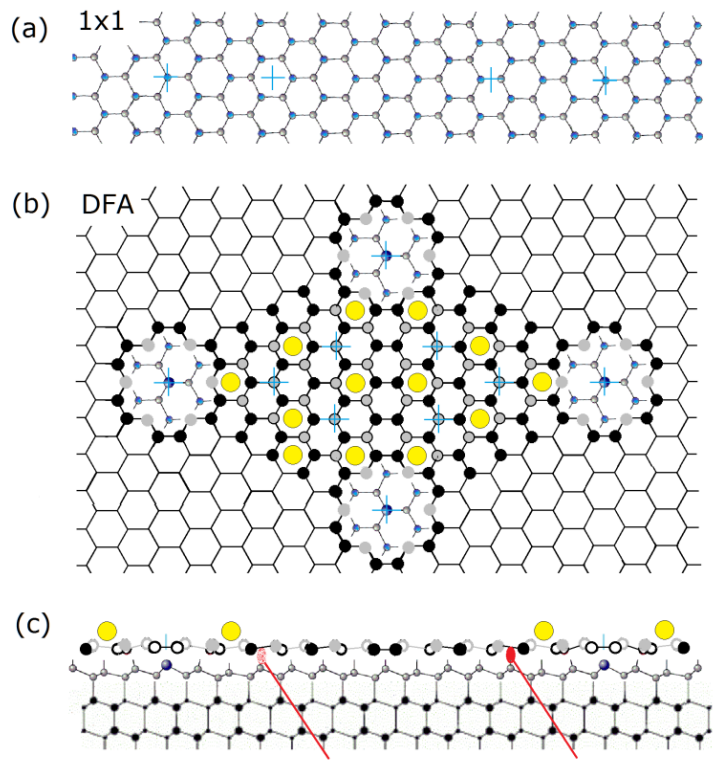

FIG. 1S: Top view of the 1x1 substrate and the proposed DFA structure in (a) and (b). The 1x1 structure shows the underlying bulk bi-layer directly below the DFA structure with registry marks of the corner hole and restatoms. The DFA side view in (c) shows a few of dangling bond charges of the substrate as periodic projections of the bond charge between the bulk Si bi-layers.

Comparisons of the DFA to the details of the most recent spin unrestricted local density calculation of the DAS structure [2] has already been provided in the main paper. An important point is that the DAS shows distortions of the substrate several bi-layers into the sample. These DAS substrate distortions will create a 7x7 pattern in subsurface region and contribute to the all the fractional order intensities which have been neglected in many interference based analyses. Without the dimers and the almost perfect nestling of the DFA into the substrate bilayers, the subsurface distortions of the DFA can be expected to be greatly reduced but still present in the DFA structure. However, the different orientations of the honeycomb atoms on each side of the unit cell relatice to the bulk staking makes one side different as shown in (c) which can explain the asymmetry observed in STM. [3] A discussed in the main paper this mirror symmetry produces the supercell structure of the 7x7 which unlike the DAS structure has full c3v with the unit cell.

In the DFA model shown in Fig 1 (b), the honeycomb atoms around the edges of the corner hole are indicated as solid grey since they are coordinated differently than the interior grey atoms and are likely displaced. (Such displacements can also be surmised from the Pattern map discussed later.) These grey atoms can have two displacement components: one downward to bond closer to the substrate atoms and one toward the center of the corner hole to relieve the stress created by the flatter $sp^2$ - $sp^3$ hybridized atoms comprising the honeycomb mesh. The blue crosses on the high (circled gray) honeycomb atoms correspond to 'restatoms' as seen experimentally as well as on the central atom in the corner hole.[4] The central atom in the corner hole is likely similar to the 'restatom' in the corner hole of the DAS model in (b) and is also likely elevated as found in the DAS calculation.[2] These distortions will together with interaction between the adatom and atom below it inevitably introduce a weak 7x7 periodicity in the subsurface atoms of the DFA as well.

Graphite, as well as the DAS and DFA structures all have similar hexagonal types of layers. The silicon layers in the DFA and DAS structures both have alternating up down atoms versus the flatten layer of various forms of graphitic carbon. Various carbon layers are well know to have different stacking sequences to provide stability. Graphitic carbon can convert to diamond which has the same lattice structure and stacking sequence as silicon, but does so only under the extreme conditions of high temperature and pressure. The stacking sequence of the bi-layer and honeycomb Si structures found under normal conditions is useful to consider since they form atop the same underlying bulk Si substrate.

A schematic of the stacking in these materials is show in Fig. 2S. This compares the stacking in hexagonal graphitic carbon, bulk silicon and a honeycomb layer. This honeycomb layer eventually forms the DFA structure after the removal of atoms to form the corner hole and the addition of adatoms. The DAS model also forms corner holes but then has sections of the top bi-layer of the bulk structure rotate by 180 degrees to create a mirror image of the other side and utilizes dimers to rebond the two sides of the unit cell together at the 'fault' boundary created by the stacking fault. In contrast the DFA creates a mirror image across the unit cell boundary by inverting the puckering of the honeycomb atoms on each side of the unit cell: atoms that were up are now down and vice versa. The formation of the corner holes in the either models produces extra atoms needed to form the adatom structures. If more/less adatoms are needed to form the overall ordered structures they can diffuse from/to step edges. [5-7]

The formation of these 2n+1 x 2n+1 structures does not form spontaneously at room temperatures but arise from heating the 2x1 cleaved surface or precipitates/nucleates at high temperatures from a molten surface layer. The local symmetry braking and rearrangement of the 2x1 structure starts at 245 °C [6] with the formation of areas of disordered adatoms. Then above 330°C , the 5x5 domains grow with an increasing amount of 7x7 which becomes predominant above ~ 600- 650°C. [7] Two kinetically distinguishable mechanisms for the conversion of 5x5 to the 7x7 have been proposed. [7] However, only a 7x7 is observed on single domain cleaved surfaces upon annealing to ~600.[8] While the formation of a 7x7 from a near molten surface is the most common way to produce the 7x7, it is unclear what happens to trace impurities and dopants during this process. This is discussed later after considering the properties of 'ideal' structures.

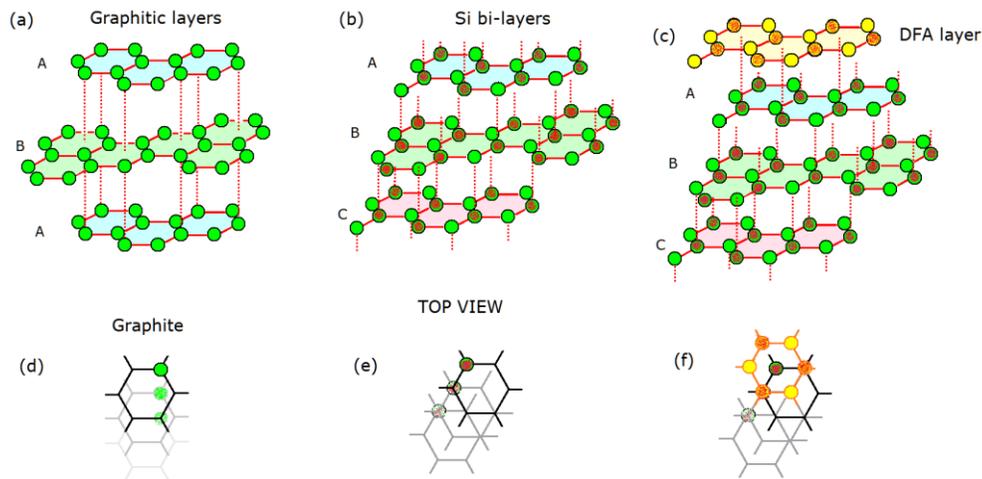

Fig. 2S Comparison of the layer stacking of (a) hexagonal graphite, (b) bulk Si (bi-layers) and (c) the honeycomb layer of the DFA structure (without adatoms or the missing atoms that form the corner holes.) For simplicity, the atoms are shown as planar but in both (b) and (c) the atoms in these layers are buckled. The layers of each are color coded to indicate their stacking sequence as shown in (d)-(f).

The DFA has a different layer stacking than either graphite or bulk silicon as shown in Fig. 2(c) which will produce three different orientations each having 3 fold mirror symmetry planes in them. In addition to this, different nucleation sites for these structures will also create larger domains of mismatched DFA or DAS structures as observed experimentally. [4-7] This has important implications in modeling the DFA structure since these substrate registry differences require that the 3 3-fold symmetry structures must be averaged in any macroscopic measurements of such structures. In contrast, most analyses of the DAS structure have exclusively assumed 6-fold symmetry.

Steps can pin the corner hole nucleation sites and may also pin the stacking direction of the DFA structure. In particular, interest in 1-D wires and related structures has focused attention on miscuts of ~1° which can produce aligned terraces only 2-3 unit cell wide. [9] Larger random terraces widths on nearly flat surface will produce random domains on a macroscopic level that will be averaged by macroscopic measurement techniques. The surface will appear to be 6 fold symmetric. Macroscopic formation of a larger single domain structure will be an exceptional case but has been sometimes observed ( as shown later in Fig. 16S.) Explicit 3- fold symmetry was first noted in the non- linear second harmonic studies of the Si111 7x7 [8] and later found in X-ray studies.[10] This difference in symmetry has important implications in modeling to extract surface structures from experimental data that will be discussed again later.

AN IMPORTANT CONSEQUENCE OF 3- FOLD SYMMETRY OF THE DFA:

A significant consequence of this particular c3v mirror symmetry for the 7x7 arises in the Patterson function analysis and the Patterson map, PM, that has been derived from the TEM diffraction intensities.[1] Here this symmetry creates 3 Harker planes as discussed in the main paper with reflection symmetry about each plane across each side of the unit cell. This creates an unusual situation in the pair correlation function of the atoms on each side of the unit cell. The Patterson function which is originally expressed in 3-D as

$$P(u,v,w) = \sum_{hkl} |F_{hkl}|^2 \, e^{-2\pi i(hu+kv+lw)}$$

In 2-D this becomes $P(u,v)$. The complete mirror symmetry across the boundaries of each supercell of a DFA structure allows this equation to be simplified to first order as:

$P_{ij}(u) = \rho(r) * \rho(-r) = \rho_i^2(r)$ since $\rho_i(r) = \rho_j(-r)$ due to the Harker planes.

$P(u,v)$ thus becomes a charge density map of the atoms along the two surface directions, $u,v$, within the unit cell boundary! Aside for the null peaks that can arise from out of phase Fourier components which cancel, this should allow the Patterson Map to be directly mapped onto the real space lattice to show most of the relative positions of the atoms. This approximation is expected to be the best in the center of the unit cell and poorest around the corner holes based on the coherence of the electron wave fronts that support interferences/diffraction or standing waves along the surface in these areas.

An overlay of the PM on primarily (but not completely) the right side of the DFA structure is shown in Fig. 3S. On the left side of the unit cell various atomic features of the DFA that coincide with the identical PM features on the symmetry reversed rights side are drawn in. Here the colored circles reflect the type of atom in the DFA as shown in Fig. 1S. Many of the upper red interior honeycomb atoms have no PM features, but the lower blue ones all exist except for those bordering the corner hole. The crosses shown represent the terminal atoms of the underlying substrate that appear in the centers of most honeycombs or under the adatoms. These substrate atoms should not show up if the substrate were a true 1x1 layer. They appear here due to the 7x7 related distortions in the substrate caused by interactions with the ad-layer atoms, particularly the adatom over the substrate atoms, as well as changes to the substrate atoms caused by the corner hole. The distortions of the substrate atoms by the adatom has been discussed in the main paper relative to the behavior of Silicene on Ag(111).

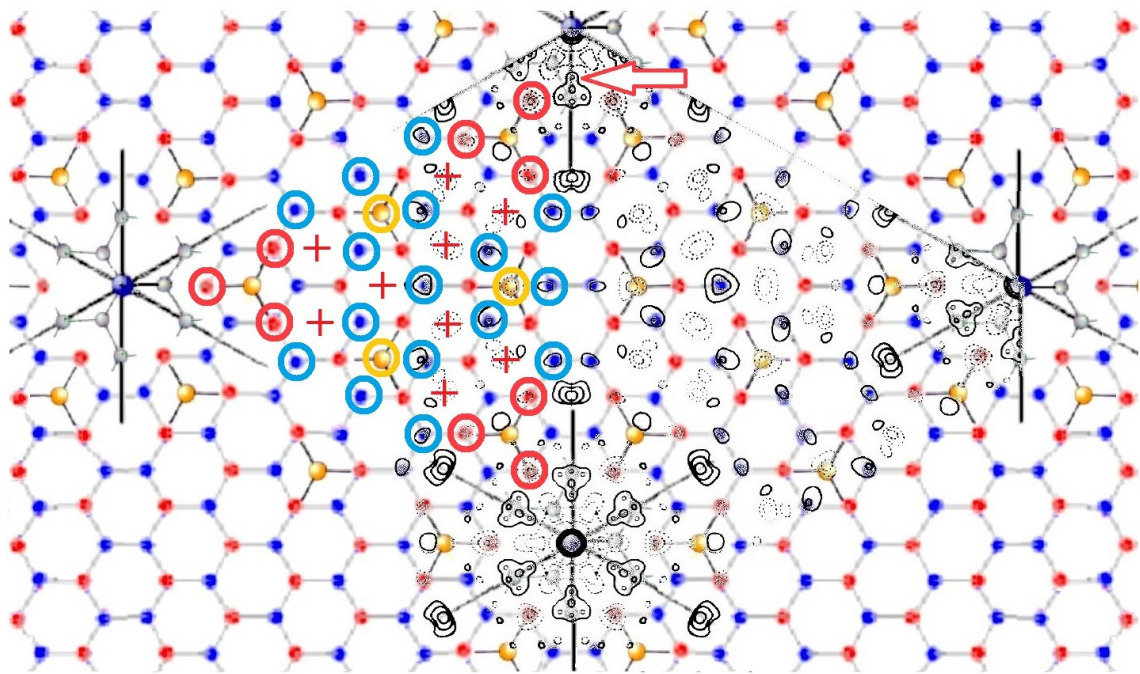

FIG. 3S. Overlay of the Patterson map created from early TEM data with the 'undistorted' DFA model. The blue and red atoms represent lower and higher atoms in the honeycomb. (Second layer atoms are shown as fainter atoms in the corner holes.) The red cross correspond to the underlying substrate atoms while the top arrow indicates the weak feature that was used to confirm the presence of DAS dimers.

The level of agreement of the PM with the atomic positions of the DFA is truly remarkable even though some atoms are not seen. Some of these differences can arise from the averaging to remove non-kinematic features and the assumption of six fold symmetry. However the replication of the PM features with the DFA suggest these are small. This may partially occur since all the diffraction arises from the same Si atoms whereas having different atoms intrinsically produce different scattering intensities and phases, etc. that will alter atom positions in a PM if one kind of atom is missing.

In regards to the upper (red) honeycomb, those that are directly around the corner hole appear but those toward the center region of the unit cell do not. These missing red PM peaks are in the center of the cell and the least distorted of these honeycomb atoms. They are likely on the same plane where all scattering centers completely interfere to produce a null. The appearance of the red atoms around the corner hole suggests they are at different heights than the internal red atoms. The blue pairs of atoms nearest the corner hole also have shifted PM peaks and are weaker probably due both lateral and vertical distortions.  The adatoms show a central structure with trigonal side features similar to that occurring in other regions of PM, e.g., some of the complex features in the corner hole.

The features inside the corner hole are complex and weak but many of them correspond to the atom position within the corner hole.  The three closest atoms to the center of the corner hole correspond to lower atoms of the bi-layer and arise as 6 probably due to the domains of the 7x7x and the averaging of the beams to make them 6-fold symmetric. The next circle of substrate atoms are 6 upper atoms in this bi-layer which this 6-fold averaging does not effect. The various distortions are probably the result of subtle distortions around the corner hole.  The butterfly features along the unit cell boundary has no counterpart and is probably an artifact which a new PM may resolve.  It is surprising how well these PM peak overlays the DFA considering that the 7x7 beams corresponding to the strong integral order beams were neglected.  However, if they are lower or of comparable intensity as the other nearby fractional order beams this neglect may have a minor effect.

As discussed later, comparing the periodic features from a Fourier transform of the 7x7 STM topograph to the kinematic diffraction pattern provides more insight as to the origins of some of the interferences arising in diffraction since STM will only detect the periodicities of the top atoms..

HOW THE FAULTED HONEYCOMB MAY EFFECT THE 2-D ELECTRONIC STRUCTURE AND ELECTRON WAVES

The FSS appears to be an unusual type of delocalized, dispersionless state arising from the interactions within the 2-D honeycomb structure.  As already shown in Fig. 1 (c) the periodic bonds arising from the CDs of the bond charges between bi-layers does project onto the location of a Si dangling bonds and into the honeycomb where the FSS is detected as discussed in the main paper.  However, these projections occur along all the other equivalent projections in three directions implying there is some other interaction that allows the FSS to be observed only in these particular locations for the 7x7.

The bonding states of the 2-D honeycomb must be largely in the plane of the surface, i.e. the $p_{x,y}$ bonds that hold it together.  Its interaction or admixing with the nearby restatom and adatom can provide greater $p_z$ character and allows it to be observed in PES and STM. For STM this means that what it detects is only the CDs that extends into the vacuum and appear between the restatom and adatom near the corner hole.  Away from the corner hole toward the interior of each side of the unit cell, the delocalized FSS can extend downward and couple/interact more with the $p_z$ charge density of DBs of the underlying substrate.  Such a spatial separation of electrons above and below the honeycomb would be favored due to reduced Coulomb repulsion. The FSS near the corner adatom, the most asymmetric of all the adatoms, my thereby reflect a perturbation of the honeycomb $p_{x,y}$ states that push it upward.

Fig. 4S shows a schematic of a hypothetical 4x4 honeycomb array that has the same honeycomb supercell arrangement with mirror symmetry across the unit cell boundary as does the DFA. It is based on an ab-initio study of freestanding planar silicene that is uniform, i.e. unfaulted, and that shows a slightly bucking of 0.44A versus the bucking in crystalline Si bi-layer s of 0.78A - the result of greater sp2 character. [13] Considering this structure from its wave properties is useful conceptually so as to create a basis to extend such concepts to the more complex DFA structure.

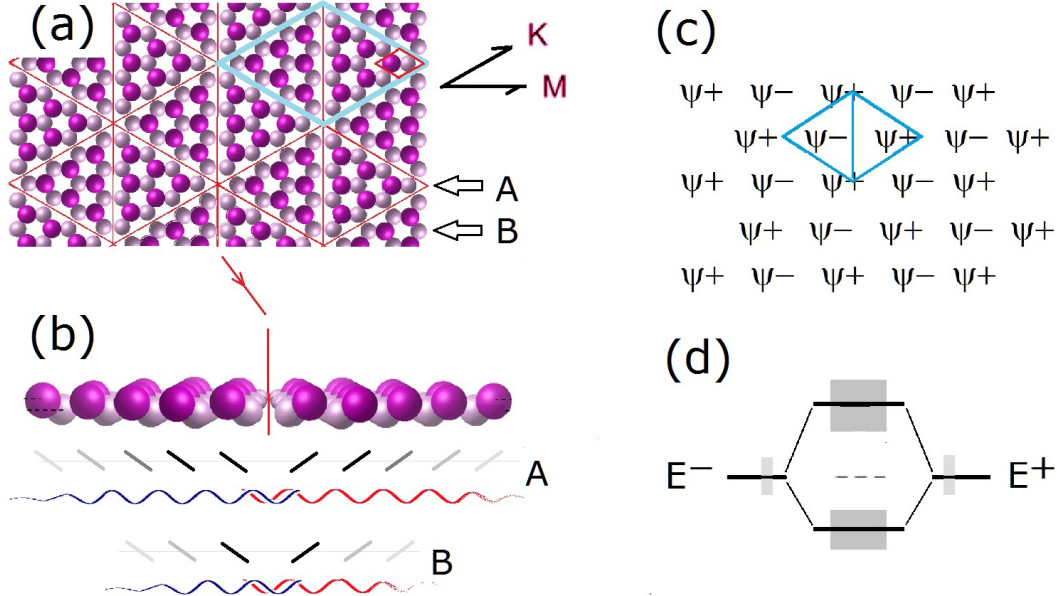

FIG. 4S: An example of a 4x4 faulted honeycomb structure in top (a) and (b) side view of one unit cell. (c) and (d) show how a generalized set of degenerate states with different symmetries can interact and form bonding and anti-bonding states for each interacting state of the original wavefunction.

In looking at the electrons as waves, the faulted structure can be viewed as having two phase reversed cells that comprise the unit cell, i.e. a supercell. For the electron waves of a unit cell to have the lowest energy and be stable, they will have integer wavelengths and the smallest number of nodes. The aliasing at the unit cell boundary changes the interferences of the waves in each supercell and becomes important. In general, the propagating electron waves nominally arising on one side of the unit cell will not match those on the other side and as a result become 'de-phased'. This de-phasing is between the wave fronts of each set of waves in any 2-D direction, and is more complex and conceptually challenging than a rectangular structure due to the triangular shape of each ordered cell.

The interferences along the wave fronts indicated as A and B in 4S (a) are supported by the different number of periodic atoms supporting these wave fronts. The electrons will find waves that support a standing wave with the lowest energy given the lattice symmetry, and they can optimize their energy by 'pushing' the atoms to positions with more suitable periodicities. Further the coherence of these waves (as well as any coherent scattering from periodic rows of atoms) along A and B will be the strongest in the central regions of the unit cell, i.e. B. This means that the central region of the unit cell will have a more well defined "coherent' wave than those along A near the corner hole. This would account for why the PM features are best defined in the center of the unit cell than near the corners. It also means that many properties of a cell will be determined by how these waves interfere at the triangular unit cell boundary making them more difficult to model as simple pair wise 1-D interferences.

In considering the two cells of this faulted system and the symmetry of the boundary between each supercell, one can envision the system to be described by symmetry related wave-functions based on the wavefunction, $\psi^+$, of one cell. The other cell of the supercell being $\psi^-$. Fig. 4S (c) shows a representation of the supercell's wave-functions based on the cells that comprise it. Without the change in symmetry each supercell would be degenerate. As shown in (d) for an odd electron system, these states can interact to produce bonding (occupied) and anti-bonding (unoccupied) states which will lower the system energy. Consider also that each supercell is not one degenerate state but each has a 'manifold' of states that can interact and split up. How the waves do this is unclear at present, but can likely be inferred from the final structures that allow standing waves, i.e., stable states of this superstructure to exist. While no calculations have yet been performed for the 7x7 DFA structure, ab-initio calculations have been performed for freestanding silicene as well as epitaxial silicene on a variety of substrates that are useful to consider. [13]

These recent calculations even though spin restricted do provide some insight as to how the energy states of the electrons in such supercells behave and respond to this supercell boundary condition. This provide another viewpoint of how the FSS may arises and its relation to features of the honeycomb lattice. Fig. 5S shows the calculated energy bands and PDOS for three different silicene structures. (b) shows the partial projected DOS, PDOS of the well established 4x4 Si structure on Ag 111 surface. This 4x4 is essentially comprised of two 3x3 cells with the same type of faulted boundaries as found in the DFA. The PDOS for the bands derived from Ag and Si states are shown and for the Si atoms are divided into the Si atoms on the top or bottom of the honeycomb. While the 'metallic' Ag 3s state plays a role in the bonding, one sees other changes in the states associated with the upper and lower atoms of the silicene honeycomb structure. Fig.5S (a) and (c) show the silicene band structure for the same silicene structure after removing the Ag substrate atoms. The colored bands in (a) show the states of the 6 atoms of this silicene while (c) shows the 12 bottom bands of the buckled structure. The solid line reflects bands for an unfaulted uniform silicene structure. The symmetry change and aliasing of the Si atoms rather dramatically change the allowed energy bands, and creates the gap at the K point of the SBZ. The filled Si band just below Ef is very similar to that found when the substrate is there (b) while several flat regions of the honeycomb bands also are found with the substrate.

Silicene on ZnS (which has a Wurtzite structure) is also a useful system to compare to since it forms a stable unfaulted lattice matched silicene structure. The dark line in (c) represents the PDOS of the silicene layer that was calculated and optimized on ZnS but then had the ZnS stripped away.

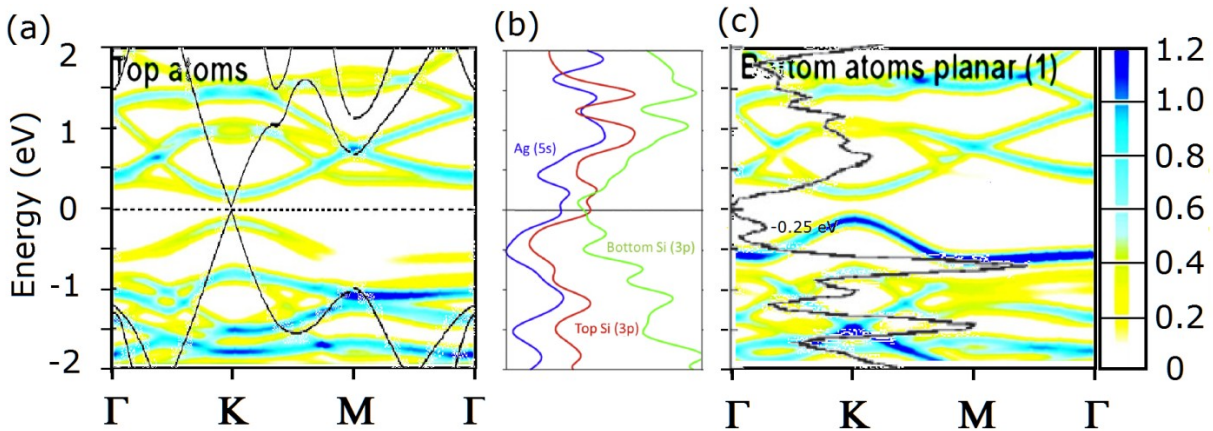

FIG. 5S Calculated energy bands (colored) for the top (a) and (c) bottom honeycomb atoms of Si in the 4x4 Si/Ag111 structure after having the Ag substrate stripped away, compared to (b) the PDOS of the 4x4 Si/Ag111 structure. The dark lines in (a) are the calculated bands for planar freestanding silicene while the dark line in (c) is the PDOS of the optimized Silicene on ZnS structure after the Zn is stripped away. (All calculations shown have been adapted from refr 13 with the author's permission.)

The calculated (complex) bands of the Silicene on the ZnS substrate show an 0.8ev indirect gap and sp3 bonding of Si atoms to both Zn and S. It is found that the interactions have removed the band narrowing at the K pt of the SBZ as seen for the 4x4√3 Si/Ag case.[13] The common underlying PDOS structure for this and the Si/Ag system in (c) arises from the interactions within the distorted honeycomb lattice. The low lying gap state of ZnS can be associated with the splitting of the Si p band for these supercell structures together with the height variations (buckling) of the Si atoms atop the substrate for either ZnS or AG. Since this structure of the silicene on Zn S has no faulted supercell, only the height variations arise. A Mulliken analysis of the combine system, shows strong mixing of the Si and Zn and S states that are indicative of covalent bonding which totally changes these bands. These calculations suggest that the honeycomb lattice of the DFA has strong intra-layer bonding that produces the FSS, and that the substrate bonds to the surface do not disrupted the integrity of the honeycomb layer as found for the case of covalent boning for the ZnS substrate. The formation of a faulted honeycomb supercell in the DFA is very similar to the faulting arising in the 4x4 Si/Ag111 system. While covalent bonding has not been proposed in the monolayer Si/Ag system, the most recent work has proposed dispersive vdW bonding.[13] This supports the idea of π-bonding or dispersive bonding of the DFA to the Si111 surface.

While the metallic screening of the Ag atoms can change the nature of bonding within the silicene structure, it is not evident from these calculations that covalent bonding arises. The existence of an energy gap for the 7x7 at low temperature[14,15] and the observation of a gap for the 5x5 at room temperature[16] is consistent with the description of the supercell states indicated in Fig 4S (c) and (d).

These details of these new bands and FSS of the DFA depend on the e-e interactions and how correlation and self-energy interactions also modify these states. It is also unclear whether spin restricted wave-functions used in most band theory have sufficient flexibility to allow these waves to 'de-phase' completely, thereby not allowing the system to relax to an even lower energy state.

An alternate way to view this dephasing as depicted in Fig. 4S (d) is from a Hartree-Fock, H-F, viewpoint. Here, instead of one wave-function that satisfies the 2-D periodic conditions (on both side of the unit cell), one uses a Slater determinant comprised of wave-functions that reflect the symmetry on each side of the unit cell. Isolated each side of the unit cell will have the same states but of different symmetries. As they interact this degeneracy will be removed. This is shown schematically in (d) where bonding and anti-bonding states arise that lower the total electronic energy. In a simple one electron picture, the reduced energy of this state (in this case the FSS at ~-0.4 eV ) corresponds to the energy saved. As shown by the grey bands in (d) each side of the unit cell starts out with its own set of degenerate states and both manifolds of states interact to form a range of bonding and anti-bonding states. (Note that this range of the bound system in (d) was taken to represent the overall band widths of the adatoms in (c).)

The important point is that the inclusion of a Slater Determinant form of wave-function in (d) will allow more flexibility and a lower system electronic energy. It also allows spin to enter into the wavefunction in a natural way via the asymmetry of these wave functions.

In this scenario not using a more complete description of e-e interactions possible within a complete H-F description (or by treating configuration interactions) may account for why LD fails to provide the DFA as the lowest energy structure in DFT calculations. The specific interactions of the adatoms in the honeycomb lattice shown in Fig. 4S is also unclear. However, the physical characteristics of the CD's of ordered or disordered adatoms all appear to be the same in STM.[7] This makes it is reasonable to assume that the adatoms all exist in a common low lying state and, may simply migrate to locations and help balance any needed or excess fractional surface charge required to satisfy the bonding in the honeycomb (as well as between the honeycomb lattice and the subsurface atoms.)

THE ATOM DENSITY OF THE 2-D DAS and DFA STRUCTURES:

Both DFA and DAS structures have corner holes and adatoms, but the details of these as well as the atoms along the unit cell boundary and in the central regions all differ. These final reconstructed 2-D layers represents an equilibrium structure after all the atoms have repositioned themselves. Fig. 6S shows only the top (visible) layers with corresponding areas distinguished by orange lines on one side of the unit cell. The line allow one to distinguished various atoms as well as count the atoms in each of the areas shown. For the DFA structure the internal triangular region containing the adatoms reflects the adatom cell. The honeycomb cell would include the next (black) honeycomb atoms around this cell. This gives a 6x6 honeycomb cell for the 7x7 lattice, i.e. corresponding to 2n for this (2n+1) x (2n+1) family of structures.

Returning to compare the atoms in Fig. 6S, the dark atoms on the periphery of the corner hole for the DFA are further away from the center than the light grey atom on this periphery. These gray atoms are in identical lattice sites as in the DAS structure. As discussed earlier both the surrounding grey and black atoms in the DFA are likely to contract around the corner hole due to the compressive stress arising from greater sp2-sp3 character within the honeycomb lattice. The atoms inside the corner hole may also be slightly shifted by their interactions with these grey and black atoms nearby.

Examining the changes in atom density for these two different structures is useful and related to stresses than can arise within these structures. Here the areas of the unit cell can be partitioned to compare the relative densities of these areas. The orange hexagon in the corner hole in Fig. 6S (b) corresponds to the corner hole partition for one entire unit cell. The inner triangle corresponds to one of the two central bi-layer partition while the surrounding 3 rectangular areas correspond to 3 of the 6 domain wall partitions between the inner triangle and the unit cell boundary. Atop this central triangular region sits the adatoms in either the $T_4$ sites or $H_3$ sites of the DAS or DFA structures, respectively. The number of fully equilibrated atoms in these partitions are shown for both structures in Table I along with corresponding statistics. The totals for DAS agree with earlier work. [5,6]

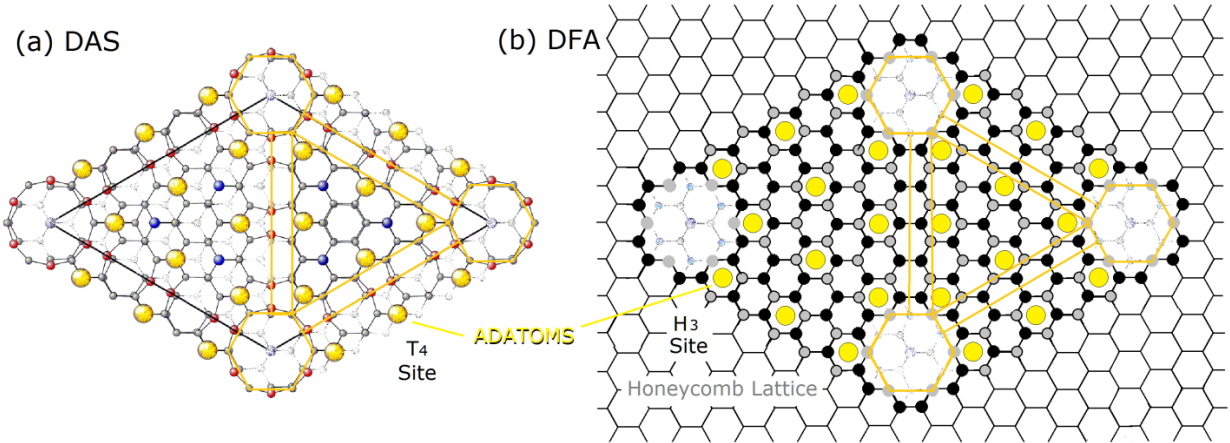

FIG. 6S Comparison of only the top layer of (a) the DAS and (b) DFA models where the fainter atoms show the underlying atom in the layer below that are not considered in the atoms above . In (a) the (red) dimer atoms are displaced from where originally as indicated by the nearby grey atoms. Due to registry differences, the adatom of the DFA turn out to be in $H_3$ sites rather than the $T_4$ sites of DAS.

TABLE I: Summary to the number of equilibrated atoms in each partition of the unit cell.

| (2n+1) x (2n+1) | 3x3, n=1 | | 5x5, n=2 | | 7x7, n=3 | | 9x9, n=4 | | Variation |
|---|---|---|---|---|---|---|---|---|---|
| Structure | DAS | DFA | DAS | DFA | DAS | DFA | DAS | DFA | |
| Corner hole | 2 | 2 | 2 | 2 | 2 | 2 | 2 | 2 | Constant |
| Domain wall | 9 | 9 | 21 | 27 | 33 | 45 | 45 | 63 | Linear |
| Central bi-layer | 3 | 1 | 21 | 15 | 55 | 45 | 103 | 95 | Quadratic |
| Adatoms | 2 | 2 | 6 | 6 | 12 | 12 | 20 | 20 | Quadratic |
| Total | 16 | 14 | 50 | 50 | 102 | 104 | 172 | 180 | |
| 1x1 | 18 | 18 | 50 | 50 | 98 | 98 | 162 | 162 | |
| Delta | -2 | -4 | 0 | 0 | 4 | 6 | 10 | 18 | |
| % | -11.111 | -22.222 | 0 | 0 | 4.082 | 6.123 | 6.173 | 11.111 | |
| Atoms/uc | 1.778 | 1.556 | 2.000 | 2.000 | 2.082 | 2.122 | 2.123 | 2.222 | |

The 1x1 unit cell of the bulk bi-layer contains 2 atoms, and for a 7x7 unit cell, the total corresponds to 98 atoms. The 'Delta' in Table I is the difference between the original 1x1 substrate and the final number of atoms for the equilibrated structures. A negative Delta indicates that there are excess atoms that must diffuse away (-) or additional (+) atoms that migrated in to form these structures. Interestingly, the same atom density of the underlying 1x1 substrate bi-layer occurs for both 5x5 structures which is also equal to the atom density of the 2x1 structure. The atom counts of the other DAS and DFA structures vary.

Several interesting features and trends for the DAS structures have been considered in interpreting STM images.[5,6] For example, the larger number of atoms needed for the larger DAS structures will require greater diffusion in from steps.[5] Similarly different densities in adjoining regions will have implication to local stresses and it's role in stabilizing or destabilizing the structure.[17] For example, the 5x5 DAS not only has the same density as the 1x1 but it's atom density in the domain wall is the same as in the central bi-layer. In the larger 7x7 DAS structure this is no longer the case, but for the 7x7 DFA structure these atom densities are equivalent. Thus internal stress may favor the 7x7 DFA structure over the 5x5 DFA all else being equal. Both DAS and DFA structures require additional atoms to create adatoms for n>2.

A detailed comparison of the atoms at the domain boundary and corner holes is shown in Fig. 7S for each 7x7 structure and provides some insight and another distinction between these structures. The DAS structure has a top layer with missing atoms along the domain boundary not present in the DFA model. The corner hole of the DFA model produces 6 extra atoms versus only 3 missing atoms for the DAS structure. This greater deficit of extra adatoms for the DFA model can provide a means to distinguish between these two models. There is also the possibility of locally pulse heating the sample to determine how many adatoms form from near perfect 2x1 or where they may go from those in a 5x5 structures.

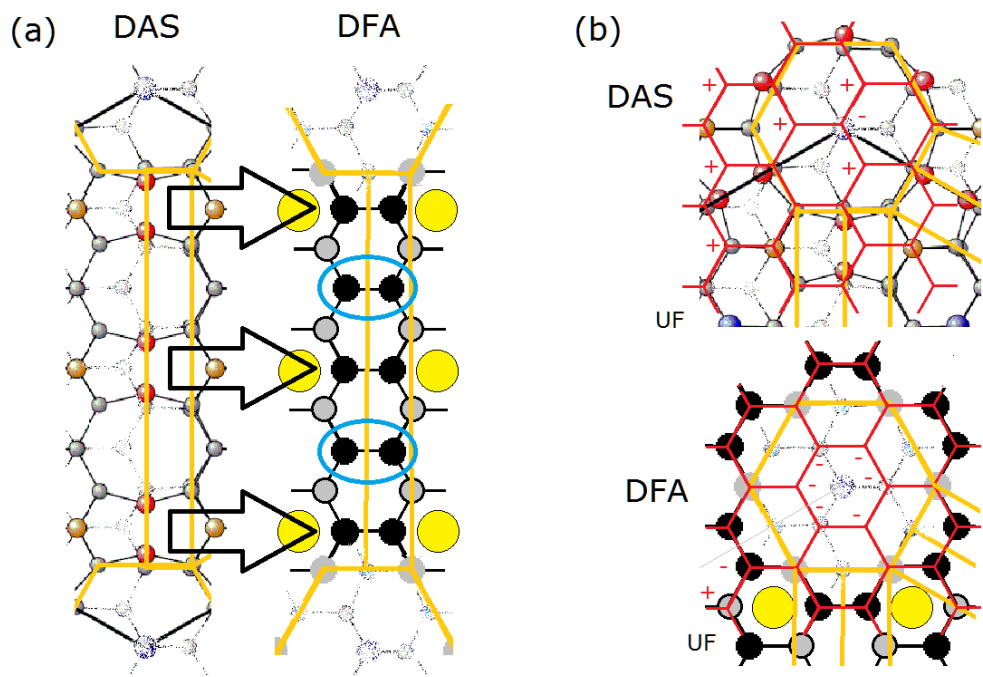

FIG. 7S :Missing atoms along the unit cell boundary (a) and corner hole (b) of the DAS and DFA.

In Fig. 7S, the DAS domain boundary has three pairs of atoms/ dimers that correspond to three pairs of atoms in the DFA structure but with different orientations. The DAS is missing two pairs of atoms as circled in blue in the DFA. The missing atom pairs in this DAS domain boundary produces 3 x (n-1) atoms to each unit cell of a DAS 2n+1 x 2n+1 structures.

Similarly the registry of the surface layer with the substrate shown in Fig. 7S (b) shows that 4 atoms in the DAS and 6 atoms in the DFA structure are no longer there and relocated. Here the + and - in the corner holes in (b) indicate whether the atoms of the former top layer were the higher or lower bi-layer atoms for either model.

It turns out that both 7x7 structures have surface layers of a higher density than the truncated 1x1 surface: the DFA being 2% more dense than the DAS structure. For the 5x5 the density is the same for both models. STM images of the conversion of the 2x1 to 5x5 show well organized regions of 5x5 without intervening gaps in its structure but does shows numerous atomic scale irregularities/defects.[6] The 7x7 under some conditions shows gaps with intervening regions of disordered atoms, and as noted earlier supports two mechanisms for a conversion from the 2x1 and 5X5 to 7x7.[6] The higher density in both 7x7 models are consistent with these proposals. An identification of the atoms in these voids formed at these nucleation sites and an accurate count of the density of adatoms could distinguish between these two models.

THE ADATOM DENSITY TEST:

There are two ways to use this information to test which of the two models occurs. Each requires starting from nearly perfect step free surfaces and exceptional perfectly ordered structures. One test starts from a perfectly flat step free surface with a perfect 7x7 surface. Then pulse anneal this 7x7 with a highly focused, uniform laser to convert small regions to non equilibrium adatom structures. A count of the adatom density can be used to infer the densities of the disrupted 7x7. This has been performed by thermal annealing of large flat surfaces and shows large regions of relatively ill-defined structures.[5]

Refining this approach one could also examine the structure of nucleation centers for the dissolution of highly perfect large islands of 7x7 surfaces. From high temperature quenching studies of Si, higher density areas are observed between 7x7 islands with densities on average of 2.22 for a 1x1 unit cell.[5] The highest density structure seen for Si 111 corresponds to the rt3 x rt3 R30 with a density of 2.33 atoms per surface unit cell.[5] For the conversion of four 7x7unit cells, there would be 24 extra atoms over the ideal 1x1 for the DFA or 16 for the DFA or DAS structures. The size of these well defined high density structures would define the starting atom density. Of course, other mixtures of adatom structures have been found to arise thereby requiring extremely well controlled annealing conditions to form just one structure. If mixed adatom structures occur, their average density could be counted in small well defined regions.

Such annealing studies for cleaved si111 2x1 surfaces also present their own set of problems in that irregular domains of different sizes can form. Domain boundaries also show additional 'debris' or additional adatoms[6] thereby complicating any counting of atoms in these irregular structures. Another issue also become relevant in such cleaved samples. Over the years two different cleavage methods have been used. Initially, notched bulk samples were cleaved by a wedge to form large flat areas for macroscopic measurements.[18-20] This allowed large optically 'flat' mirror like areas were possible that were required for macroscopic analysis of mostly flat, single domain cleaves. However, later STM studies[6] used wafers with tabs cut in them that was broken off or fractured. These revealed mixed domains of various buckled 2x1 structures.[6,21] Published optical images of such fractured wafer samples show more glassy-like regions indicative of a high density of steps that would indicate unusual stress field during fracture.[6] The formation of such cleavage domain structures from fractured samples may have significant implications to the nature and structure of the 2n+1 x 2n+1 structures and what forms after annealing. This is discussed further later.

DEPENDENCE OF BONDING ON THE SIZE OF THE DFA STRUCTURE:

Considering the difference in bonding for the DFA and DAS models, one can speculate on the relative stability of the DFA structures as a function of unit cell size as done for the DAS structure.[22] This is not only different due to the differences in bonding in these two structures, but because of how the different local structures and their bonding vary with unit cell size. For example, the bonds in the central bi-layer region grow quadratically with the size of the unit cell, n, while the domain walls grow linearly. And the corner holes have a fixed contribution to the total system energy.

Fig. 8S shows the 2-D DFA lattice with the domain boundary and central bi-layer partitions indicated, but now with the underlying substrate bi-layer. The dangling bonds of the substrate are drawn on only the left side of the unit cell for simplicity. As noted in the main text the interdigitation of p-orbitals allows a type of π-bonding, D-PB, to occur between the substrate and 2-D DFA layer. Considering Fig. 8S, there are three general bonding configurations that are shown in the insert. These are for the silizene structure, the double chain structure (along the domain boundary) and the asymmetric chain structure in the interior, central bi-layer partition. These interior D-PB will be identified as ID-PBs.

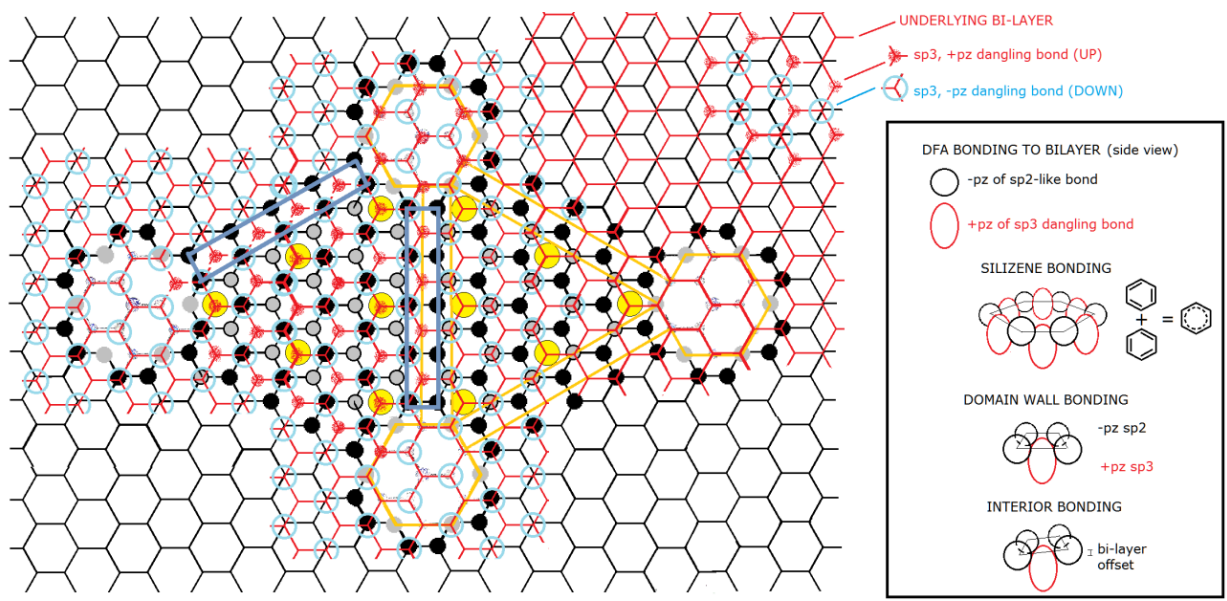

FIG. 8S: DFA 2-D layer shown atop the underlying Si bi-layer. The up facing substrate dangling bonds at this interface are show as solid red circles on the left side of the unit cell. The blue circles are the down facing dangling bonds of the top layer. The insert shown the interfacial configuration of the different structures that can create interfacial bonds.

In a very simplified model, the bonding of the structure in Fig. 8S consists of the piecewise contributions of these three types of bonds plus the adatoms' bonding contribution. (This neglects the energy gained by having extended interactions.) The total number of bonds as a function of bond type is shown in Table IIS (a). For the smallest structure the bonding around the corner hole as well as domain boundary ID-PB bonds dominates. Here, the interior bi-layer $\pi$–bonds contribute a small bonding component but grows quadratically with unit cell size as does the much stronger covalent bonds of the DAS structure. In exploring the total bonding of the family of DFA structures the weaker interior ID-PBs are used as a reference value for the bond strengths of the different components.

TABLE II: Different types of interfacial DFA interactions/ bonds and their respective density.

| BONDS | 3x3 | 5x5 | 7x7 | 9x9 | 11x11 |
|---|---|---|---|---|---|
| Periphery ID-PB | 2 | 6 | 8 | 10 | 12 |
| Interior ID-PB | 1 | 6 | 30 | 56 | 90 |
| Silizene | 3 | 3 | 3 | 3 | 3 |
| Adatoms | 2 | 6 | 8 | 10 | 12 |
| ATOM S/ U.C. | 16 | 50 | 102 | 172 | 260 |

| BOND DENSITY | 3x3 | 5x5 | 7x7 | 9x9 | 11x11 |
|---|---|---|---|---|---|
| Periphery ID-PB | 0.1250 | 0.1200 | 0.0784 | 0.0581 | 0.0462 |
| Interior ID-PB | 0.0625 | 0.1200 | 0.2941 | 0.3256 | 0.3462 |
| Silizene | 0.1875 | 0.0600 | 0.0294 | 0.0174 | 0.0115 |
| Adatoms | 0.1250 | 0.1200 | 0.0784 | 0.0581 | 0.0462 |

Calculations of the bonding of the DAS as a function of size ( n) shows a gradual decrease in bonding structures larger than the 7x7 and steeper decrease for those smaller. [22-24] The covalent sp3 bonds of the dangling subsurface bonds in the interior of the DAS unit cell are expected to be stronger that the interior $\pi$-bonds of the DFA structure, but both will increase quadratically with unit cell size. For the DAS structure these stronger internal covalent bonds are offset by the strain introduced by the dimers which grows linearly with unit cell size. In contrast, the weaker interior p-bonds of the DFA limits the energy benefits of growing larger. This feature of bonding with larger DFA size can account for the nucleation and 3-D growth seen frequently, [6,25] as well as the difficulty in forming larger ordered 2n+1 x 2n+1 structures as seen by STM at elevated temperatures.[12,26,27]

The inter-layer bonds at the interface that hold the intra-layer bonded DFA structure to the surface exist in an extended, repeated structure along the unit cell boundary and suggest delocalized $\pi$–bonding. Such delocalized bonding is hypothetical but can be based on the concepts derived from how bonding and charge rearrangement arises in the DAS model. [4,22-24] Each of these bonds in the DFA corresponds to one of the downward facing dangling bonds of the sp$^2$ and thereby should have an upward facing +p$_z$ charge density that interact and is referred to as an interdigitated $\pi$–bonding, D-PB. As is the case for the rest atoms, such 'dangling bonds' should also become shared and if partially unoccupied become filled by the electrons from the highest lying states, e.g., those of the adatoms. The asymmetry of the FSS site near the adatom may enable these FSS +p$_z$ orbital to become fully compensated and shared by the electrons of the adatom and the delocalized (silizene) bonds around the corner hole. These would be coupled to the p-bands of all such +$p_z$ sites, not just the local FSS state, so as to form the broad band of states seen in AR-PES discussed in the main paper. The unusual coupling of the FSS to these corner hole adatoms, would then account for its observation by STM.

THE STABILITY OF THE DFA STRUCTURE WITH UNIT CELL SIZE:

One can also use these bond counts or bond densities for the unit cell to estimate the relative stability of the DFA structures as a function of unit cell size as done for the DAS structure.[22] The lower section (b) of Fig. 9S shows the bond densities of the various local interactions for the different bonding components to the DFA structure. Since the corner holes frequently form whether or not they support a full 2n+1 x 2n+1 structure or fragments thereof , they are considered here to be the most stable of these pi- bonded structures in the DFA structure. This is not surprising if one considers silizene as a linear p-bonded chain that wraps around to form a cyclic system to become a resonance structure that strengthens the π-bonds. Using the interior ID-PB as the weakest bond strength, one can weigh the relative strengths of the other bonding contributions to this bond as follows: the silizene as 4x, the periphery IDPBs as 3x, and the adatoms as 3x. Summing these up give a net bonding shown by curve 1 in (a) of Fig.9S. Varying these weights differently can give rise to a broader peak as shown by curve 2.

While this is an extremely simple, qualitative analysis that makes many assumptions in relative bonding contributions, it demonstrates that the DFA structure can lead to stable 5x5 and 7x7 structures. Regardless of these simplifications one still expects the DFA to be less stable than the DAS for larger unit cell sizes, simply because the substrate π–bonds inside the cell are the weakest bonds of the DFA structure and they will dominate for larger unit cell sizes as noted in Table I.

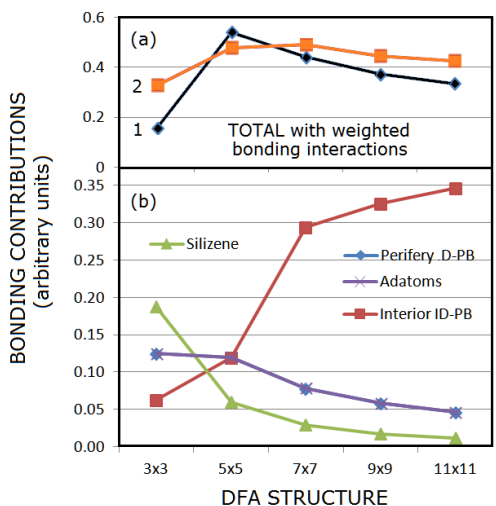

Fig. 9S. Graph of the bond densities for each type of local interface structure in (b) and a weighted total of their contribution to the total bonding in (a). This assumes the interior ID-PBS have the weakest bonding, silizene the strongest and the adatoms and periphery D-PBs are in between.

THE INTERACTION AND BONDING BETWEEN THE CORNER HOLES:

As noted in the main paper, theoretical work has examined the energies of these structures and predicts 3x3, 5x5, 7x7, etc structures with the 3x3 being the most weakly bound.[22-24] However, experimental results[6] do not show the full development of ordered 3x3 structure, not even isolated units of 3x3 cells. Instead, at most two corner holes link up and form a dumbbell like structure [6] and in some case only one corner hole surrounded by an arrangement of adatoms form [27] which are yet to organize and/or form a second corner hole.[6] As a result the 'bonding' of two corner holes is another interesting question.

In the last section the relative stability of these different structures for the DFA model was estimated based on simply counting the number of different types of atoms present in the honeycomb in various locations that could provide D-PB bonding. Here, the bonding of the corner holes is considered for these two structures.

Fig. 10S shows the unit cell boundaries of (a) the DAS and (b) DFA models as a function of n (size) for each (2n+1) x (2n+1) structure. In (a) all the DAS structures have the same sp3 like covalent bonds and add dimers as it grows larger. The 3x3 DAS structure lacks a restatom in between the adatoms, that accepts electrons from the adatom to reduce the system energy, but still has a 'restatom' in the corner hole allowing the 3x3 to be an odd electron system and have a half filled state at $E_F$ ( just as the other ideal 5x5 and 7x7 would.) There are no restatoms on the top ad-layer to allow charge redistribution from the adatoms which results in larger Coulomb repulsion between the adatoms. Overall as predicted [22-24] the strain also increases with increasing larger DAS family member. As discussed later the 5x5 is unusual in that experimentally it does not show a metallic surface at room temperature.[6]

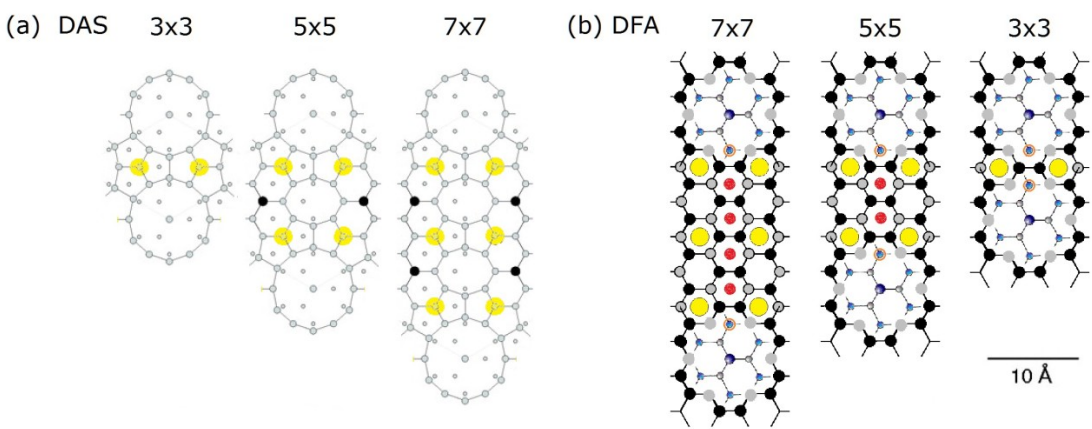

FIG. 10S The structure of the atoms between the corner holes for the DAS (a) and (b) DFA models as a function of size. The adatoms are shown as yellow in both while in the DAS the dark solid circles represent restatoms. The red and orange circles in (b) represent the original dangling bonds, i.e., upward facing $p_z$ orbitals ('dangling bonds') from the substrate between the centers of the corner holes. As discussed in the main paper these effectively act like spines that help 'bond' the honeycomb 2-D lattice to the surface.

In contrast, the DFA structure relies on D- PB's along the domain walls of these structures to form an extended 2-D mesh of bonds that holds the corner holes together. The red circles along this domain boundary represent the upward facing DB's from the underlying substrate along this domain boundary. The orange circle bordering the corner hole represents one of 6 DBs that stabilize the corner hole to help form the silizene structure. The charge redistribution for the DFA can occur not only to the 'restatoms' but also to the honeycomb atoms that interact with the DB. The number of fully available DB's ( those not involved with the bonding around the corner hole) decrease to zero as n decreases to 3. The D-PB that occurs for the 3x3 involves only the pairs of orange dangling bonds bordering the domain wall that are already interacting to stabilized the silizene structure. As a result the bonding of the 3x3 must decrease even more than shown in the hypothetical total energy curves shown in Fig. 6S earlier. Again this is consistent with experimental observations that full 3x3 structures do not form. [6,12]

TOPOGRAPHIC EVIDENCE FOR THE PRESENCE OF DIMERS:

Another factor in favoring the DAS structure in early work was the smaller holes or depressions along the unit cell boundary that was visible in most high resolution STM images. Early work modeling the superposition of atomic charge densities for the DAS model found evidence that this reflected a physical gap associated with the occurrence of the dimers.[28] However, this can also be the result of the decaying tails of the charge densities for the adatoms in either the DAS or DFA structure.

High quality STM topographic images achieved by Koehler [6] are shown in Fig. 11S (a) for the 5x5 structure along with line profile along A-B in (b). This is taken from an image of mixed 5x5 and 7x7 phases where the topographic features of the adatoms of both structures are identical aside from their different arrangements on the surface.

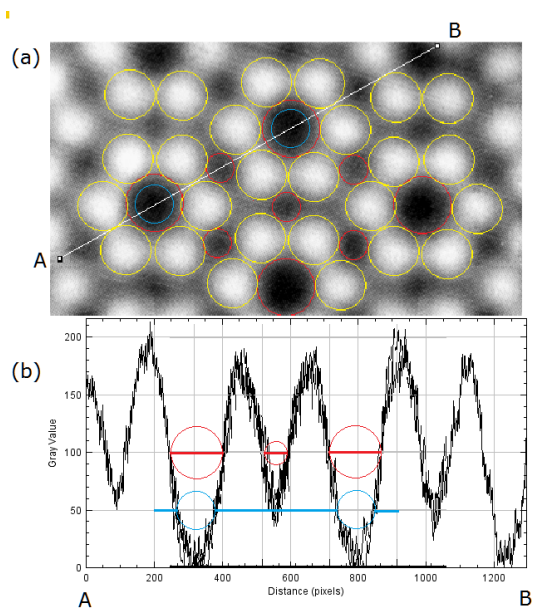

FIG. 11S: STM image (a) and (b) the corrugation height along A-B for a 5x5 surface.( +2V, 1nA). The peaks from 225 to 1050 have been superimposed on a second scan. The yellow circles represent equi-

density adatom CD contours as do the small and large red circles. Blue circle represents a the CD contour corresponding to the lowest point within the smaller hole on the unit cell boundary.

The yellow circles in (a) corresponds to a periphery of uniform charge density around each adatom. The horizontal red line in (b) is drawn along the saddle point between the adatoms. The large red circles and lines below corresponds to the edge of the corner hole while the blue line and circles the midpoint of the corner hole. The smaller red circle in (b) turns out to be half the diameter of the larger one. Both the yellow and red circles appear to touch, forming a contour of constant charge density. The blue circles corresponds to a an even lower contour that reflects the charge density around the sides of the corner hole at the lowest level of the smaller depression. The center of the corner hole is still lower and the lowest region accessible by the probe tip.  Overall, based on these space filling spheres, the smaller depressions along the unit cell boundary can be accounted for by the decaying charge density of the adatoms and their non-uniform spacing along the unit cell boundary.  Confirmation of this can be observed in more well resolved low temperature STM topographs.[29]

Fig. 12S shows a comparison of the 7x7 topographs at T= 0.3K [29] and 298K [6] of the 7x7 surface and compares them to the 7x7 DFA model. At low temperatures improved contrast occurs between the adatoms and the small depression to make them appear more  rhombohedral in shape. The blue rhombuses have been drawn in a similar region on all three images.  In (a) and (b) the shape is independently defined by the regions of similar grey scale while in (c) it is define by the bisector of the bonds between atoms along the unit cell boundary.  The improved contrast in (a) make the shape of this small depression more closely resemble the scalloped contour of the neighboring adatoms. While simulated STM images of both models may be useful, they are unlikely to distingush between the DFA and DAS  models since the CDs of predominantly the adatoms will dominate either STM image.

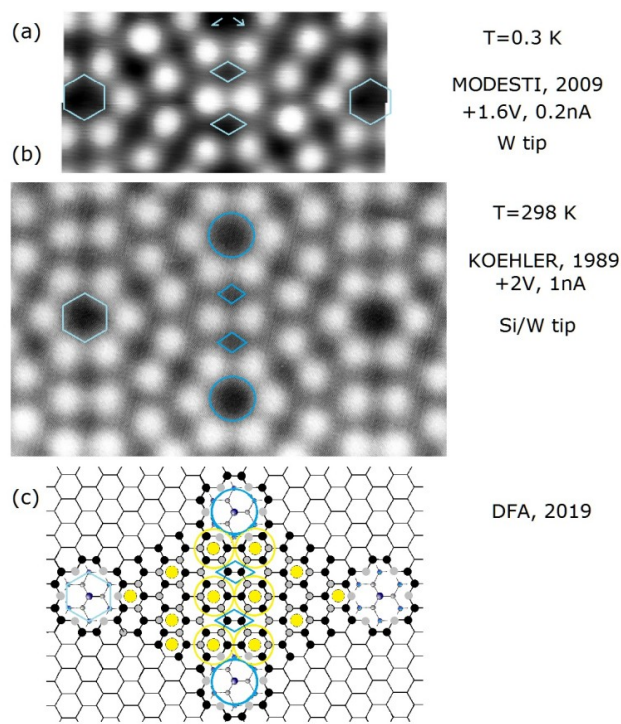

(a) T=0.3 K
MODESTI, 2009
+1.6V, 0.2nA
W tip

(b) T=298 K
KOEHLER, 1989
+2V, 1nA
Si/W tip

(c) DFA, 2019

FIG. 12S: STM topographs compared to the DFA model show in earlier figures. The small horizontal offset midway in (a) has been made to correct for a small change in the tip during the scan. The blue overlays are described in the text.

In these low temperature measurements heavily n (As) doped Si(111) wafers were used having a resistivity of less than 0.005 ohm-cm or dopant levels of ~$10^{-19}$/cm$^3$.[22] Concerns of sample charging and band bending during STM measurements of such semiconductor surfaces have generally lead to the wide use of highly or heavily doped samples in most STM experiments, even those at room temperature. The role of intrinsic or extrinsic (contamination based) dopants has been largely neglected in most all STM studies of Si111 reconstructed surfaces.

The low temperature STM image also shows additional features that distinguish it from the room temperature images. Namely, at low temperature the STM image shows random variations in the adatom heights by ~20% as well as ~0.2 A variations in the spacing between pairs of adatoms along the unit cell boundary. These differences are complicated by two factors: the high doping levels used and non-equilibrium carrier effects[29,30] that can shift the energies of the various surface bands. The latter is unlikely as all adatom states will likely shift the same, but doping effects should be more local. These lateral distortions may also reflect a Jahn-Teller distortion[29] associated with electron-electron correlation effects, or the effect of local dopants in the near surface region whose fields are perturbing the adatom charge density and bonding. High resolution LT STM topographs under flat band conditions would resolve this.

Another low temperature STM experiment[31] has avoided using highly doped samples as well as non-equilibrium carrier effects by employing sample illumination to create carriers and maintain flat band conditions at low temperatures.[32] Fig. 13S shows the results of an STM scan at 5K during illumination.[31] The sample dopant level is 1000X less that that used in Figure 12S (a). In the middle of the scan the illumination is monetarily turned off and then resumed. Surprisingly, imaging of the 7x7 surface resumes with the same substrate registry as before implying that the tip has not changed. While the 7x7 pattern is clear, this image is does not appear as highly resolved as in Fig. 13S (a) which was taken at an even lower temperature of 0.3 $^{\circ}$K. (This may also be a result of how these grey scale images are scaled and processed for display.) Never the less, in addition to reduced adatom intensities some adatoms are missing. Both of these variations are indicated by the adatoms circled in blue.

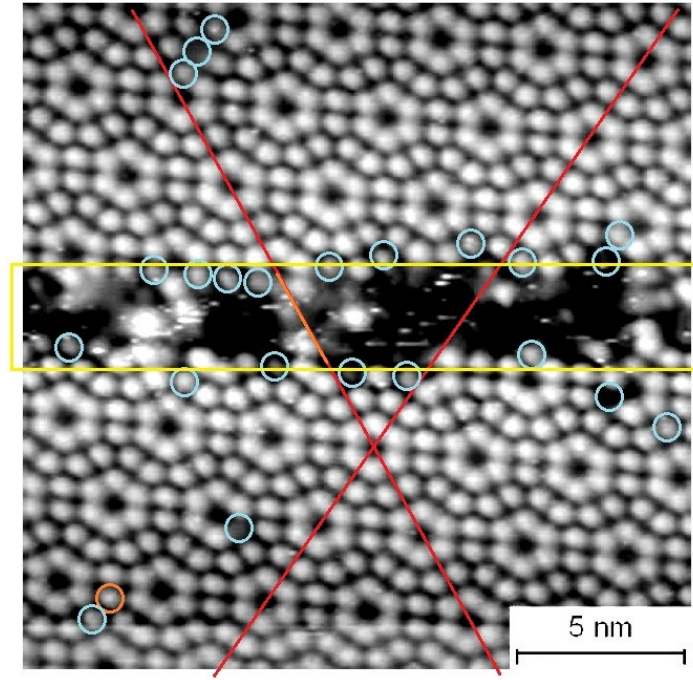

FIG. 13S: STM image of the 7x7 at 5 °K under illumination and flat band conditions with a time period of interrupted illumination as approximately indicated the yellow lines.[31] Blue circles indicate adatom irregularities relative to more uniform neighboring adatoms . (STM image acquired at 1nA, +1.5eV.)

If one assumes that the adatom height differences in the illuminated areas are derived from uniformly distributed dopants, then the number of dopant defects seen here when flat band conditions exist should be less than one in the entire image. ( 5 per 2 unit cells / 1000, times the 35 unit cells in Fig 13S or <1.). This reduction in possible dopant altered adatoms is about right, except when the carriers are no longer in equilibrium. A larger number of adatom irregularities occur when the illumination is terminated and initially when illumination is resumed. It would appear that during these on/off cycle non-equilibrium carrier dynamics arise and the adatoms become more sensitive to deeper lying dopants due to reduced free carrier screening.[29,30] Similar variations in adatoms features are also observed in room temperature STM images that are discussed next.

DOPANTS AND DEFECTS ON Si111 2n+1 x 2n+1 RECONSTRUCTED SURFACES:

The perfection of the reconstructed Si111 surfaces studied has greatly increased since the advent the STM. Even the first STM image of the 7x7 consisted of only one unit cell after many hours of scanning the surfaces.[3] As UHV conditions and compatible UHV materials were used, STM UHV system became cleaner and 'hardened' over time: perfection of the prepared surfaces gradually improved. Nevertheless defects in the 7x7 have been widely observed [3-7, 12, 25-27] and studied [33-35] but are not fully understood. To focus more attention on the possible role of dopants and defects on these reconstructed surface and in the near surface region, a variety of measurements are presented. As discussed later sample defects appear to have played a role in X-ray diffraction studies [36] as well as TEM HR atomic imaging work. [37] Other studies of defects formed from a various foreign atoms or adsorbed layers on silicon have been widely examined and more completely understood but are not discussed here.

The most widely observed defect on the 5x5 and 7x7 surfaces arise from disordered adatoms that have different local arrangements or mixed arrangements.[4-7,37] These typically occur below ~600 $^{\circ}$C. In many cases the adatoms simply fill a different set of lattice points on a bi-layer or honeycomb lattice which merge into larger ordered 5x5, 7x7, or even larger integrated structures.[7] The disordered adatoms (and possible contaminants) are seemingly pushed and corralled into a small area between ordered regions [6] or between larger ordered regions of exceptionally flat surfaces.[5] STM can only tell indirectly of dopants and defects below the surface.

One type of defect discussed here is a local defect that appears to manifest it's presence by modifying the appearance of adatoms observed by STM. These appears to depend on the doping level and how the sample was 'processed' to form ordered an 5x5 or 7x7 structures. Adatom vacancies in Si 111 structures have been theoretically studied [33-35] but have yet to be clearly identified with defects observed experimentally. The presence of defects can be important in creating corner holes while dopants can remove or add an electron locally to change the electronic structure of one or a few unit cells.

As demonstrated by early MEIS studies, [39] the highest quality 7x7 surfaces were found to occur with Shiraki cleaned samples or samples prepared for semiconductor processing. After system 'bake-out', a thorough out-gassing of the sample holder is performed. After system pressure returned to normal, a flash-off of the native oxide to 1100 $^{\circ}$C is performed, and with it all the contaminants atop the oxide. Then the sample is left to cool down usually by itself. The trade off in this has been between a fast cool down to minimize heat transfer that can produces thermal drift ( critical in the STM) as well as outgassing versus a slow cool down. The alternative approach that e seem to produce exception 7x7's is a slow cool downs that allow greater surface atom diffusion but at the expense of potentially greater contamination and thermal drift from heat transferred to the STM. Vacuum pressures typically spike during these short anneals but after many repeated studies these gradually reduced as the UHV system is more thoroughly outgassed. Use of sample transfer capabilities[7] will avoid thermal drift of the STM . This along with cryogenic pumping and scrupulous system bake-outs and outgassing improved the perfection of these surfaces. Some of the most perfect 7x7 published to date were achieved from 111 wafers (native oxide covered) that were thorough cleaning in ethanol and distilled water, flash heated to 1200 $^{\circ}$C , annealing at 900 $^{\circ}$C followed by a slow cool down of 1 $^{\circ}$/sec all under pressures below $10^{-10}$ millibar, exceptional UHV conditions.[40] Even these show some of the same adatom defects discussed here.

In addition to surface perfection, semiconductor properties are known to be altered by impurity and dopants which can be accidently transferred to the sample.[39] For example, simply holding a Si wafer with stainless steel tweezers makes the entire wafer unusable for semiconductor processing. As a result, doping and defect or contamination disordered surfaces may occur during cleaning that alter the properties of the surfaces produced.

Sputter cleaning of Si is problematical as back-sputter occurs which deposits metal contaminant from the stainless steel UHV chamber as well as residual carbon onto the sample. Also boron exists within the chamber from the boro-silica windows used in UHV chambers. Such boron will also will dope the sample.[39] Boron chamber contamination has been found to change cleaved n-type samples to p- type after simply annealing a cleaved sample.[32] It is likely that boron left around the sample /sample holder diffused onto the sample during annealing. Use of cryogenic sample holders and cryogenic pumping greatly reduces residual contamination so long as the sample holder is optimized to heat only the sample.

As noted earlier the formation of the 5x5 and 7x7 from the cleaved Si111 surface also has complications due to the different cleavage approaches used. Cleavage variations are evident from the earliest optical studies of cleaved surfaces [18] to more recent STM studies.[6-21] Making matters worse, most all STM measurements use highly doped samples to minimize charging effects and local band bending associated with the tip.

The effect of dopants on the surface electronic structure and surface properties of notched and wedge cleaved Si111 has more recently been studied and reveals that the apparent dopant density is ~50% higher than the expected bulk dopant distribution.[41] Thus the cleavage plane may be partially defined by regions having a higher dopant density where local stress fields may exist. The equilibrium distribution of dopants in the last atomic layers of a Si111 surface, and in particularly after sample annealing, is largely unknown.

FEATURES OF THE 5 x 5 SURFACE:

The first observation of a 5x5 was on a Si111 sputtered surface after an anneal to ~600 $^{o}$C.[42] Over a number of years the thermal transition of the 2x1 was investigated by several independent leading researchers,[43-47] finding only a 7x7 as observed by LEED: no 5x5 features were observed. The earliest STM report of the 5x5 structure was based on Si-Ge epitaxy work [49] which deposited 500 Angstroms of Ge on Si111 that was sputtered back to the original interface and annealed at 650 $^{o}$C to form a Si-Ge alloy. This was established by LEED to have a 5x5 pattern with nearly equal Si and Ge concentrations as determined by Auger studies.[49] STM measurements of this particular type of 5x5 surface shown in Fig. 14S and display three different adatom heights. Two for the adatoms on each side of the unit cell (usually associated with 7x7 filled state images) and a third that was slightly fainter than either of the other two adatoms. These reduced adatom heights are circled in Fig 14S. Given the uncertainty of the mixed Si- Ge surface composition, the lower adatom height could be Silicon atoms substituted for Ge in the top layer or a carbon impurity for either SI or Ge. These local adatom defects parallel a recurring type of adatom defect discussed next. It should also be noted that in an analysis of the 2x1 conversion to the 7x7, evidence for an intermediary charged vacancy state was inferred from the variation of the transition temperatures with the observe work function changes.[46]

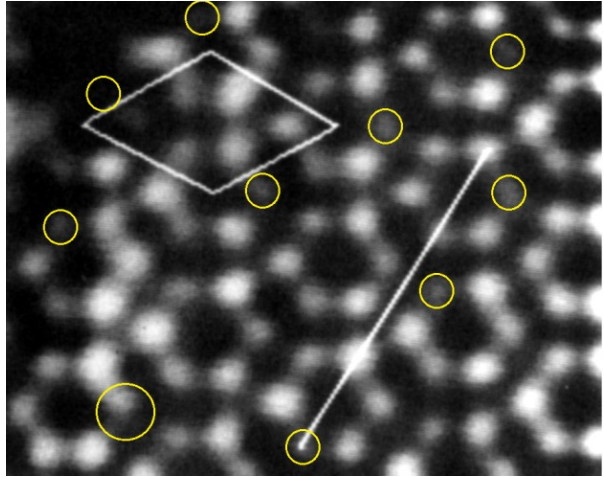

FIG. 14S: STM topograph (-0.5eV, 1nA) of a 5x5 formed from a sputtered layer of Ge on Si111 Ge-Si annealed to 650°C. [41]

These unusual lower intensity adatom features are also observed after heating cleaved surfaces to form the 5x5 or 7x7. [50] This is in addition to areas of disorder or contamination frequently observed.[5,11] Such contamination is more typical during annealed cleaved samples since the sample holder for cleavage is more rigid and less isolated which can prevent full out-gassing before annealing. Never the less, there are still only a handful of STM studies of the formation of the 5x5 from the 2x1 surfaces, and by far most all are from fractured 111 surfaces.

In these STM studies using fractured samples, the STM images suggest that many of these same local adatom defects exist up to temperatures of at least 600°C.[5] Also it should be noted that the first STM measurements of a cleaved Si111 2x1 surface [16] were performed on a wedge cleaved sample, but then later work [6,21] used fractured wafer samples which were less problematical to handle and anneal. These wedge cleaved 2x1 studies showed few surface states in the gap than found in later studies of fractured samples. The most recent 'fractured ' 2x1 studies show a larger number of complex states in the gap.

Fig. 15S shows an early scan of the 5x5 formed after a mild anneal of a fractured 2x1 surface. [50] Based on other work by these authors[5] this image is comparable to 5x5 images taken after a 20 or 80 sec. anneal at 420°C or a 100 sec anneal at 410°C. Of these, the latter show a 50-50 mix of individual areas of 5x5 and 7x7. [6] The 5x5 in Fig. 15 show many defects with many of the smaller ones localized primarily around the corner hole. The higher white features were attributed to contamination from the thermocouple during annealing.[6] In this image there is no apparent preference for corner hole defects to be located on any particular side of the unit cell as predicted for charged vacancy defects on the 7x7 surface. [34]

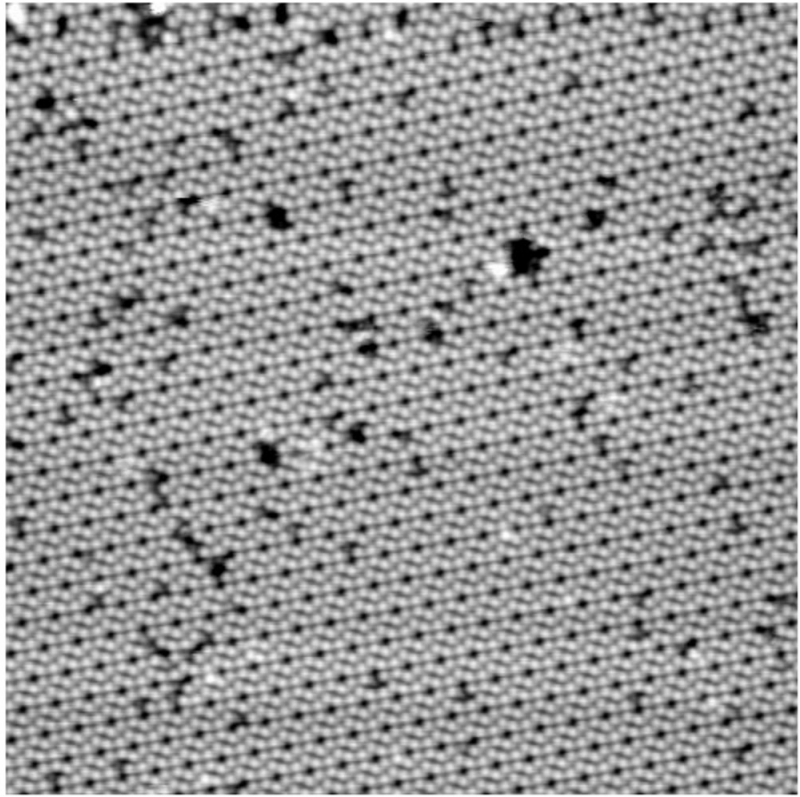

FIG. 15S. STM topograph of a 5x5 surface ( +1.5V, 1na) obtained after mild heating of a 2x1 structure. [50] Similar defects were also observed in 900sec anneals at 330 °C. [6]

Fig. 16S shows an close-up of one of these corner hole defects noted earlier [6] and indicates a different charge density in some adatom locations, as if the adatom is missing or carbon has been substituted for a silicon atom. Calculations of adatom defects [33-35] show larger shifts and changes in the filled state DOS than in the empty state but do not resolve what is observed. Since cleavage may propagate along planes with a statistically higher dopant density than would be statistically expected, this particular local adatom defect may reflect intrinsic dopants at these sites or nearby below.

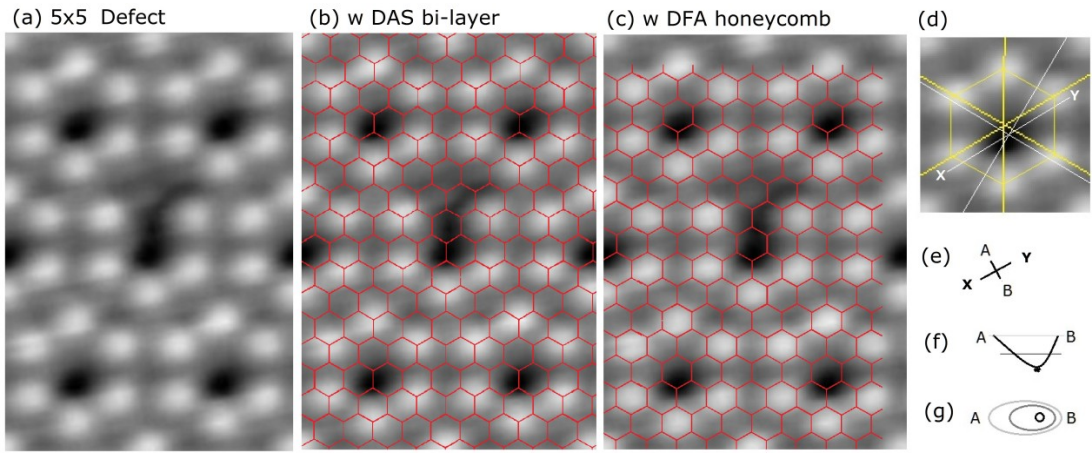

FIG. 16S (a) shows an enlarged area of a frequently observed defect of the 5x5 structure. A DAS bi-layer is overlaid in (b) and DFA honeycomb in (c). This is consistent with some type of substitution at this particular site for either model. The tip allows reasonable imaging of the adatoms but an asymmetry of the tip shank exists as indicated in (d)- (f) that complicates interpretations.

Bias dependent STM measurements of this same 5x5 structure [6] also reveals another type of local adatom defect around the corner hole unlike those shown in Fig. 15S and 16S. Fig. 17S shows simultaneously acquired scans of both (a) the empty states and (b) filled states. [5] Here the empty state topographs has equally intense adatoms with one adatom (circled in blue) slightly shifted. The corresponding filled adatom state is missing. Within a simple one electron picture this can be accounted for by a different atom that largely maintains/supports the adatom empty energy band but is missing an electron, e.g., a dopant. Such bias dependent adatom features were also observed in early STS measurements of the 7x7 and revealed in published figures,[4] but like many of the complex aspects of these surfaces, were never fully understood. Even recent filled state topographs of the 7x7 show such missing adatom CDs, [30-31] Such defects may go unidentified since the majority of Si 111 topographs are made by tunneling out of the empty states and can miss anye filled state CD anomalies.

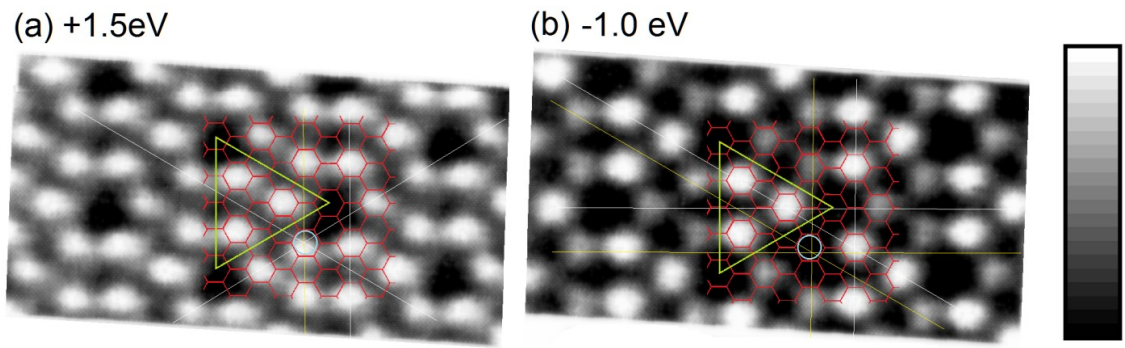

FIG. 17S: BIAS dependent images of the 5x5 surface formed by mild heating of the cleaved 2x1 surface.[6] The blue circles indicate a 'different' adatoms, most prominently seen in the filled state image.

Further heating to 630° C produces the 7x7 as shown in Fig. 18S which also contains many of the same corner hole defects (and others ) as found in the 5x5. One expects that at higher anneal temperatures near and above 1100° C ( used to produces highly perfect7x7's), massive diffusion and redistribution of surface and near surface atoms, dopants and contaminants will occur. It is assumed that annealing removes all defects. The question remains open whether such local defects found at these lower annealing temperatures may play a role in stabilizing or nucleating the 5x5 and 7x7 structures formed thereafter. Hypothetically, at high temperatures a small concentrations of dopants could be driven to the surface to help stabilize a surface reconstruction. The slow annealing between 900-600K ( as used by many) could allow any excess dopants to diffuse back into the bulk now that they have assisted in balancing the electron count needed to form the reconstruction.

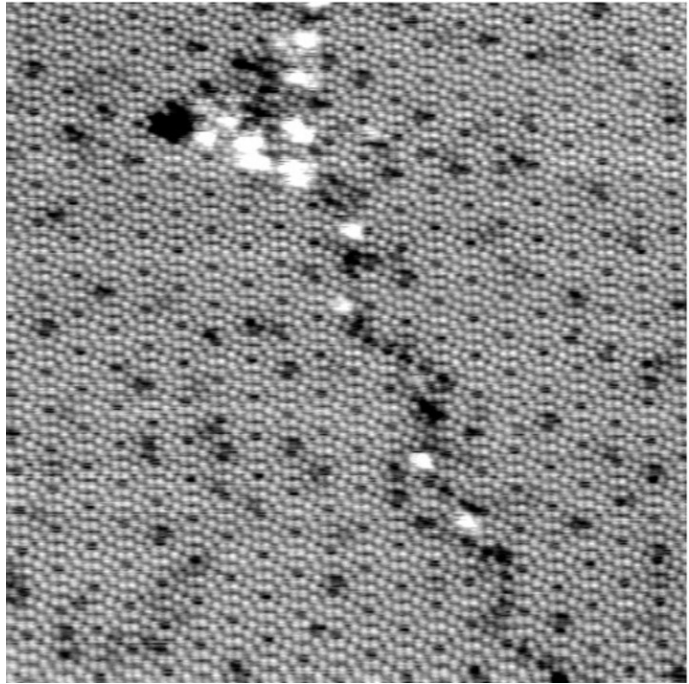

FIG. 18S. STM image of the 7x7 as converted from the 5x5 in Fig. 16S by annealing to 630°C. [6,50] The fault line indicates a misfit of the 7x7 in two regions, while the triangle is depleted of adatom and like has contaminants.

One reason for local defects or dopants to play a role in forming the 5x5 structure or even the 7x7 structure arises from the energy gained by creating an even electron unit cell and an energy gap near $E_F$. A significant energy gap of ~ 250meV is experimentally observed at room temperature for the 5x5,[6] in contrast, to the small 40-80 meV energy gap found for the 7x7.[14,15] The larger gap of the 5x5 cannot be overcome by thermal energy alone and the formation of carriers that populate the partially filled band as believed to occur at room temperature for the 7x7. Thus, defects and dopants that change the number of electrons, even locally, from even to odd even, and vice versa can change the 5x5 system energy and drive different local structures. Again, in numerous early studies of wedge cleaved single domain 2x1 samples, the 5x5 was not observed in annealing studies.[42-46] The conversion of the 2x1 to 7x7 starts at 210 °C and is higher on some cleaves by what was believed to be the presence of steps [43-45] and possible charged defects.[46] This is in contrast to the behavior of fractured samples studies[6] by STM where after annealing to 410 °C produces large areas of 5x5 that cover half the surface !

For example, second harmonic optical studies of wedge cleaved 'hyper pure' Si111 [8] found a uniform conversion from the 2x1 surface starting at 245 °C as the sample was slowly annealed. ( The wedge cleavage method used in these studies is described in more detail in another publication.[47] ) The 3 fold mirror symmetry planes of the 7x7 was detected at 275 °C and the 2x1 signal disappeared by ~500 °C, 174 seconds after its initial decay upon heating. The 2x1 signal showed a very uniform decay of the unique 2x1 signal with no breaks characteristic of any other phase. This however is expected since the c3v symmetry planes of the 2-D 5x5 and 7x7 structures are identical and the occurrence of either will remove the 2x1 SHG signal. Never the less, only the 7x7 was ever observed in LEED again reflecting the conclusions of at least 5 other earlier temperature dependent studies of the direct 2x1->7x7 conversion on single domain cleaved samples. [43-46] Certainly this does not exclude the formation of much smaller domains of 5x5's that go undetected in LEED. The 5x5 domains seen from fractured samples[6] as shown in Fig. 15S should be easily detected by LEED.

The domains of 5x5 found by STM typically cover half the surface after annealing to 410 $^{\circ}$C producing mixed 5x5 and 7x7 domains of comparable domain sized ( ~1000A ).[6] Under these conditions the 5x5 should have been observed by LEED in the SHG and earlier studied, but they were not. It was also found that the step free wedge cleaved samples in both the SHG work and earlier studies [43-46] had the lowest conversion temperature to the 7x7. The higher conversion temperatures found in the STM studies of fractured surfaces [6] is clearly characteristic of the poorer quality wedge cleaved samples. The difference found between wedge cleaved single domain surfaces versus fractured multi domain 2x1 surfaces used in most STM studies of the 5x5 cannot be ignored.

ADDITIONAL EVIDENCE FOR THE ROLE OF DEFECTS.

Many of these defects can also be surmised from the High resolution TEM [51] and early x-ray [52] studies that originally helped to confirm the DAS structure of the 7x7. Defects cannot be excluded in the first TEM diffraction study [1] since only diffraction was used to characterize these thinned samples. Since these experimental methods have improved considerably since their first application to Si surfaces, these early results are considered in further detail. This is particularly important in view of more recent improved studies using either of these methods.

X-RAY STRUCTURAL ANALYSIS of the 7x7:

Takayanagi's pioneering work on the 7x7 DAS structure[1] involved an analysis of the diffraction intensities to optimize the atomic positions using the structure factors for numerous model geometries and variations therein. TEM diffraction conditions were selected to minimize multiple scattering effects to allow a kinematic analysis of the diffraction intensities. This involved rotating the sample 8$^{\circ}$ and averaging beam intensities under the assumption that near 6 fold symmetry was occurred. The atom positions were then optimized to fit Takayanagi's diffraction intensities to provide his 'optimized' DAS model. Takaganagi's TEM samples were thinned samples (~300 A thick) heated in a UHV TEM with only diffraction intensities used to characterize the structure and perfection. However, Takayangi used the STM preferred method of oxide flash to prepare his amples and study them with a He- cooled cryogenic shiled around his sample in a TEM. As discussed earlier the 6-fold symmetry he observed my reflect the presence of various 3-fold symmetry domains formed and averaged by macroscopic sampling.

X-ray diffraction from the 7x7 surface is weak but its true kinematic scattering that makes analysis more certain in principle. However, the construction and operation of a UHV compatible chambers, windows and diffractometers required for x-ray measurement is very difficult. The first x-ray diffraction analysis was performed using the structure factors and a rigorous fitting procedure that quantitatively measured the agreement of the structure factor for each model and variations therein.[52] Given the weak x-ray signals from the 7x7 surface, equivalent beams were averaged using Takayanagi's 6-fold symmetry assumption. Such an assumed symmetry also reduced the complexity in calculating the structure factors in comparing to the diffraction experiments. Never the less some atoms had to be constrained in such modeling and as a result some adatoms and the atom below them were moved as pairs.

These samples[52] used for this first x-ray analysis were prepared by sputtering and annealing to 850 $^o$C and showed residual carbon contamination of 0.5%. Data acquisition was done at a base pressure of  3x $10^{-10}$ Torr  and required  48 hours due to the low x-ray intensities.  This base pressure exposes the sample to ~17 mono-layers of vacuum contaminants and ~2 mono-layers of atomic hydrogen if a hot filament or ion source exists in the chamber, i.e., an ion gauge is on.   In addition, at these pressures residual $H_2O$ in the chamber is known to react with the surface and is difficult to detect.[47] Evidence is presented that the 7x7, in particular the adatoms, have been effected by these measurement conditions.  Additional x-ray studies were later performed to study the truncation rods [53] as well as check the original in plane measurements [10] using the same chamber but with an upgraded UHV monochromator.  These later x-ray studies likely  took as long or longer times to collect the required data given the very low intensity regions of the truncation rods that need to be measured accurately.

Twenty years later new x-ray measurements of the 7x7 were performed  as a baseline for studies of the wetting layer in the growth of Pb on Si111. [54,55]  These measurements used a third generation (undulator based)  x-ray light source at the APS of the Argonne National Lab that was more highly collimated than previous sources and estimated to provide $10^4$ x greater intensity.  These factors  not only improved the data acquisition time but the data quality by reducing the background tails of integral order peaks as well as diffuse scattering to more clearly measure diffraction intensities, including the weaker fractional order peaks.

This new xray study [54,55] was ion-pumped with a $LN_2$ cooling stage  to improve the quality of the Si111 7x7 surfaces. Cleaning  procedures involved high temperature oxide flash off as used in STM studies. The fractional order beam profiles were used to quantify the degree of surface perfection which varied with preparation  conditions. The Debye Waller factors derived in the fitting of this new data were also found to depend on sample quality: [55]  the surfaces  with the sharpest fractional ordered peaks were found to require smaller Debye-Waller factors than those found in the earlier [52] x-ray studies. This is reasonable as increased Debye Waller factors will emulate the reduced elastic scattering from structural disorder. Such observation thereby calls for the use of an independent measurement of the Surface Debye temperature based on the temperature dependence of the intensities. Instead the first x-ray studies relied on fitting diffraction beam profiles and relative intensities, the latter of which is structure and defect dependent. This first study was also done years before positron reflection studies found a much lower Debye temperatures  for the adatoms than anyone ever expected. [56] Bulk Debye temperature of crystalline Si are 535 $^o$K which corresponds to a mean square vibrational amplitudes,$<\mu^2>$  of  0.25A at room temperature [57,58] whereas, the first X-ray simulation [52] used   $<\mu^2>$ of 1.5A for the first two layers of atoms  and 0A ( i.e. rigid atoms) for the dimers.  Early  surface diffraction measurements determined surface Debye temperatures of 440-420 $^o$K [59,60] or $<\mu^2>$ of 0.38-0.37A .[59]  A  more accurate  study using reflection ( not diffraction) high energy position studies[56] revealed a value of 0.14A  for the adatoms.

A visual comparison of the TEM and these new X-ray diffraction intensities show some differences that may account for variations in the details of the DAS structures derived from them. This is not unexpected since the sample processing conditions were different and the diffraction in the TEM measurements is not purely kinematic.

The most recent x-ray data[54,55] taken with a superior x-ray source and with state of the art 7x7 sample preparation likely represents improved data. One of the surprising differences found in the author's analysis of this new data for the DAS model was that the compression of the atoms forming the 'dimer' bond was much less ~ 2.89A than that found earlier, ~2.49A. This corresponds to a significantly weaker dimer bond found earlier or than found in LD calculations. This new data, however, confirmed the unusually high adatoms height above the surface (even higher but within overlapping error bounds) as found in the first x-ray study.[52]

To show a comparison of the original TEM and two x-ray data sets, the intensities of a few rows of diffraction peaks have been extracted from the reported data and plotted for comparison. Fig. 19S (a) and (b) show a profiles of the original TEM and later x-ray diffraction intensities for two rows of peaks along the (0k) directions. The peaks selected are shown in (c) which are from the new X-ray diffraction intensities and calculated intensities for the DAS structure [60] using the original earlier Debye temperatures. ( It is should be noted that comparative intensity plots of experiment and calculated intensities that use half circles, provides an illusion of smaller intensity differences than their areas since the eye focuses on the linear differences at the boundary of these half circles. As a result the areas of these circles are used in comparing relative intensities in (b) and (d).)

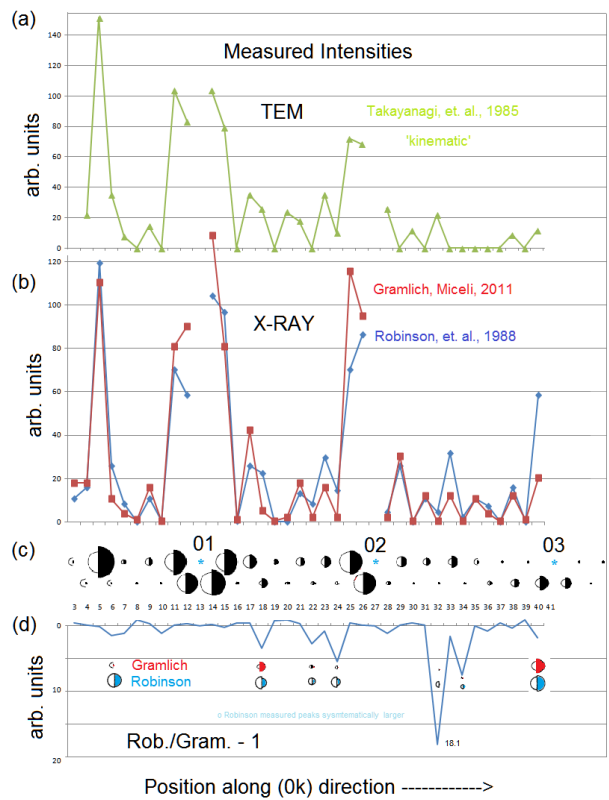

FIG. 19S: (color online) In plane diffraction intensities measure (a) by TEM under' kinematic' conditions, (b) x-ray measurements by Robinson, et. al. [52] and later by Gramlich and Miceli.[54,55] The order of the peaks shown in (c) are taken from a swath of the latest diffraction intensities [55] along the (0k) direction that include the next row of diffraction spots. Panel (d) shows $I_R / I_M -1$, a measure of how much lower the intensities are for certain peaks in the latest x-ray measurements.

In comparing the two sets of x-ray data, the newer data shows many of the second row peaks to be of significantly lower intensity than those of the earlier data. A plot showing these systematic low intensities is given in panel (d). Given the fact that the 7x7 in the early x-ray work was formed from sputter cleaning and only mild annealing of the surface, one expects this to be a less perfect surface than the later work which employing the standard STM cleaning protocols used to form more perfect 7x7's. This would also account for the need for large $<\mu^2>$ 's in these first xray measurements to reduce certain diffraction intensities to agree with the DAS simulated structure factors.

In further support of such disorder in the earlier work, it is noted that each of these higher intensity peaks occur in the direction of the rows of adatoms, thereby allowing adatom disorder to lessen destructive or constructive interference in these directions. This adatom disorder may also explain the differences between the optimized structures of this early x-ray structure and most recent calculated structure discussed in the main paper.

The mean square vibrational amplitudes,$<\mu^2>$, are useful to consider in more detail. The $<\mu^2>$ 's used in the early x-ray refinement for the bi-layer and adatoms were cited as 1.5± 0.6A, and for the dimers as 0.0 ±0.4 A. This choice thereby strongly suppresses the scattering contributions from the adatoms and atoms in the top bi-layer while the dimers atoms have no thermal suppression of their scattering. Bulk silicon has a room temperature Debye temperature of 531± 4° K or an RMS of ~0.25A. [56,57] One would ordinarily assume that the dimers have the bulk Debye temperature. The $<\mu^2>$ of the top layer of atoms used to optimize the fit meanwhile are 5 to 3.75x that of the bulk would scale these $<\mu^2>$'s to Debye temperatures [59] of 535/2.35 - 535/1.94 or 223 - 275° K, respectively. This represents a very large departure from the RHEED measured surface Debye temperatures of 420°K [59] or a $<\mu^2>$ of 0.37A. A most recently measurement of $<\mu^2>$ found 0.14A or a Debye temperature of 709° K.[56] These are high values but are expected to be more accurate than surface diffraction determinations since they are performed using positrons in a reflection mode. These authors conclude that because the positrons do not sample into the bulk, they primarily see the adatom's vibrations.[56] The x-ray Debye temperatures utilized were clearly unrealistic and, as noted by the authors, [52] selected to optimize the fit to the measured intensities. It is now clear that many diffraction peaks in this early work were affected by defects and likely compensated by the unphysical Debye-Waller factors used for different surface atoms.

Despite these unusual parameters and questions about surface perfection in earlier results, the overall intensity of all these x-ray diffraction patterns remain very similar. A central question is how do various periodic features in the these structures reflect themselves in the diffraction patterns. Clearly the use of rigid dimers has accentuated it's contributions to the calculated x-ray diffraction pattern used to match data.

In general, the different periodicities and atom displacements within this complex 7x7 structure will modulate one another to produce the complex variations observed. Taking a Fourier transform of the topograph will reveal the power spectrum arising from the different periodic features in the topographs [61] and may help understand the origins and various contributions to such diffraction pattern. Fig. 20S (a) shows an STM topograph of an average quality 7x7 surface and a close-up showing the quality of the STM image and atom resolved features.[61] The power spectrum of the Fourier transform of the topographs in (a) is shown in (b) which was then drift corrected. The diffraction pattern of the most recent x-ray study [54,55] that used similar STM clean procedures is shown in (c). (d) shows a recent high quality LEED pattern.[62]

The higher intensity of 0, 3/7 family of peaks, the so called pseudo 2x2 features, in the FT power spectrum is similar to the strong peaks seen in the xray pattern, but has an even better match to the peaks in the LEED pattern.  This includes  row of  nearly equal intensity peaks between the 0 3/7 and 3/7 0 features. The xray and LEED pattern show similar 3 fold symmetry and probably have different admixtures of different domains.

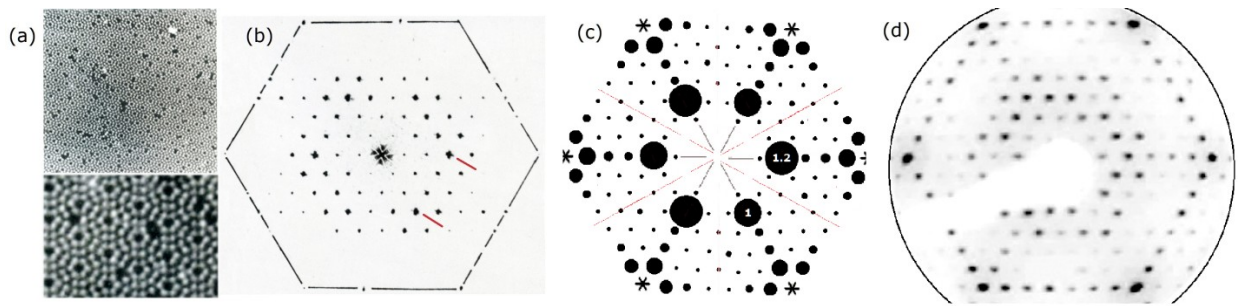

FIG. 20S:  Comparison of real and reciprocal spaces images of the si111 7x7 surface. (a) shows the large area STM topograph ( -2V, 1nA)  together with a close-up of the level of detail and noise in this topograph.  (b) shows the power spectrum of the large area image in (a) from 1988, (c) shows the x-ray diffraction intensities from 2011 and (d) is a LEED pattern obtained at ~125 eV in 2015 [62] all plotted on the same scale.

In the FT in (b) there is a  slight asymmetry between the vertical and the horizontal spot patterns that can be associated with edge effects in transforming the rectangular STM image even though the edges of the image are feathered  to reduce 'edge ringing".  Averaging the image right and left and up and down  produces near  perfect 6 fold symmetric image as seen in many of the older LEED patterns and in the early TEM and xray studies. The strong $1/d^2$ fall off in the FT power spectrum from the origin in (b) is expected [61] and leads to the weaker features for larger periodicities. The residual noise in the topographic data of Fig. 20S prohibits a simple correction to the power spectrum to relate relative intensities of the small and larger wavelength Fourier components.  Interestingly, the large area STM image shown in (a)  has comparable defect densities as in the images of the 5x5 or the 7x7 formed from by further annealing the 5x5 found in earlier. [6] ( See Figs.15S and 18S.)

Fourier analysis of such real space images with improved STM tips and the higher resolution possible in today's STMs may help distinguish the origin of certain diffraction features. By comparisons to more penetrating x-ray data one may be able to distinguish what diffraction features, if any, may be associated with possible dimers. Improved STM tips that can resolve the restatom[63] may also show whether and how the restatoms may contribute to the interference features found in the diffraction patterns.

The diffraction data intensities so far discussed arise from in-plane scattering and reflects predominantly the in plane geometry of the surface atoms, whereas the diffraction measurements along a truncated lattice rod provides depth information. This later was studied in 1991 on the same Brookhaven A16X x-ray beam line used in the first x-ray study but now with a UHV diffractometer.[53] Studies of these truncation rods were first performed in 1986 on 7x7 samples that had been encapsulated by a layer of amorphous Si, $\alpha-Si$ to allow the long data acquisition times required to measure these rods.[64] Since a 7x7 pattern was observed it was assumed that part of the underlying atomic arrangements of a clean 7x7 was intact.(namely the stacking fault of the 7x7.) The truncated lattice rods of the 7x7 were also examined in 2011 [54,55] using the much brighter APL x-ray source. A comparison of these three different truncation rod measurements are enlightening and again reveal a trend in the likely structural defects present in these three different truncation rod measurements.

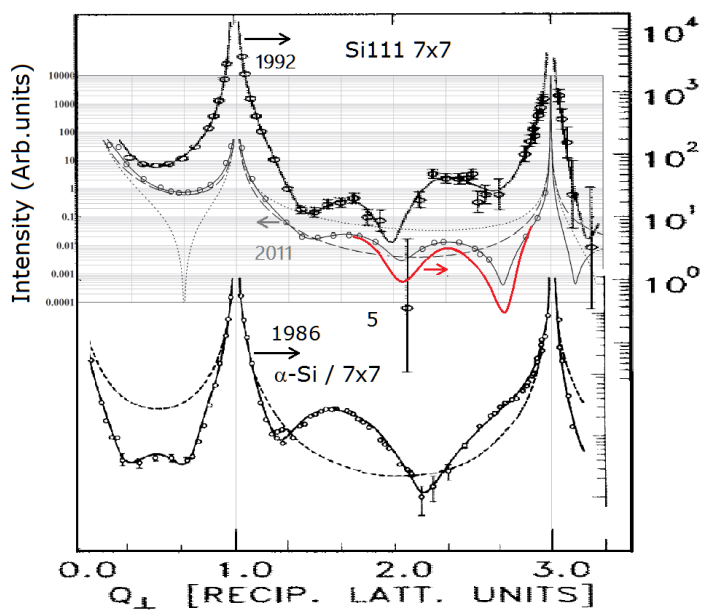

FIG. 21S: X-ray diffraction Intensity along the lattice rod up to the 33 refection from the 7x7 in 1992,[53] 2011 [54] and in 1986.[64] The layer dependent spacing fit to the data points is shown by the solid lines through all sets of data points. In the 2011 plot the dotted and dashed lines represent the two different terminations of an otherwise ideal Si111 surface while the red curve is plotted on the same scale as the 1992 and 1986 results. The error bars for the 2011 data are within each circle.[55] The 1986 curve at the bottom is from an amorphous Si capped 7x7 surface.

The results of these measurements are shown in Figure 21S. What is evident is the improved signal to noise in the data and reduced scatter of the data points using the brighter light source in 2011. The solid line through the data points represent the fit to a model structure of either sets of data as done earlier. The adatom height based on this original 1991 data [33] was 1.58 ± 0.2 A whereas the fit for the new 2011 data [33] is 1.80 ± 0.2Å. In the latter, only the spacing of the atoms in the first underlying bilayer were required to achieve this near perfect fit. The first bi-layer atoms was expanded by 0.08 ± 0.03Å and the lower bi-layer atom by 0.02 ± 0.03Å. Again, neglect of these small bilayer distortions out of the surface plane would appear reasonable.

The overall background differences between these two data sets shown in Fig. 21S reflects whether the instrumental / geometric correction factors were made to the data or in the calculated profiles.[55] While this new x-ray model provides a admirable fit to the data, more data points are needed near the low intensity regions to refine this fit, for example, the notch at 2.66 RLU. This feature is a key difference in these two data sets and cry out for more detailed measurements now possible with even brighter x-ray sources ! Never the less, the large adatom height found in this new measurement again agrees with other experimental determinations. In addition the longer dimer bond also found in this later x-ray study again raise key questions about the structure and reliability of the dimers found earlier.

The comparison of the truncations rods arising for the amorphous Si encapsulated 7x7x surface in 1986 [64] to the later 7x7 rods shows and interesting trend. The low intensity rod structure between 2-3 RLUs suggests that the 1991 sample structure is somewhere in between the 2011 and 1986 rod profiles. An increasing degree of contamination and/or defect formation in these three x-ray studies is a reasonable possibility. Reiterating, the 1989 studies required long exposure of the clean 7x7 surface under moderate UHV conditions for data acquisition which corresponds to ~ 17 Langmuir exposure to contaminants; and if the ion gauge was left on, to ~ 2 monolayers exposure to atomic hydrogen. The later x-ray 1991 studies were performed on the same beam line and vacuum conditions as in 1989. Although the data acquisition time was never reported for the 1991 studies, one expects long measurement times due to the low signals along the truncation rods, as indicated by the large error bars between 2-3RLU.

As discussed earlier the observation of a 7x7 pattern does not mean that the surface structure is the same as when it is formed in UHV. In particular, for the amorphous Si encapsulated 7x7, some periodicity of the corner holes will likely remain after the adatoms or even the stacking fault are chemically modified as is observed in low temperature Si epitaxy studies on the 7x7 by STM.[7] The adatoms are the most likely point of chemical modification since they have the highest lying electronic states at and near $E_F$. Thus the trends observed in these three truncation rod profiles suggest that the most recent 2011 7x7 x-ray study likely has the best 7x7 surface and the early 1985 like the encapsulated surfaces are more defective. The 1991 x-ray profile appears to be somewhere in between. One must conclude that the 1991 rod truncation study likely has significant defects as did the original 1989 x-ray study.

A new analysis of this and the original x-ray data is underway to understand how this new data may affects the refinement of the in-plane arrangement of the atoms as well as whether the DFA structure(s) fits any model.[60] To date this uses the same assumption of these earlier models including the same Debye Waller factors and 6 fold symmetry.  Currently, the original optimized Takayanagi model produces a 30% better fit to this new x-ray data than the DFA model shown in Fig. 1S. Using the original x-ray parameters for the DAS model, this new data still provides a satisfactory fit to the DAS model with a $\chi^2$ of 4.9. The 'optimized' fit found for the original x-ray data [31] was much better with a $\chi^2$ fit 1.6 , but as noted earlier, was based on a less perfect surface and very unphysical values of $<\mu^2>$ for the adatoms and dimers..

These new x-ray measurements and analysis [60] are also being used to consider and optimize the DFA model. The simplest DFA modeling assumes 6-fold symmetry to compare to the original x-ray analysis and after simulated annealing found the best $\chi^2$ of 8.2. The resulting structure is physically unrealistic, but however, suggests other variations to the DFA model not shown in Fig. 1S. One of the clear limitations in this present preliminary analysis is that 6 fold rotational symmetry has been assumed. As discussed earlier this may not be significant for the DAS but is for the DFA which has explicit 3 fold symmetry.  It should be noted that later work by the authors of the original x-ray work [10] later confirmed that the 7x7 did not have 6-fold symmetry and  concluded:

*"This asymmetry will result in modifications in the structure models of Si(111)7x7 to include p3m1 symmetry in place of the current p6mm. "*

As already noted, such 3-fold symmetry was found earlier in the SHG work on the 7x7 and can be seen in the LEED pattern shown earlier in Fig. 20S.

HIGH RESOLUTION TEM IMAGING OF THE 7x7:

Image reconstruction using HR-TEM is a very complex technique which has also evolved and dramatically improved over the past 30 years.[65] There are many problems and pitfalls in applying this technique since the real space image from the diffracted electrons must be reconstructed after all phase information from scattering is lost.  Some of the peculiarities of interference effects found in the previously discussed Patterson analysis can also occur here as well.  In addition, imaging occurs with 'imperfect' electron optics lenses and very weak signals.  In this approach one also has to worry about interferences between the front and back surfaces, i.e. the role of sample thickness as well as secondary electron scattering - particularly important where defects or domains exist.  Thinning of TEM samples usually produces bowled regions which create preferentially aligned steps. Nevertheless, pioneering work was performed on the Si111 7x7 surface that produced plan view, phase contrast HR TEM images consistent with the DAS model.[51]

These HR TEM studies [51] were performed with one of the first UHV TEMs, the JEM-2000FXV, operated at 200 kV (~ resolution 2 Å). The vacuum around the specimen during cleaning was in the $10^{-8}$ Torr range, and during observation 'in the $10^{-10}$ Torr range'.  In comparison current HRTEM technology [65], both experimental capabilities and theoretical analyses, have greatly improved.  In retrospect, there are ample reasons to consider the conclusions reached on the limited and highly processed original TEM data as preliminary. As the authors originally state:

*"in the future, with better data, it may be possible to improve on our method and use lower symmetries. The second point concerns our neglect of focus differences between the top and bottom surfaces."*

Unfortunately, the shift in funding agencies from silicon surfaces to other materials when improved HR TEM capabilities were realized, prohibited further work on the 7x7 surface to resolve these questions. [66]

The following discusses some of the currently known shortfalls of these early plan view HRTEM measurements to image the surface atoms of the 7x7 based on current more recent HRTEM methodology. [64] Based on the data obtained, and known interference effects possible, one can speculate on why these early methods validated the DAS structure for the 7x7.

The original HR TEM data was very noisy and had to be extensively averaged.[51] Also In this early HR TEM work two defocus conditions were used. The 35nm defocus yielded experimental features suggestive of the DAS structure while the 135nm defocus condition did not. No rational why one defocus condition was favored over others or why the 35nm condition was preferred or unique, except that it provided features consistent with the DAS model. Simulations of the widely accepted DAS structure for both defocus conditions [51] each showed reasonable correspondence to their defocused experimental TEM image. However only the 35nm defocus bore any resemblance to the expected 7x7 DAS structure. In addition, none of the other existing adatom models were considered to understand the specificity of this early HRTEM modeling to adatom or structural variations. As the case for the original Patterson function analysis of the 7x7 TEM diffraction intensities [1], the DFA model was unknown and never considered.

When viewing surfaces in conventional plane view phase contrast mode, HREM mode, whether on or tilted off zone, the bulk lattice constant dominates the contrast, top–bottom surface scattering interactions arise and significant image processing is required. [51, 65] In current "state of the art" HRTEM analyses, many more features are considered including: (1) a full imaging series covering a wide range of focusing conditions, (2) more detailed evaluations of inelastic scattering, (3) evaluations of front and back contributions, i.e. sample thickness, and (4) simulations of alternative models. [64] Unfortunately, such considerations were never pursued for the 7x7. [66]

As discussed next, the periodic features in the DAS or DFA structure, such as the adatoms and atom columns, can also produce additional local minimum as well as local maximum in the TEM image due to the nature of interference and fringing from these scattering centers. These have implications to image interpretation thereby making full HRTEM simulations a necessity, not to mention the afroementioned considerations necessary for a successful analysis. [65]

The original data [51] was very noisy and had to be averaged: first, by three fold rotational averaging, and then by averaging using 6 fold as well as translational symmetry. Such averaging has to be detrimental since it is now clear that the 7x7 exhibits 3m symmetry. [8,10,62] The TEM images also shows regions where aberrations are evident, thereby causing further degradation of the information. Fig. 22S shows a portion of the least averaged data that used three fold averaging. A honeycomb lattice is superimposed on it. This represents a region of the published image with the lowest apparent aberrations. The corner hole-like features that define the 7x7 unit cell are also relatively weak and noisy, but evident.

The blue tick marks along the center of the unit cell shown in Fig. 22S show unequal spaces along the DFA honeycomb where no adlayer atoms occur, but do occur in the substrate layer below as in the DAS model. As shown on the right side of Fig. 22S with blue lines, many of these features along the unit cell boundary correspond to periodic repeats or fringes/ghosts of other periodic features. In this case it appears to be from the adatoms. The wave fronts from these adatoms when collected over a limited angular range can produce fringing from each other along the direction of their wave vector, analogous to the periodic fringe features of Young's two slit experiments [67] or the ghost features that can occur in Patterson analyses. [68] Such fringe effects are complex and can have many other contributions. Such fringe/ ghost features in the TEM image will complicate determining the positions of the atoms along the domain boundary for either DAS or the DFA structure.

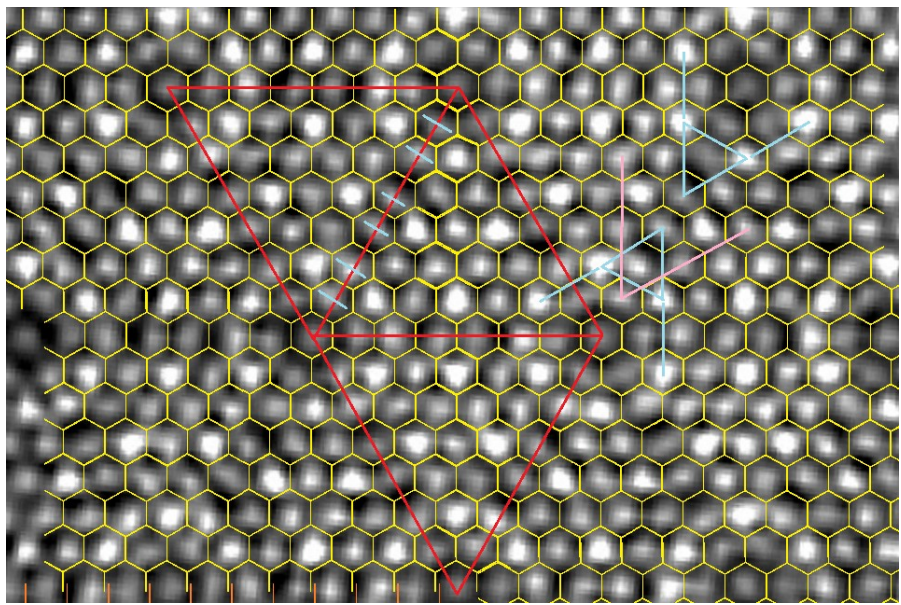

FIG. 22S: Three fold averaged High Resolution STM image for 200 KeV ( $\lambda$=0.25 pm, / .00025nm ) with a 35nm defocus. [51] ( Used with permission of the authors.) Here, the intensity scale has been inverted to aid visual comparisons to prior STM topographs, making atoms or atom columns light. The bottom portions indicates the shifted cells toward the image edge indicative of aberrations in this image.

A more sensitive color rendition of this grey scale image is shown in Fig. 23S. Variations in this image can be associated with noise, but discernible systematic variations are also found. These arise in the atom pair like features along the domain boundary separating the two side of the unit cell and can be counted. These variations when compared to the DAS model are found to be of a non-statistical nature. These features are circled and color coded as being of nearly equal intensity ( pink) or asymmetric intensity (white) pairs. Statistical noise should produce a 1: 3 ratio of the edge atom features to the center atom features. However, the unequal edge atom features occur 74 % of the time. The equal features occur 35% of the time relative to the unequal edge atoms which is close to the statistically expected 33%. Since the unequal features occur randomly around one side of the corner hole, it is unlikely they are caused by ghost features noted earlier. These particular defects as well as possible ghost peaks will impact the structural determination by HRTEM and most certainly will complicate the HRTEM validation of dimers that exist in the DAS model along the unit cell boundary.

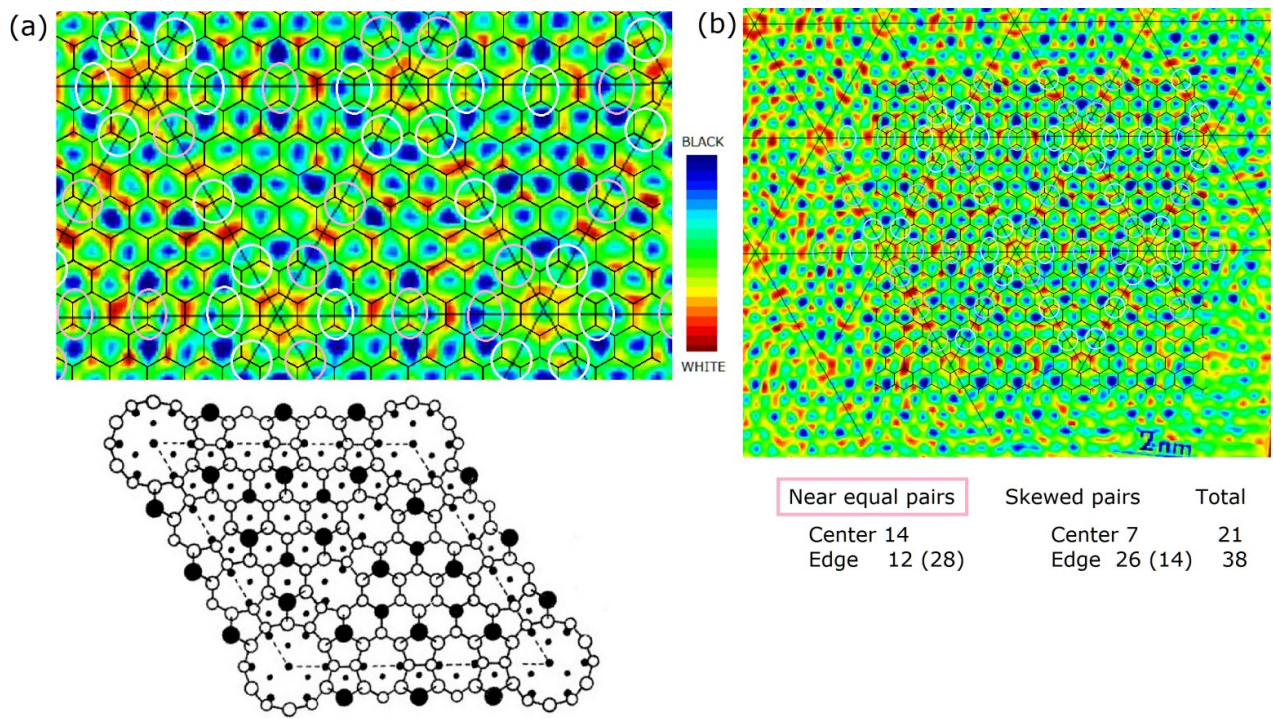

FIG. 23S: Color mapped HR TEM image of the three fold averaged data [51]. (a) is an enlarged region of (b) where all pair-like features of dimers (expected in the DAS model) are circled. White circles indicate skewed pairs and pink circles indicate near equal pairs. The statistics are shown below image (b) where the number in parenthesis reflects the expected count for each type of pair.

This occurrence of unusual features near the corner hole in these HRTEM images parallels what is observed and shown earlier in STM images of missing adatom near the corner hole (Fig. 13S-16S) and ) or missing filled state charge density in these adatom (Fig. 17S). Organic contamination from the vacuum chamber or boron from observation windows, are likely present in these early experiments and can be mobilized during annealing when these structures are formed. Thus, these samples may not reflect an ideal pure 7x7 surface but one which has been stabilized by induced impurities or defects in the surface region, now apparent around the atoms near the corner hole. Any symmetry averaging of such data will incorporate these irregularities into the model of an assumed 'ideal' 7x7 structure.

These measurements and analyses were repeated in a new microscope [69] with better signal to noise but the same 6 fold as well as translational averaging was used, and in addition, statistical filtering .[70] This produced a much cleaner image that is shown in Fig. 24S. Again such averaging throws away symmetry features of p3m1 symmetry by assuming p6mm. In this figure one can also more accurately pinpoint the various Youngs / ghost fringes associated with other strong periodic features on the sample and is shown in Fig. 24S.  Here,  the tick lines with on the left dark regions which are periodic and can account for the darkened areas along the domain boundary. The corresponding lines on the right correspond to opposite phase (light) features. Either constructive or destructive interference can contribute to any feature along the unit cell boundary, particularly when a faulted (antiphase) boundary occurs.

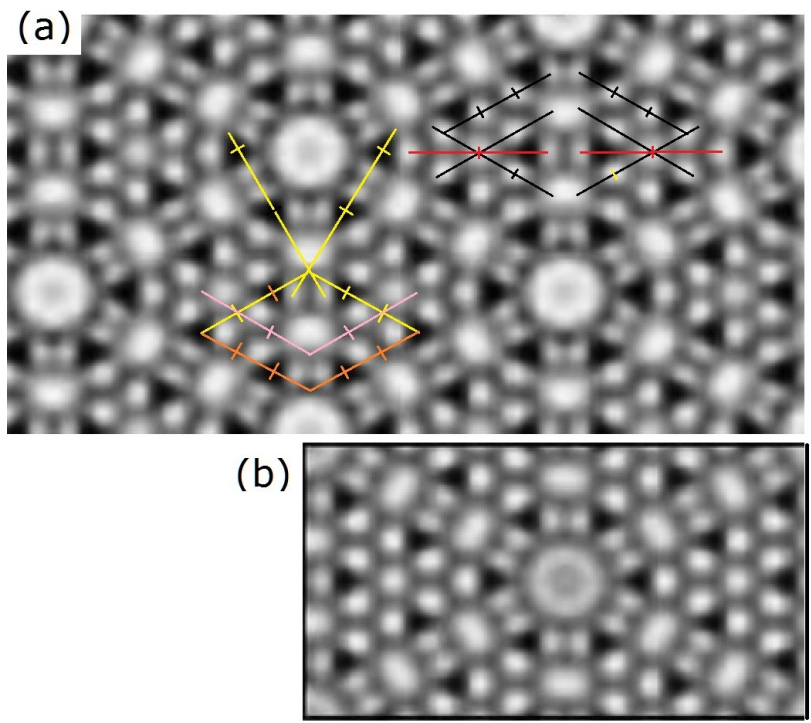

FIG. 24S: The fully averaged HRTEM image (a) and a simulated image in (b)  (used with the permission of the authors.)  In (a) the left lines represent periodic dark features that can be projected onto the region of the domain boundary  while those on the right reflect  projections of lighter features - the two phases from collective interference. (b) represents a simulated image based on the DAS model.

HRTEM image simulations shown in Fig. 24S (b) confirms the overall features found earlier [51] but now with improved signal to noise. A color mapping of this image is shown in Fig.25S and exhibits these features more clearly.  Here the DFT calculated DAS structure is superimposed on the processed experimental data in (a). The inset in (b) again shows the author's [69] simulated/calculated DAS HRTEM image.  Here, the higher density features along the domain boundary correspond qualitatively to the 'calculated dimers' in the simulation and as expected in DFT calculations, but they differ quantitatively. Using the centroid of the observed features to denote the atom center, one finds unequal atom pairs  in the  TEM experiment of 2.92 and 2.24 A.  In contrast, the TEM simulation gives fixed dimer bond distances of 2.32A versus Takayanagi's dimer of 2.56A. [1]  The DF calculations show atom pairs separated by  2.44 and 2.34A, very close to a bulk 2.35 Si - Si bond distance.   Then in 2011, the improved x-ray data indicated a DAS dimer bond length of 2.89A. The large dimer bond near the corner hole of Fig 25S could be real or arise  from possible defects that appear to arise around the corner hole.  However, given all the aforementioned issues in the present HR TEM analysis, extracting atomic positions along this unit cell boundary must be considered an over interpretation of the data, including whether DAS type of dimers even exist at all !

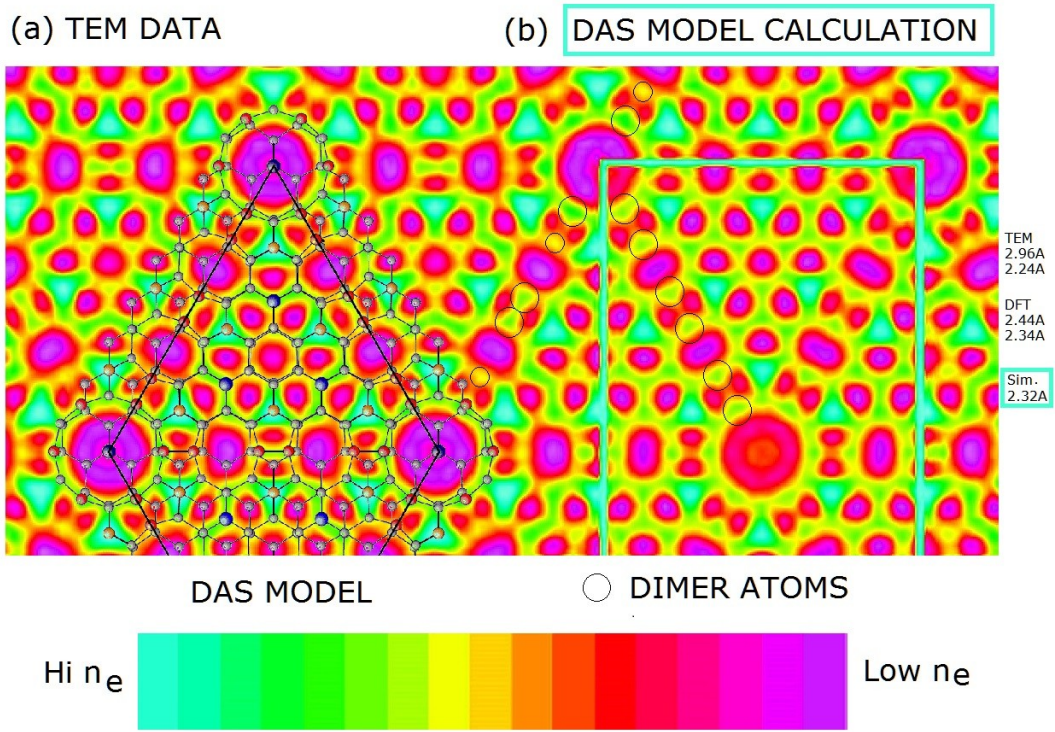

FIG. 25S: (Color online) Color mapped fully averaged HRTEM experimental image with (a) the DF calculated DAS structure [2] overlaid as well as a simulation of the HRTEM image for DAS . The atom pair bond lengths from the TEM image, DF calculations and from the simulated image are indicated on the right for comparison. The circles along each unit cell boundary show regions of higher density.

It is instructive to compare this early Si111 7x7 HR TEM analysis to more recent extensive multi-slice HRTEM plan view phase contrast analyses of the c(6x2) SiTiO3 surface.[65,71] This surface has some interesting features which mimic those of the 7x7 surface. Fig. 26S shows some results of this HRTEM study with top and side views of the structure in (b) as found from STM studies and widely accepted.[72] These HRTEM studies considered 41 defocusing conditions and also modeled the effects of front and back interference, i.e., sample thickness in the simulated TEM images.

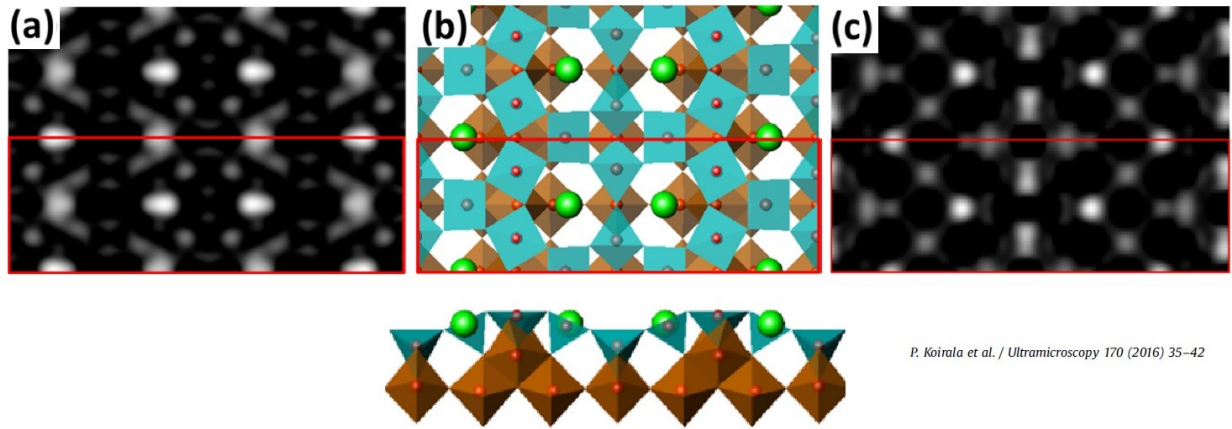

FIG. 26S: HRTEM simulated images of the c(6x2) SrTiO3 surface[65] in (a) and (c) compared to the accepted surface structure in (b).[72] (a) uses a 4.51 nm thick sample and 6 A defocus. 41 defocus slices were obtained in -1.05nm increments on the TEAM 0.5 instrument (FEI Titan-class) at the National Center for Electron Microscopy. Simulations also include an $<\mu^2>$ vibrations of 0.4 Å .

This structure is interesting in that two unit cells possess similar reflection symmetry/anti-phase zig-zagged arrangement of atoms down the center of Fig. 26 (b) as is found in the unit cell of the 7x7 structures. The arrangement of the Ti atom features shown in (b) is experimentally observed and found in the simulated HRTEM images of (c) where (c) is for the thinnest sample simulated ( 4.51 nm) using a 6A defocus. However, in (a) a extra feature between ' Ti' atoms is observed corresponding to periodic features of these Ti atoms. Under certain defocusing conditions this extra feature disappears to reveal predominantly the Ti atoms, but it also produces the light and dark alternating structures along this 'anti-phase' boundary'. These features in (a) suggests atom pairs, i.e., dimer-like features. However, there are no known dimers in this particular structure! [71] Instead, the interference features from the complex phase interference for this anti-phase arrangement of atoms would appear to produce different ghost fringes that produce these additional dimer-like features in either (a) or (c).

Thus, it is possible that the ghost features from interference from the adatoms of the 7x7, in addition to possible defects in these corner hole atoms, may create the appearance of "dimers" in these early HR TEM results. However, in defense of the presence of dimers there are recent STM measurements at T=78 °K that also show features suggestive of a dimer.

RECENT MEASUREMENTS OF THE SURFACE CHARGE DENSITY ALONG THE DOMAIN BOUNDARY:

Fig. 28S shows scanning tunneling spectroscopy, STS, images of the 7x7 (a) obtained at 78 °C using a cluster atom tip and lock-in techniques [62] and (b) those calculated [73] using a realistic cluster atom tip. However, the calculated STS image is based on the differential tunneling from a reference profile of the 7x7 at 1.75 eV where the two sides of the unit cell appears largely flat. This image shown has been contrast enhanced since it has the smallest range of intensity/DOS of all the calculated images. It also occurs over a very narrow energy range ( ± .081 eV). This means that the calculated or measured DAS STS images are susceptible to any errors in how it is derived or measured.

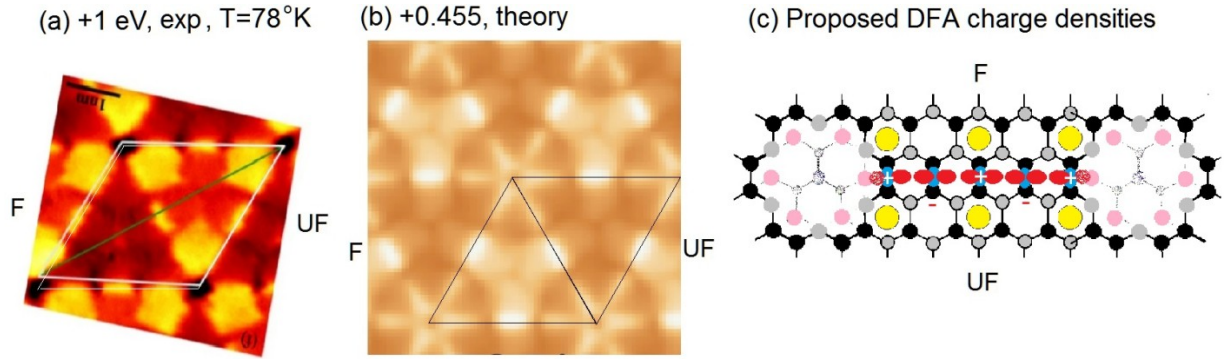

FIG. 27S: (a) Drift corrected STS image of the 7x7 at +1 eV using lock in techniques [63] compared (b) calculations of an STS image at +0.455 eV using the reference topological profile taken +1.75 eV.[73] (c) shows a schematic of the DFA structure along the unit cell boundary where the blue or red lobes represent filled/bonding and anti-bonding states of the atom pairs. The + ( along the cell boundary) and - indicates possible changes in phase or small undulations in the heights of each pair. The faulted and unfaulted sides of the unit cell are indicated in all.

These two images have many similar characteristics but are different. The +0.455 eV calculated image if corrected by x1.6 - 1.7 for self- energy corrections, corresponds to a corrected value of 0.730-0.77 eV, which is close to the experimental value. As noted earlier, non-equilibrium carrier dynamics [47] at low temperature can create band bending that may shift the experimental scale up by ~0.25 eV. Thus it would appear that these two images can be directly compared.

The calculated DAS image has charge density built up along the boundary of the unit cell that parallels the location of the dimers in the DAS model. However, the empty state image of dimers should generally not reflect the positions of the dimer bonding charge as the empty dimer state should be orthogonal to the filled dimer state! A more likely alternative is that the calculated dimer like charge density in (b) reflects a complex combination of states/waves producing this CD and may have little to do with the dimer bond charge per se.

The experimental image also shows the same central feature along the unit cell boundaries, but differs markedly from the calculated image around the corner hole. Here, an inversion of the contrast between the faulted and unfaulted sides of the unit cell with three fold symmetry occurs. Again this may arise from the very small changes in the tunneling arising and the manner by which these different images were created. However, the 3 broad experimetnal features around the corner holes on the unfaulted side of the unit cell are where a corresponding delocalized empty (anti-bonding) FSS is likely to occur.

The CD for the DFA model can be considered as follows. The (blue) bonding and (red) anti-bonding CD's associated with the DFA atom pairs shown in Fig. 28S (c) can account for the experimentally variations of the CD's along this boundary. This was discussed earlier more generally in terms of the small depressions in the CD along this boundary that arises from the experimental CD contours.  Here, the pairs of atoms of the honeycomb reflect their (filled) bond charge (in blue) while the empty bond charge (in  red) are rotated 90° to maintain their orthogonality to the honeycomb pairs. As the Patterson analysis suggests, these different pairs along this boundary also appear to be phased differently relative to each other and/ or pushed slightly  up and down as indicated by  + and -  due to their lateral interactions and small standing wave phase changes of these states with the adatom states.

This can account for any features arising from the atom pairs of the DFA along the unit cell boundary as also seen in Fig. 27 (a).  As found for the monolayer silicene structures such irregularities in these atom pairs will likely decompress the smaller buckling arising from increase $sp^2$ hybridization in the interior regions of the honeycomb.

The DFA atom pairs next to the corner hole can be part of the silizene CD  that is not only localized around the corner hole atoms but also part of a delocalized anti-bonding FSS state in this region. This experimentally observed unoccupied CD arises on the unfaulted side of the unit cell and is opposite to the filled FSS state which arises predominantly on the faulted side.  Such differences in the location of the occupied FSS and its unoccupied FSS counterpart minimizes its overlap and is expected in wave-mechanics.

One balance,  there is strong evidence that the DAS model based on early experimental work presents ample ambiguities when the totality of results are examined in detail.  Many important structural and surface state features calculated for the DAS model simply fails to explain several experimentally measured structural features as well as one of the observed surface states. The DFA structure appears to offer an explanation for these deficiencies. What remains to be understood is the nature of the interactions arising at the 7x7 surface that favor this unusual bonding and the departure of the DFA structure from a covalent bulk-like surface structure.

-------====-------